\documentclass[aps,pra,showpacs,twocolumn,superscriptaddress, nofootinbib,10pt]{revtex4-2}

\usepackage{amsfonts} 
\usepackage{amsmath} 
\usepackage{amssymb} % for math symbols
\usepackage{amsthm} % for theorems
\usepackage{graphicx}
\usepackage{dcolumn}
\usepackage{xkcdcolors}
\usepackage{tabularx}
\usepackage{array}
\usepackage{epstopdf}
\usepackage{xcolor}
\usepackage{mathrsfs}
\usepackage{mathtools}
\usepackage{booktabs}
\usepackage{comment}
\usepackage{cancel} %to strikeout
\usepackage{ragged2e}
\usepackage{physics}
\usepackage{hyperref}
\usepackage{pgfplots} % for plots in latex
\pgfplotsset{compat=1.18} % Or your current version
\usepackage{eqparbox} % fancy plotting
\usepackage{url}
\usepackage{cleveref}
\usepackage{tikz}
\usepackage{listings} % for intro python code

\usepackage[caption=false]{subfig} % for figures

\theoremstyle{definition}

\usepackage{silence} % silence a warning
\hypersetup{colorlinks=true, citecolor=orange, urlcolor=blue, linkcolor=magenta}

\newlength{\temp}
\newcommand{\FixedSize}[2]{\makebox[#1][l]{\ensuremath{#2}}}

\graphicspath{{./images/}} % explicit the graphic path

\usepackage{silence} % silence a warning
\begin{document}

\title{Renormalizable graph embeddings for multi-scale network reconstruction}

\author{Riccardo Milocco}
\email[Corresponding author:]{diego.garlaschelli@imtlucca.it}
\affiliation{IMT School for Advanced Studies, Piazza San Francesco 19, 55100 Lucca (Italy)}
\affiliation{ING Bank N.V., Bijlmerdreef 106, 1102 CT Amsterdam (The Netherlands)}
\author{Fabian Jansen}
\affiliation{ING Bank N.V., Bijlmerdreef 106, 1102 CT Amsterdam (The Netherlands)}
\author{Diego Garlaschelli}
\affiliation{IMT School for Advanced Studies, Piazza San Francesco 19, 55100 Lucca (Italy)}
\affiliation{Lorentz Institute for Theoretical Physics, Leiden University, Niels Bohrweg 2, 2333 CA Leiden (The Netherlands)}

%\date{\today}

\begin{abstract}
In machine learning, graph embedding algorithms seek low-dimensional representations of the input network data, thereby allowing for downstream tasks on compressed encodings. 
%When the input data is a graph, embeddings can be obtained for global network properties as well as for individual nodes. 
Recently, within the framework of network renormalization, multi-scale embeddings that remain consistent under an arbitrary aggregation of nodes onto block-nodes, and consequently under an arbitrary change of resolution of the input network data, have been proposed. 
Here we investigate such multi-scale graph embeddings in the modified context where the input network is not entirely observable, due to data limitations or privacy constraints.
This situation is typical for financial and economic networks, where connections between individual banks or firms are hidden due to confidentiality, and one has to probabilistically reconstruct the underlying network from aggregate information.
We first consider state-of-the-art network reconstruction techniques based on the maximum-entropy principle, which is designed to operate optimally at a fixed resolution level. 
We then discuss the limitations of these methods when they are used as graph embeddings to yield predictions across different resolution levels. 
Finally, we propose their natural `renormalizable’ counterparts derived from the distinct principle of scale invariance, yielding consistent graph embeddings for multi-scale network reconstruction. We illustrate these methods on national economic input-output networks and on international trade networks, which can be naturally represented at multiple levels of industrial and geographic resolution, respectively. 
\end{abstract}

\maketitle

\section{Introduction}

Complex networks offer a solid framework for modeling a wide range of systems that support societally relevant phenomena, from economic activities to ecological processes \cite{2005_ScaleFreeSelfSim_Song, 2023_LPRecSysTransData_Yilmaz, 2018_MultiScale_Unfolding_GarciaPerez}. 
Basically, any set of pair-wise interactions can be represented via a graph, by properly defining the constituents as nodes and their connections as edges. For instance, the Input-Output network (ION) \cite{2015_WION_Cerina} represents flows (links) between sectors (nodes) of an economy, the World Trade Web (WTW) \cite{2010_WTW_Fagiolo,squartini2011randomizingI,squartini2011randomizingII,mastrandrea2014reconstructing} considers countries as nodes and international trade flows as edges, and food webs \cite{caruso2022fluctuating} represent flows of energy among species in an ecosystem.
Network science has developed a general toolkit for analysing and modeling such diverse real-world networks, often in synergy with several field-specific disciplines. Over the last few years, these interactions have expanded to the domains of machine learning and artificial intelligence. 
In particular, similarities between certain classes of network models introduced in statistical physics and so-called graph embedding techniques have clearly emerged \cite{2020_impossibility_of_low_rank_red_Seshandhri,2020_LPCA_Chanpuriya,2021_symLPCA_Chanpuriya,2024_MSNE_milocco}. 
While typically developed with different goals in mind, these techniques try to encode the observed structure of a given input network onto a low-dimensional latent space, such that e.g. the original data can be compressed, downstream operations can take place in the latent space representation, and the original system becomes endowed with a simple (random) model from which alternative but realistic instances of the network can be sampled.

Recently, an important question has been raised in the context of Network Renormalization \cite{gabrielli2025network}, namely how to consistently transform the description of a network upon a change in the chosen resolution level -- which is ultimately arbitrary or dictated by data availability, thereby not necessarily reflecting any intrinsic property of the physical system.
In particular, real-space \cite{2018_MultiScale_Unfolding_GarciaPerez} (geometric) and dual-space \cite{2023_LaplRG_Villegas} (Laplacian) approaches to network renormalization have been proposed, respectively generalizing Kadanoff's \cite{1966_Kadanoff_RG} and Wilson's \cite{1971_Wilson_RG} approaches from regular lattices to heterogeneous networks.
In the framework of random graph models, network
renormalization leads to the requirement of invariance of a network model with respect to the chosen
resolution level \cite{2023_MSNR_Garuccio,gabrielli2025network}. In particular,
since the constituent nodes of a network can be aggregated arbitrarily onto super-nodes or `blocks', one can formulate a rigorous requirement of invariance of the functional form of the probability distribution of a random graph model under arbitrary node aggregation \cite{2023_MSNR_Garuccio,2024_RecMSM_Lalli,avena2022inhomogeneous,2024_MultiScaleNetRec_Ialongo}. In the context of models with independent edges, this requirements leads to a unique multi-scale model (MSM) for the  probability of connection between nodes and to specific rules for the renormalization of the associated model parameters \cite{2023_MSNR_Garuccio,2024_RecMSM_Lalli}.
The consequences of this result for graph embedding techniques used in machine learning have been recently highlighted: namely, while state-of-the-art node embedding models provide no guarantee of the consistency of the vector parameters representing nodes in latent space as the network resolution is varied, the `multi-scale' network model mentioned above allows for the formulation of an invariant node embedding method \cite{2024_MSNE_milocco}. For the first time, this method allows for the extraction of node embedding vectors that, when a set of nodes is merged into a block, transform additively into their vector sum over the set.
Convenient consequences of this result are the interpretability of the (otherwise obscure) vector sum in latent space and the possibility of connecting embeddings corresponding to different resolution levels -- without the need to run the embedding algorithm anew.
When applied to input-output networks defined at different taxonomic levels of the industrial classification of economic sectors, multi-scale embeddings have shown improvement over alternative approaches \cite{2024_MSNE_milocco}.

Building on these recent results, here we address the problem of consistent multi-scale node embeddings in a modified context, i.e. 
when data limitations or confidentiality imply that the input network is not entirely observable. This is the typical situation e.g. in financial and production networks, where privacy requirements make the individual connections unobservable, and only aggregate information is available. 
For example, accessing `microscopic' firm-to-firm networks would reveal trading partners and potentially the business strategies of each actor in the economic system. As a result, publicly available network data are often coarse-grained versions of the underlying microscopic networks, where the constituent firms have been aggregated either geographically into regions or countries \cite{2002_Gleditsch} or industrially into sectors \cite{OECD_website}. 
Reconstructing the underlying firm-to-firm network from partial or aggregate data remains an open challenge, and several network reconstruction methods have been proposed for this purpose \cite{2018_Review_RecMeth4Net_Squartini,2021_ReconstructingNetworks_Cimini,2024_Rec_Supply_Chain_Mungo}.
The importance of this challenge lies in the fact that the (reconstructed) networks are used for downstream tasks such as stress tests, measurements of systemic risk, or simulations of the propagation of shocks in the economy: if coarse-grained networks are used for these tasks, the estimated systemic risk can over- or under-estimate the risk calculated at the finer scale at which the actual stress propagation takes place \cite{2022_ESRI_Diem}. 

Our goal is proposing node embedding methods that are both consistent upon node aggregation and operable when the input graph is not fully observable, thereby connecting graph embedding methods with network reconstruction techniques. We will initially consider state-of-the-art network reconstruction methods \cite{2018_Review_RecMeth4Net_Squartini,2021_ReconstructingNetworks_Cimini,2024_Rec_Supply_Chain_Mungo}, investigate their (lack of) consistency under node aggregation, and finally recast these methods into a multi-scale framework \cite{2024_MSNE_milocco} where the fitted model parameters renormalize consistently upon coarse-graining. 
In particular we start with two widely used reconstruction models: the Configuration Model (CM) \cite{2004_StatMecNet_Park} and the fitness-induced Configuration Model (fitnCM) \cite{2004_Fitness_WTW_Garlaschelli, 2015_CimiModel_Cimini}. Both models belong to the family of Exponential Random Graphs (ERGs), whose functional form is derived from the principle of maximum entropy (i.e., minimum bias) under given constraints, which is optimal for a fixed resolution scale. Specifically, the constraint for the CM is the degree sequence (number of connections per node), while for the fitnCM, it is the total number of links.
Consequently, the fitnCM is more privacy-oriented, as it depends on a single global constraint and ascribes all the node-level heterogeneity to some observed fitness values (the so-called `fitness ansatz'). These two models are widely applied in the field of network science \cite{2018_Review_RecMeth4Net_Squartini,2021_ReconstructingNetworks_Cimini,2024_Rec_Supply_Chain_Mungo}, motivating our choice to use them as benchmarks.
%One drawback of the flexibility in network modeling is that, even when describing the same generative process, there is no unique way to define the nodes involved. By recalling the ION, one researcher may have access to the data at the most granular level (e.g., National Industry), while another to a more aggregated classification (e.g., Industry). These two graph representations exhibit different topological properties, but it is worth mentioning that the coarse-grained network is \textit{uniquely} derived from the microscopic one by defining a partition $ \Omega$ of microscopic nodes into blocks. This procedure can arbitrarily iterate, yielding a \textit{multi-scale} representation of the original network with \textit{nested} partitions.
We will show that, since the CM and its fitness-based variant (fitnCM) operate at a single resolution level (just like most network models), they provide an optimal reconstruction only at that level. If nodes are grouped into blocks, the parameters of these models must be re-estimated, because the parameters obtained at different hierarchical levels ($ \ell = \left\{0, 1\right\}$) are not explicitly unrelated to one another. We refer to this class of models as single-scale models (SSMs).
%To lack of generalizability of SSM has prompted the network science community to extend the Kadanoff's seminal work on renormalization in lattice systems \cite{1966_Kadanoff_RG} to graphs domain. One notable result is the renormalizable hyperbolic model \cite{2018_MultiScale_Unfolding_GarciaPerez}, an Exponential Random Graph (ERG) that supports renormalization, but only through a series expansion and under the constraints of homogeneous partitions. To overcome these limitations, the Multi-Scale Model (MSM) \cite{2023_MSNR_Garuccio} was specifically design to be generalizable at multiple scales and applicable to arbitrary partition of the microscopic nodes.  In this framework, the scalar parameters of each block are uniquely derived by summing the parameters of its constituent nodes—a principle referred to as the renormalization rule. For this reason, the summation can be interpreted as the analogue of the aggregation process, but in the parameter space.
Next, adapting the multiscale approach derived in \cite{2024_MSNE_milocco} for the fully observable case, we introduce the multi-scale counterparts of the above models, denoting them as the degree-corrected multi-scale model (degcMSM) and the fitnessMSM (fitnMSM) respectively.
%While the fitnMSM has already been analyzed in \cite{2023_MSNR_Garuccio}, this work marks the first application of degcMSM to real-world datasets. Specifically, all t
These models, which are derived from the principle of scale invariance, are first evaluated at the native scale where parameters are identified (the microscopic level) and then at coarser scales, through the renormalization (summation) rule on the parameters themselves. This evaluation allows a fair comparison between the effects of the two underlying principles (maximum Shannon Entropy and scale-invariance) from which SSMs an MSMs are derived. %Unlike ERG models, the MSM framework does not guarantee that the maximum of the likelihood corresponds to the degree-corrected solution (cfr. the CM \cite{2011_AnalMax_Squartini}). Therefore, we also display the results for the MSM parameters being estimated through the maximum likelihood (maxlMSM). However, this variant requires the knowledge of the full adjacency matrix, which is not compatible with the network-reconstruct settings, we have discussed so far.

It should be noted that community structures (intended as partitions of nodes into sets of densely interconnected ones) are widely encountered in various networks, including those emerging in neuroscience and social sciences \cite{2008_InfoMap_Rosvall, 2008_Louvain_Blondel}. Consequently, the methodologies introduced in this paper are applicable to a broader range of domains than the ones (Input-Output Network and World Trade Web) analyzed here. At the same time, we stress that the concept of `super-nodes' or blocks used here is more general than the one of communities (as it does not require any notion of dense subgraphs), and this extended generality is one of the main advantages of the approach we are presenting. For instance, one is often forced to change the resolution level adopted to represent the input network according to some externally imposed taxonomy of nodes (e.g. the industrial classification of firms or sectors in an economic network), irrespective of whether this taxonomy has anything in common with community structure, topological node similarity or proximity, and so on.

The rest of the paper is organized as follows.
In \autoref{sec:Graph_Renormalization}, we present the multi-scale setting, namely the formal definition of a graph and the coarse-graining procedure.
In \autoref{sec:Methodology}, we introduce the Configuration Model, the fitnessCM, the degree-corrected Multi-Scale Model (MSM), the  maximum-likelihood MSM and the fitnessMSM, alongside with their renormalization (summation) rule. 
In \autoref{sec:DatasetDescription}, we describe the ING Input-Output Network and the World Trade Web datasets, as well as the \textit{coarse-graining} process to generate the \textit{higher-scale} representations. 
In \autoref{sec:Results_and_Discussions}, we present and discuss the multi-scale results, focusing on network and machine-learning performance metrics.
Finally, all the technical details supporting the main text are reported in the Supplementary Material \cite{SuppMat}.

\begin{figure}[tbp]
    \centering
    \includegraphics[width=\linewidth, trim = {3cm 2.3cm 2.8cm 2.1cm}, clip]{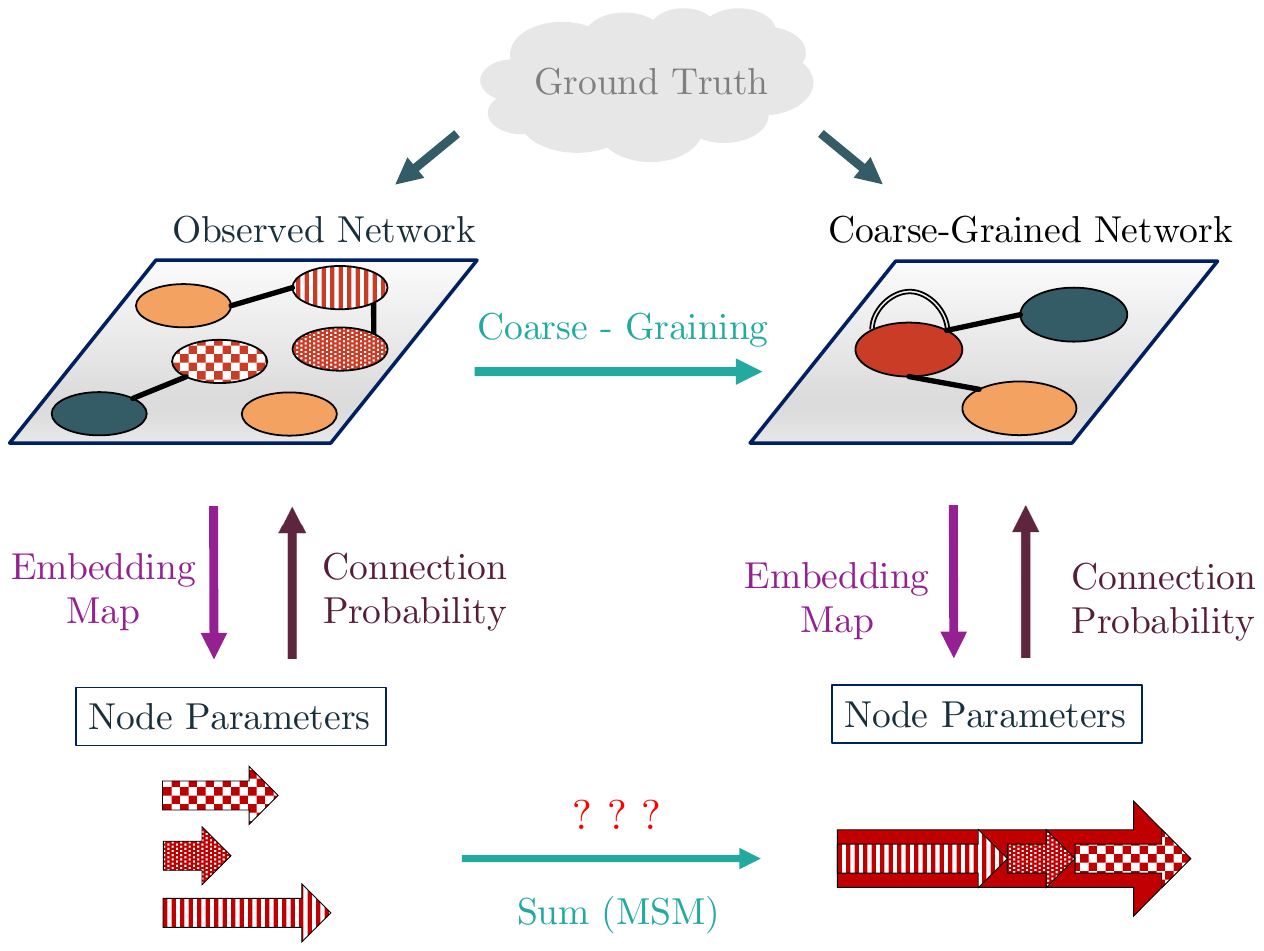} 
    \caption{The ground truth appears at multiple scales, depending on resolution of the dataset. Nevertheless, The macro-scale representation (shown in the plots on the right) can be uniquely obtained by coarse-graining the network at the microscopic scale. \textit{But how (scalar) embeddings are connected across different scales?} On the left, one finds the learning procedure of the microscopic node embeddings, whereas on the right, the macroscopic counterparts. To better highlight the point, we focused on the nodes with a custom pattern (mesh, dots, stripes) that are grouped in the red-filled block-node. In single-scale models, the micro-parameters cannot be directly used to calculate the macro-ones (indicated by the red question marks). In contrast, the multi-scale model overcomes this limitation, as the macro-scalars are the sum of micro-embeddings (see the alignment on the right).}
    \label{fig:MSNE_ResQuest}
\end{figure}

\section{Graph Renormalization}
\label{sec:Graph_Renormalization}
In this section, we formally define how to represent a graph in mathematical terms and introduce the coarse-graining procedure. For clarity, quantities associated with the microscopic level are formally denoted using the subscript $\ell = 0$, while those corresponding to coarser (aggregated) levels are indexed by $\ell > 0$. Those quantities may be also referred to as an $ \ell$\textit{-quantity}. For instance, \textit{2-vectors} and \textit{2-nodes} identify the node parameters and block-nodes at level $ 2$, respectively. Furthermore, when a statement holds for any level $ \ell \geq 0$, we adopt the notation $ \xi_{ij} := \xi^{(\ell)}_{i_{\ell} j_\ell}$ to refer to pair-wise components of a generic quantity $ \xi$.

Let us consider a binary undirected graph with $ N_0$ 
{\raggedright
microscopic nodes $\mathcal{V}_0$ (indexed as $ i_{0} \in [1, N_{0}]$).
Their connections (called edges or links) between nodes are defined as
$$ \mathcal{E}_0 := \left\{(i_0, j_0) : i_{0} \in \mathcal{V}_0, j_{0} \in [i_0, N_{0}], a^{(0)}_{i_0 j_0} = 1\right\}$$ 
where $ a^{(0)}_{i_0 j_0} = 1$ indicates the presence of an edge between nodes $i_0$ and $j_0$, whereas $ a^{(0)}_{i_0 j_0} = 0$ otherwise.} 
This graph can be represented by an $ N_{0} \times N_{0}$ adjacency matrix $ \mathbf{A}^{(0)}$ which is symmetric ($ a^{(0)}_{i_0 j_0} =  a^{(0)}_{j_0 i_0}$), due to the undirected nature of the interactions. We do not consider multiple edges between the nodes, but we do allow for self-loops, which appear along the diagonal of the adjacency matrix ($a^{(0)}_{i_0 i_0} = 0, 1$). 
% Moreover, $ a^{(0)}_{ij} := a^{(0)}_{i_0 j_0 }$ as it is clear from the superscript at which level the nodes are regarded. 

\subsection{Coarse-Graining}

To derive the coarse-grained representation of the 0-graph, we first assume that each 0-node belongs to a single, unique block. This is achieved by defining a \textit{non-overlapping, arbitrary} partition $ \Omega_{0}$ that maps the 0-nodes ($\mathcal{V}_0$) to a new set of block-nodes
\begin{equation*}
    \mathcal{V}_1 := \left\{i_{1} :=  \Omega_{0} (i_{0}) \quad \forall i_{0} \in \mathcal{V}_0 \right\}.
\end{equation*}
Secondly, two blocks are connected if at least one edge exists between their internal nodes. This relationship formally defines the coarse-grained entries $a^{(1)}_{i_1 j_1} $ as
\begin{equation}
    \label{eq:coarse-graining_01}
    a^{(1)}_{i_1 j_1} = 1 - \prod_{i_{0} \in \Omega_{0}^{-1}(i_{1}), j_{0} \in \Omega_{0}^{-1}(j_{1})} (1 - a^{(0)}_{i_0 j_0})
\end{equation}
where $ \Omega_{0}^{-1}(i_{1})$ denotes the set of 0-nodes mapped to the block-node $ i_{1}$ (and analogously for $ j_{1}$). In practice, the operation performs a \textit{logical OR} operation over the potential 0-links connecting two blocks. In this context, the missing 1-links reflect the non-existing connection for all the 0-pairs, while an active 1-edge may be obtained in multiple 0-cofigurations.
Furthermore, the mapping $ \Omega_{0}$ is \textit{surjective} but not \textit{injective}, since a block-node $ i_{1}$ is populated by one or more 0-nodes. 

We allowed for self-interactions at every order as $ i_{0}$ can be equal to $j_{0}$. The feature is consistent with the emergence of self-loops at the coarse levels, which can arise either from self-loops on constituent nodes or interactions between different internal nodes. Following these steps, we derived the coarse-grained adjacency matrix $ \mathbf{A}^{(1)}$ which is binary and symmetric, alike the 0-graph, but its dimension are reduced to $  N_{1} \times N_{1}$ where $ N_{1}$ are the total number of blocks at level $ \ell = 1$.

This lumping procedure can be repeated $ \ell + 1 (\geq 1)$ times by introducing a partition $ \Omega_{\ell}$ that groups the $ \ell-nodes$ into $ N_{\ell + 1}$ block-nodes
\begin{equation*}
    \mathcal{V}_{\ell+1} := \left\{i_{\ell + 1} := \Omega_{\ell}(i_{\ell}) \, \forall i_\ell \in \mathcal{V}_{\ell} \right\}.
\end{equation*}
Since these partitions $ \left\{\Omega_{\ell}\right\}_{\ell \geq 0}$ are nested (non overlapping), one can compose them to obtain a direct mapping of 0-nodes to $ \ell$-nodes:
\begin{equation}
    \label{eq:direct_composition}
    \Omega_{0 \to \ell} := \Omega_{\ell} \circ \, \cdots \, \circ \Omega_{0}.
\end{equation}
As a result, the $ (\ell+1)$-adjacency matrix ($\mathbf{A}^{(\ell + 1)}$) can be obtained from the $\mathbf{A}^{(0)}$ either by coarse-graining level-by-level (see \autoref{eq:coarse-graining_01}), or in a one-step fashion through
\begin{align}
    \label{eq:coarse-graining_ell}
    a^{(\ell + 1)}_{i_{\ell + 1} j_{\ell + 1}} 
    &= 1 - \prod_{i_{\ell} \in \Omega_{\ell}^{-1}(i_{\ell + 1}), j_{\ell} \in \Omega_{\ell}^{-1}(j_{\ell + 1})} (1 - a^{(\ell)}_{i_{\ell} j_{\ell}}) \\
    \label{eq:coarse-graining_0ell}
    &= 1 - \prod_{i_{0} \in \Omega^{-1}_{\ell \to 0}(i_{\ell + 1}), j_{\ell} \in \Omega^{-1}_{\ell \to 0}(j_{\ell + 1})} (1 - a^{(0)}_{i_{0} j_{0}})
\end{align}
where $ \Omega^{-1}_{\ell \to 0} := \Omega^{-1}_{0} \circ \dots \circ \Omega^{-1}_{\ell}$ is the inverse map of the direct composition \autoref{eq:direct_composition}.

The \textit{nested} set of partitions $ \left\{\Omega_{\ell}\right\}_{\ell \geq 0}$ can be uniquely described by means by a dendrogram, as shown in \cite{2023_MSNR_Garuccio}. A ``horizontal'' cut of the dendrogram yields a partition at a fixed scale. For example, in the WTW case, this corresponds to aggregate countries that are located within a specific geographical distance. On the other hand, cutting the dendrogram at different heights yields multiscale clusters, such as a state and a continent. 

In many applications, the context of the problem directly suggests a partitioning of the nodes into appropriate blocks, without the need for an explicit distance matrix. A possible example is the classification of economic sectors based on the NAICS taxonomy \cite{NAICS_website}. This system assigns longer codes according to a more detailed description of the industries: for example, the two-digit code 72 corresponds to Accommodation and Food Services, and it includes Full-Service Restaurants associated with $722511$.

However, when such natural partition is available, one may rely on alternative methods to obtain a partition of the 0-nodes. These include applying any standard non-overlapping community detection algorithm, e.g. Louvain \cite{2008_Louvain_Blondel,2023_LaplRG_Villegas} or even assigning 0-nodes to communities at random.

\section{Methodology}
\label{sec:Methodology}

In the previous sections, we emphasized that a single \textit{phenomenon} can be analyzed at several resolution levels. Moreover, since each element of the adjacency matrix $ a_{ij}^{(\ell)} \in \left\{0,1\right\}$, every edge $ (i_\ell,j_\ell)$ can be modelled as a Bernoulli random variable. Omitting the dependence on the level $ \ell$ (see \autoref{sec:Graph_Renormalization}), the adjacency matrix pairs are modeled as
\begin{equation}
    \label{eq:a_ij_Bernoulli_rv}
    a_{ij} =
    \begin{cases}
        1 & \qquad p_{ij} \\
        0 & \qquad 1 - p_{ij}
    \end{cases}
    \qquad \forall i \leq j.
\end{equation}
The models used in this study are the fitness Configuration Model (fitnCM) \cite{2015_CimiModel_Cimini}, the Configuration Model (CM) \cite{2004_StatMecNet_Park} (CM), the degree-corrected MSM (degcMSM) and the maximum-likelihood MSM (maxlMSM), and the fitness multi-scale model (fitnMSM) (\cite{2023_MSNR_Garuccio}). Among these, we selected the CM as the representative of single-scale models. To the best of our knowledge, only the MSM framework is renormalizable over arbitrary node partitions.

\subsection{Configuration Model} 
\label{sec:CM}
The Configuration Model (CM) \cite{2004_StatMecNet_Park} estimates the probability $ p_{ij}$ that two nodes $ (i,j)$ are connected, assuming that the empirical degree sequence $ \left\{k_{i}\right\}_{i \in [1, N]}$ is known. Unlike many other network embedding methods \cite{2023_ZooGuide_Baptista}, the CM relies exclusively on node degrees to learn its parameters.
% This is a typical ``spam vs. non-spam'' classification problem \cite{2021_LogReg4SpamNonSpam_Khanday}, but applied to objects (i.e., the links) that do not carry explicit features. 

By maximizing the Lagrangian in \autoref{SI:sec:CM}, one obtains the following connection probability
\begin{equation}
    \label{eq:CM_probability}
    p_{ij} = \frac{x_{i} x_{j}}{1 + x_{i} x_{j}}
\end{equation}
where the parameters $ x_{i} := e^{-\theta} \geq 0$ are determined by maximizing the likelihood function, excluding self-loops:
\begin{align}
    \label{eq:CM_likelihood}
    \mathcal{L}(\vec{x}|\mathbf{A}) &= \sum_{i < j} a_{ij} \ln(x_{i} x_{j}) - \ln(1 + x_{i}x_{j})
\end{align}
where $ \vec{x} := \left[x_{i}\right]_{i \in [1, N]}$.
At the stationarity point, the parameters satisfy the following systems of equations 
\begin{equation}
    \label{eq:CM_maximum_point}
    0 \stackrel{!}{=} \partial_{i}\mathcal{L}(\vec{x}|\mathbf{A}) \Leftrightarrow 
    \langle k_{i} \rangle 
    := \sum_{j (\neq i)} p_{ij} 
    = \sum_{j (\neq i)} \frac{x_{i} x_{j}}{1 + x_{i} x_{j}} \stackrel{!}{=} k_{i},
\end{equation}
where the initial conditions are sampled uniformly at random, as the optimization problem admits a unique solution. The optimization is performed over the \textit{structural inequivalent} nodes -that is, nodes with distinct degrees (see \autoref{SI:sec:StructE_not_StatE}). As a result, nodes sharing the same degree are assigned the same parameter $x_i$. This establishes a one-to-one correspondence between observed degree $k_i$, and parameter $x_i$, implying that the number of independent parameters equals the number of unique degrees in the network.

A key point to emphasize is that the parameters $ \vec{x}$ are fixed such that the expected degrees (left-hand side) match the observed degrees (right-hand side). Nevertheless, the solution is equivalent to the maximum likelihood estimate \autoref{eq:CM_likelihood}, which assumes the knowledge of the full adjacency matrix. Therefore, the degree sequence serves as \textit{sufficient statistics} for the Configuration Model, enabling a privacy-oriented procedure to learn the parameters. Additional details on the derivation can be found in the appendix.

\subsection{fitness Configuration Model}
In many real-world context, the degree sequence is not publicly available due to \textit{privacy concerns} that restrict the access to node-level information, e.g. a person or a firm, \cite{2004_Fitness_WTW_Garlaschelli}. However, global network metrics, such as the total number of links, can be often acquired with fewer limitations. To adapt accordingly the CM, one assumes that each hidden parameter $ x_{i}$ is proportional to a node-specific observable $ \phi_{i}$ referred to as ``fitness''. This assumption leads to the so-called ``fitness ansatz'', namely
\label{sec:fitnCM}
\begin{equation}
    \label{eq:fitness_ansatz}
    x_{i} \equiv \sqrt{\delta} \phi_{i}.
\end{equation}
In other words, it states that the heterogeneity of the parameters is captured by the observable $ \phi_{i}$ scaled by a global parameter.

Substituting this relationship into \autoref{eq:CM_probability}, the fitn Configuration Model (fitnCM) probability reads
\begin{equation}
    \label{eq:fitnCM_probability}
    p_{ij} := \frac{\delta \phi_{i} \phi_{j}}{1+\delta \phi_{i} \phi_{j}}
\end{equation}
where the $ \delta$ parameter is determined by requiring the expected number of links $ \langle L \rangle$ to match the observed link count $ L$, namely
\begin{equation}
    \langle L \rangle := \sum_{i < j} p_{ij} = \sum_{i < j} \frac{\delta \phi_{i} \phi_{j}}{1+\delta \phi_{i} \phi_{j}} \stackrel{!}{=} L.
\end{equation}
As with the CM, this constraint also corresponds to the maximum likelihood estimation of the model parameters.

The choice of the fitness observable $\phi_{i}$ depends on the specific domain. For example, we set the fitness vales as the strengths (total exchange money) of a sector, formally $$\phi_{i} := s^{undirected}_{i} := \sum_{j (\neq i)} \frac{w_{ij} + w_{ji}}{2},$$ where the $ w_{ij}$ denotes the monetary flow from node $ i$ to $ j$ \cite{2015_CimiModel_Cimini}. On the other hand, for the WTW, fitness are identified with the Gross-Domestic Product of a country $ i$, i.e. $\phi_{i} \stackrel{!}{=} GDP_i$ \cite{2023_MSNR_Garuccio,2004_Fitness_WTW_Garlaschelli}. In both cases, the $\phi_{i}$ are \textit{additive} as the strength, or GDP, of a group of nodes is the sum of its memebers' strengths, or GDP.

\subsection{Multi-Scale Models}
\label{sec:MSM}
A natural approach for modelling a phenomenon represented at multiple scales ($ \left\{\mathbf{A}\right\} := \left\{ \mathbf{A}^{(\ell)} : \ell \geq 0 \right\}$) is through \textit{renormalizable} models. These models retain the same functional form throughout the levels (cfr. \autoref{SI:Inconsistency_CM}), whereas their parameters renormalize. Among others (see \cite{2018_MultiScale_Unfolding_GarciaPerez}), the ideal candidate should allow for arbitrary and heterogeneous partitions $ \left\{\Omega_{\ell}\right\}_{\ell \geq 0}$ of the 0-nodes, not being restricted to a specific aggregation procedure. To address these challenges, the fitness multi-scale model (fitnMSM) was proposed in \cite{2023_MSNR_Garuccio}. In this essay, we build upon that framework by equipping it with node parameters $ \vec{x} := \left[x_{i}\right]_{i \in [1, N]}$ (see \autoref{eq:CM_probability}), which can be determined either by enforcing the degree sequence (yielding the degree-corrected MSM, or degcMSM) or maximizing the likelihood (yielding the maximum-likelihood MSM, or maxlMSM). Finally, since the CM does not account for the self-loops, we also discard them in the MSMs, we introduce in the following sections.

\subsubsection{Degree-Corrected MSM}
\label{sec:degcMSM}
The degree-corrected MSM (degcMSM) defines the probability that two nodes are connected as
\settowidth{\temp}{$1- e^{-\frac{1}{2}   x_i  ^2 - w_{i}} \quad $}
\begin{equation}
    \label{eq:MSM_pij}
    p_{ij}:= p(x_{i}, x_{j}, w_{i}) =
    \begin{cases}
        \FixedSize{\temp}{1- e^{- x_i, x_j}} i \neq j\\
        \FixedSize{\temp}{1- e^{-\frac{1}{2}   x_i  ^2 - w_{i}}}  i = j
    \end{cases}
\end{equation}
where $ x_{i} \in \mathbb{R}_+, w_{i} \in \left[ \left. -\frac{1}{2} x_{i}^{2}, \infty \right)  \right.$ are seen as the propensity of creating external connections and self-loop interactions, respectively. 
As for the Configuration Model, we estimated the parameters based on the degree sequence $ \vec{k} := \left[k_{1}, \dots, k_{N}\right]$ (no information about the self-loops) \cite{2011_AnalMax_Squartini}.
Specifically, we forced the expected degrees to coincide with the observed one, namely 
\begin{equation}
    \label{eq:MSM_enf_deg}
    \langle k_{i} \rangle := \sum_{j (\neq i)} p_{ij} \stackrel{!}{=} k_{i}.
\end{equation}
Since the expected degree $ \langle k_{i} \rangle$ is a one-to-one (monotonic) function of the parameter $ x_{i}$, each observed degree corresponds to a unique value of $ x_{i}$. That is, two different nodes sharing the same observed number of neighbors have the same parameters.

\subsubsection{Maximum-Likelihood MSM} 
In \autoref{sec:CM}, enforcing the degrees (method of moments) and maximizing the likelihood gave the same parameters $ \vec{x}$ (\autoref{eq:CM_maximum_point}). However, for the MSM this is no longer true, since the MSM is not an Exponential Random Graph model. Therefore, we decided to report the expected values of the maximum-likelihood MSM (maxlMSM) for a direct comparison with the degcMSM. However, this is only a ``theoretical'' comparison, as one should have the observed adjacency matrix, which is not available in the contex of network reconstruction. Still, it is an intriguing result to see how the maximum-likelihood solution relate with the method of moments one.

The maxlMSM uses the same connection probability of the degcMSM (\autoref{eq:MSM_pij}), but the parameters are found by maximizing the likelihood
\begin{align}
    \label{eq:MSM_likelihood}
    \mathcal{L}( \vec{x} |\textbf{A}) 
    &= \sum_{i < j} a_{ij} \ln(p_{ij}) + (1-a_{ij}) \ln(1-p_{ij}) \\
    &= \sum_{i < j} a_{ij} \ln(1 - e^{-x_{i}x_{j}}) - (1-a_{ij}) x_{i}x_{j}.
\end{align}
where  $\mathbf{A} = \left\{a_{ij}\right\}_{ \left\{i \leq j\right\} }$ is the adjacency matrix of the observed network.
The gradient of \autoref{eq:MSM_likelihood} reads
\begin{align}
    \label{eq:MSM_grad}
    \partial_{x_i} \mathcal{L} = \sum_{j (\neq i)} \left(\frac{a_{ij}}{p_{ij}} - 1\right) x_j,
\end{align}
which vanishes at the maximum point, i.e. the optimal value for the $ \vec{x}$. 
As described in \autoref{SI:sec:StructE_not_StatE}, at the optimal point, nodes with identical neighbors (same neighborhood, or $ N(\cdot)$) have the same parameters. For example, focusing on nodes $ i,j$, $ \mathcal{N}(i) \stackrel{!}{=} \mathcal{N}(j) \leftrightarrow x_{i} \stackrel{!}{=} x_{j} $. This implies that nodes with the same degree, which share the same parameters for the degcMSM, have different maxlMSM parameters if they are connected to different nodes. However, having the same neighbors, also implies the same parameters under the degcMSM.

\begin{center}
    {\textit{Optimization Details}}
\end{center}

The initial conditions for all the models have been chosen at random from the uniform distribution $ \mathcal{U}[0,1]$ and used Scipy \cite{2020_SciPy_Virtanen} to find the maximum of the likelihood. Before setting on this, we explored ways to automatically find good initial points using Particle Swarm Optimization (PSO) \cite{2018_PySwarms_Miranda}, but the results were the same of starting with random initial conditions. Further investigations on initialization strategie are beyond the scope of this paper.

\subsubsection{Fitness MSM}

The fitness Multi-Scale Model (fitnMSM) \cite{2023_MSNR_Garuccio} extends the general MSM framework (see \autoref{SI:sec:derivation_MSM_probability}) to address the issue of data confidentiality—a topic discussed in more detail in \autoref{sec:fitnCM}. As with the fitness Configuration Model (fitnCM), the fitnMSM uses the fitness ansatz (\autoref{eq:fitness_ansatz}), where a single global parameter ($ \delta $) controls the total number of links $ L $. Differently from the fitnCM, the fitnMSM requires the fitness to be \textit{additive}, that is, the fitness of an aggregated block must equal the sum of the fitness values of its individual nodes (\autoref{SI:eq:MSM_sumX}). Luckily, the fitness measures previously adopted for fitnCM—such as GDP or node strength—are indeed additive. Hence, the difference between the fitnCM and fitnMSM lies in the functional form of the model, not in the nature of the fitness values.

Applying the fitness ansatz, the connection probability \autoref{eq:MSM_pij} becomes
\settowidth{\temp}{$1- e^{-\frac{\delta}{2} \, \phi_{i}^{2}} - \eta w_{i}$}
\begin{align}
    p_{ij} = 
    \begin{cases}
        \FixedSize{\temp}{1- e^{-\delta \, \phi_{i} \, \phi_{j}}} \qquad i \neq j\\
        \FixedSize{\temp}{1- e^{-\frac{\delta}{2} \, \phi_{i}^{2}- \eta w_{i}}} \qquad i = j
    \end{cases}
\end{align}
where $ \eta \geq 0, w_{i} \in \left. \left[ -\frac{\delta}{2} \, \phi_{i}^{2}, \infty \right. \right) $ control the self-loops and $ \delta \geq 0$ for the total number of links. 
Since our goal is to compare this model with the fitnCM (not including the self-loops), the $ \delta$ is estimated with the method of moments, namely
\begin{equation}
    \label{eq:tot_link_MoM_WTW}
    L \stackrel{!}{=}  \langle L  (\delta) \rangle := \sum_{i < j} p_{ij}(\delta).
\end{equation} 
As discussed for the degcMSM (\autoref{SI:sec:derivation_MSM_probability}), fitting the $\delta$ with the method of moments ($\delta_{L}$) does not necessarily return the same value as the maximum-likelihood estimate ($\delta_{\mathcal{L}}$). In both ION and WTW datasets, we observed that ( $\delta_L > \delta_{\mathcal{L}}$ ), meaning that the likelihood-based model underestimates the actual number of links. Therefore, we used $\delta_L$ (method of moments), as it fixes the empirical link count as the fitnCM.

\subsection{Renormalized Models}
In this section, we introduce the renormalization procedure applied to the previously discussed models. The multi-scale models (MSMs) are inherently renormalizable by design (see \autoref{sec:MSM}). In contrast, for the CM and fitnCM, renormalizability is not inherent and must be imposed externally (see \autoref{SI:Inconsistency_CM}).

\subsubsection{Summed MSM}
This class of models is defined by enforcing the following two conditions, namely
\begin{align}
    \label{eq:self-cons_scale-invariance_conds}
    \begin{cases}
        \mathbf{P}_{sum}^{(\ell)} &\stackrel{!}{=} \mathbf{P}_{cog}^{(\ell)} \\[1ex]
        \mathbf{P}^{(\ell)} &\stackrel{!}{ \equiv } \mathbf{P}^{(m)}
    \end{cases}
\end{align}
$ \forall \ell \geq 0, m \geq 0$, which are jointly referred to as the \textit{scale-invariance} (or \textit{self-consistency}) properties. They require that the renormalized probability $ \mathbf{P}_{sum}^{(\ell)}$ equals the \textit{coarse-grained} $ \mathbf{P}_{cog}^{(\ell)}$ with the same functional form (\cite{2023_MSNR_Garuccio}, \autoref{SI:sec:derivation_MSM_probability}) for all the scales.

We characterize the \textit{fitted} vectors at finest level $ 0$ as $ \left\{x_{i_0}, w_{i_0}\right\}_{i \in [1, N_{0}]}$. Moreover, we define $\left\{x_{I}, w_{I}\right\}_{I \in [1, N_{\ell}]}$ as the block-parameters at the level $ \ell > 0$ where the block $ I$ corresponds to a group of original nodes via $ I := \Omega_{0 \to \ell}(i_0)$. Assuming that the block-node parameters are obtained by \textit{summing} the lower level ones, namely
\begin{equation}
    \label{eq:MSM_sum_graining_rule}
    x_{I} = \sum_{i_0 \in I} x_{i_0} \textnormal{ and } w_{I} = \sum_{i_0 \in I} w_{i_0},
\end{equation}
the \autoref{eq:self-cons_scale-invariance_conds} is automatically satisfied due to the model's functional form \autoref{eq:MSM_pij} (see \autoref{SI:sec:derivation_MSM_probability}). In other words, this additive coarse-graining rule ensures that the structure of the model remains unchanged across all levels.

As a result, the connection probability between block nodes ( $I$) and ( $J$ ) at level ( $\ell$ ) can be computed by simply inserting the summed parameters from \autoref{eq:MSM_sum_graining_rule} into the original MSM probability function:
\settowidth{\temp}{$1- e^{-\frac{1}{2}  x_I  ^2 - w_{I}} \;$} 
\begin{equation}
    \label{eq:MSM_pIJ}
    p_{IJ}:= p(x_{I}, x_{J}, w_{I}) = 
    \begin{cases}
        \FixedSize{\temp}{1- e^{-x_Ix_J}} I \neq J\\
        \FixedSize{\temp}{1- e^{-\frac{1}{2} x_I^2 - w_{I}}}  I = J.
    \end{cases}
\end{equation}
% As said, the single-scale models (SSM) are not self-consistent, so one should fit the model for each different level, as if the network was generated by a different generative process (even if we know this wasn't the case). Thanks to this feature, the MSM has a lower computational complexity with respect to the SSMs (see \autoref{SI:sec:Algorithmic_Complexity}).

\subsubsection{Summed CM}
\label{sec:sum_CM}
Once the microscopic $ \vec{x}^{(0)} := \left\{x_{i_{0}}\right\}_{i_{0} \in [1, N_{0}]}$ have been estimated, the Configuration Model (CM) lacks a built-in renormalization procedure. As such, two possible strategies can be followed: ``coarse-graining'' the microscopic probabilities (\autoref{eq:MSM_pIJ_cg}) or by refitting the parameters at that level $ \ell$. Both solutions suffer from a similar drawback due to the lack of scale-invariance of the CM (cfr. \autoref{eq:self-cons_scale-invariance_conds}). Specifically, the first method yields a coarse-grained probability matrix $\mathbf{P}_{\text{cog}}^{(\ell)}$ that can no longer be expressed in the original CM form using group-level parameters. Hence, the resulting model ceases to be a proper Configuration Model (see \autoref{SI:Inconsistency_CM}).
On the other hand, refitting the CM at each level $ \ell$ leads to $ \ell$-vectors that are unrelated with the $\ell-1$ parameters. From the model's perspective, each level arises from a distinct generative process, thereby losing connection of the parameters across scales.

To address these drawbacks, we used the same renormalization (summation) scheme as in \autoref{eq:MSM_sum_graining_rule} which leads to the connection probability among (I,J) blocks:
\begin{equation}
    \label{eq:CM_Psummed}
    p_{IJ} = \frac{x_{I} x_{J}}{1 + x_{I} x_{J}}
\end{equation}
This procedure satisfies the second condition in \autoref{eq:self-cons_scale-invariance_conds} by preserving the functional form of the CM at the block level. While the first condition (equivalence between renormalized and coarse-grained probabilities) is still not functionally enforced, we proceed pragmatically to determine at which extent this affects higher-level expectations.

\subsubsection{Fixed (renormalized) fitnCM}
Similarly to the CM, the fitnCM lacks scale-invariance properties. Therefore, we implement a renormalization scheme whereby $ \delta^{(\ell)} \stackrel{!}{=} \delta^{(0)}$ (the latter parameter was fixed at the microscopic level $ 0$), while the block-fitness $ I$ is obtained by summing the inner fitness $ \left\{\phi_{i}\right\}_{\left\{i \in I\right\}}$ (see \autoref{SI:sec:CM}). Note that this approach is equivalent of applying the sum rule of the CM.
Finally, the higher-level connection probability among (I,J) blocks is the following:
\begin{equation}
    \label{eq:fitnCM_Psummed}
    p_{IJ} = \frac{\delta^{(0)} \phi_{I} \phi_{J}}{1 + \delta^{(0)} \phi_{I} \phi_{J}}.
\end{equation}

\section{Data}

\begin{table*}[t]
    \parbox{.45\linewidth}{
    \centering
    \begin{tabular}{c|c|c|c|c}
        \toprule
        Level $ \ell$ & $ N_{\ell}$  & $ L_{\ell}$ & $\rho_{\ell}$ \\
        \midrule
        0 & 972 &  136887 & 0.29\\
        1 & 647 &  93659  & 0.45\\
        2 & 303 &  30861  & 0.67\\
        3 & 87 &  3364   & 0.89\\
        \bottomrule
        \multicolumn{4}{c}{} \\
        \multicolumn{4}{c}{} \\
    \end{tabular}
    \caption{Table for the ION datasets reporting, for each level $ \ell$, the number of nodes ($ N_{\ell}$), the \textit{reduced} number of nodes, the total number of links and the density.\label{tab:ION_n_nodes_edges}}
    }
    \hfill
    \parbox{.45\linewidth}{
    \centering
    \begin{tabular}{c|c|c|c|c}
        \toprule
        Level $ \ell$ & $ N_{\ell}$  & $ L_{\ell}$ & $\rho_{\ell}$ \\
        \midrule
        0 & 182 & 9993  & 0.61\\
        1 & 152 &  7305 & 0.64\\
        2 & 112 &  4822 & 0.67\\
        3 & 92 &  2728  & 0.65\\
        4 & 62 &  1332 & 0.70\\
        5 & 32 &  400 & 0.80\\
        \bottomrule
    \end{tabular}
    \caption{Table for the WTW datasets reporting, for each level $ \ell$, the number of nodes ($ N_{\ell}$), the \textit{reduced} number of nodes, the total number of links and the density.\label{tab:WTW_n_nodes_edges}}
    }
\end{table*}

\subsection{ING Input-Output Network}
\label{sec:DatasetDescription}
ING Bank N.V. regularly reports the economic transactions of all ING clients for different years. 
We focused on firm-to-firm payments in the year 2022 between ING bank accounts, filtering out the transactions involving \textit{individual} or non-Dutch clients, flows to or from non-ING account and the internal transfers within the same firm, i.e. self-payments. Since ING is the largest bank in The Netherlands \cite{2024_SP_GLobal_Banks_Ranking}, the dataset covers a major portion of the national market. More precisely, we selected the year 2022 to simplify the numerical calculations and to avoid distortions in the data from the aftermath of the COVID-19 pandemic. However, the procedure may be easily extended for other time intervals, such as multi-year spans or quarterly snapshots. 

At the firm-to-firm (f2f) resolution, the dataset consists of $ N_{f2f} \approx 3.4 \cdot 10^{5} \text{ nodes}, L_{f2f} \approx 4 \cdot 10^{6} \text{ links}$, yielding a density $ \rho_{f2f} \approx 3.5 \cdot  10^{-5}$. This classifies the network as \textit{large and sparse}. To analyze the data at a higher level, we aggregated the firms according to their NAICS (North American Industry Classification System) codes, and set an edge among two sectors if there was \textit{at least one} payment among firms belonging to one of them (see \autoref{eq:coarse-graining_ell}). Then, to focus on the production structure of this graph, we excluded sectors such as ``Public Administration'' (92), ``Finance and Insurance'' (52), ``Management of Companies and Enterprises''/``Holdings'' (55) as they are not directly related with a product/service. In particular, ``Public Administration'' fluxes include taxes and fees;  ``Finance and Insurance'', loans, that are not part of the supply chain of a product; and ``Management of Companies and Enterprises''/``Holdings'' include business entities that own shares in multiple firm. Lastly, we mapped, for simplicity, the 6-digits NAICS codes to integers, i.e. $ 111110 \to 1, 111120 \to 2, \cdots$.

Our work is the first application of the MSM to the \textit{multiscale structure} derived from the ION. Therefore, we focused on the \textit{economic relationships} among the sectors, discarding both the directionality and the volume of monetary flow. In this context, each edge is undirected (reciprocated) and binary. For example\footnote{The $i,j$ refers to the observed $ 0$-nodes, namely $ i := i_{0}, j := j_{0}$}, if $ w_{ij}$ refers to the total amount of money sent from $ i \textnormal{ to } j$ at level $ 0$, we \textit{reciprocated} the transaction by setting $ w_{ij} \to w^{'}_{ij} := \frac{w_{ij} + w_{ji}}{2}$ \cite{2010_WTW_Fagiolo}. Furthermore, an edge was regarded as active ($ a_{ij} = a_{ji} = 1$) if $ w^{'}_{ij} > 0$ or absent otherwise. This results in a symmetrical and binary edge, whenever there was \textit{at least} one flow between two sectors. At this stage, the ION consists of $ N_{0} = 972$ sectors and $ L_{0} = 1.4 \cdot  10^{5}$ links, yielding a network density of $ \rho_{0} \approx 0.29 $, roughly 4 order of magnitudes higher than $\rho_{f2f}$.

% In the following, we will ease the notation using the definitions $ a^{(0)}_{ij} := a_{i_0j_0} := a^{(0)}_{i_0j_0} $ as one explicit $ \ell$ suffices to know the scale of the other quantities. 

\subsection{Coarse-Graining The ION}
In the previous section, we constructed the \textit{binary undirected} adjacency matrix $ \mathbf{A}^{(\ell = 0)}$ representing the interactions among the 6-digits sectors (\textit{0-nodes}). Here, we describe the coarse-graining procedure to obtain the \textit{multi-scale} unfolding of $ \mathbf{A}^{(0)}$.

At first, we grouped together the 0-nodes that share the same first $ 6 - \ell$ digits, with $\ell \in [0,4]$. For instance, the sectors $ 111191, 111199$ are merged in the same block $ 11119$ starting from $ \ell = 1$.

Secondly, we set an edge among two $ \ell$-nodes if there existed \textit{at least one} connection between any of their constituent 0-nodes (see \autoref{eq:coarse-graining_0ell}). Our procedure uses a \textit{one-step} approach to easily generate every level directly from the $ \mathbf{A}^{(0)}$, without passing through the intermediate scales. The latter procedure is also allowed by the MSM. 

Note that from $ \ell = 4$ onward, the coarse-grained graphs become \textit{fully-connected}, limiting a meaningful \textit{statistical} modelling up to $ \ell = 3$. For technical details on the evolution of the number of nodes $ N_{\ell},$ links $ L_{\ell}$, and density $ \rho_{\ell} := \frac{2L_{\ell}}{N_{\ell}\left(N_{\ell} - 1\right)}$, refer to \autoref{tab:ION_n_nodes_edges}.

\subsection{World Trade Web}
As a second application, we analyzed the World Trade Web (WTW) tracked by the Gleditsch dataset \cite{2002_Gleditsch}, which reports bilateral trade flows (imports and exports) for every country. We focused on the year 2000 (the most recent one), and excluded the missing states in the BACI-CEPII GeoDist \cite{2011_GeoDist_Mayer}, whose distances are required for the coarse-graining procedure. After this filtering step, the graph has $N_{0} = 185$ 0-nodes. Although we illustrate the method for the year 2000, the methodology applies to any other period of time.

For every pair $( i, j)$, the dataset reports the \textit{exported} volume $ w_{ij}$ and the \textit{imported} $ \omega_{ji}$ one (all in USA dollars). Since the reported export from $ i$ to $ j$ ($ w_{ij}$) generally differ from the import from $ j$ to $ i$ ($ \omega_{ij}$), we redefined $ w_{ij} \leftarrow \frac{w_{ij} + \omega_{ji}}{2}$ \cite{2010_Evo_of_WTW_Weighted_Networks_Fagiolo} as the averaged amount of trade from $ i \textnormal{ to } j$. 

Successively, as for the ION, we \textit{symmetrized} the transaction matrix by mapping every weight $ w_{ij} $ to the average flow between the two directions, namely $ w_{ij} \to w^{'}_{ij} := \frac{1}{2} \left( w_{ij} + w_{ji} \right)$, i.e. the average flow between the two directions. By renaming $ w^{'}_{ij}$ as $ w_{ij}$, the resulting weights are symmetric, i.e. $ w_{ij} = w_{ji}$. Given the high link \textit{reciprocity} of the transaction reported in the Gleditsch dataset \cite{2023_RecEcon_DiVece}, this undirected representation is a sound approximation. Then, we \textit{binarized} the import-export matrix to construct the adjacency matrix $ \mathbf{A} := \Theta(\textbf{W})$ where the $ \Theta(\cdot)$ is the Heaviside function.

\subsection{Coarse-Graining of WTW}
To coarse-grain the level $ 0$, we used the inter-country geographical distances \cite{2011_GeoDist_Mayer} to iteratively merge closer nodes via \textit{single-linkage agglomerative clustering} as in \cite{2023_MSNR_Garuccio}. This procedure returns a dendrogram whose \textit{leaves} are the 0-nodes, whose \textit{branching points} are the block-countries, and whose branch \textit{heights} equal the single-linkage distances between sub-clusters. By cutting this dendrogram at 18 successive levels $\ell\in[0,17]$ we obtain partitions $\Omega_{\ell}, \ell \ge 0$, such that the number of block-countries are approximately $ N_{\ell} := N_0 - 10 \cdot \ell$. For $ \ell \geq 7$ the coarse-grained graph become fully connected. Therefore, a meaningful statistical modelling is possible up to $ \ell = 6$. See \autoref{tab:WTW_n_nodes_edges} for the evolution of the number of nodes, links, and network density.

\section{Results and Discussions}
\label{sec:Results_and_Discussions}
In this section, we apply the previously introduced models to the ION datasets. The corresponding results for the WTW dataset are provided in the Appendix.

\subsection{Scale-Invariance Evidence}
\label{sec:ScaleInv_NetMeas}

\newlength{\subfigurewidth}
\setlength{\subfigurewidth}{0.45\linewidth} % Adjust the 0.45 to whatever fraction of the line width you prefer

\begin{figure}[t]
    \centering
    \subfloat[Level 2: fitness-based models.
        \label{fig:ING_sum_vs_cg_pmatrix_fitmodels}]{\includegraphics[width=\subfigurewidth]{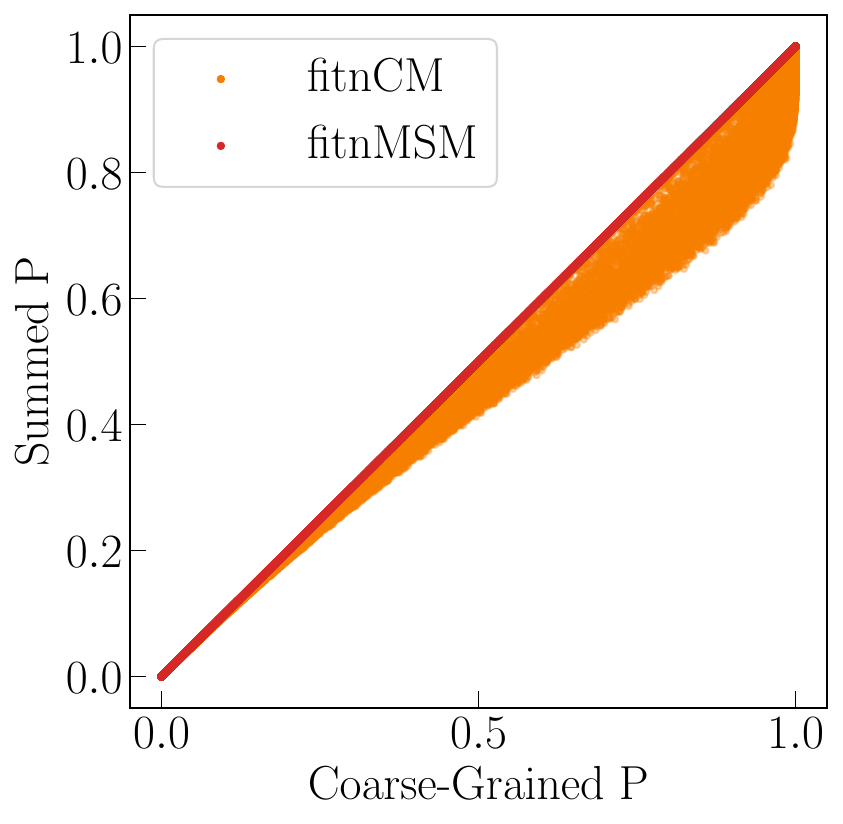}}
    \subfloat[Level 2: degree-based models.
        \label{fig:ING_sum_vs_cg_pmatrix_localmodels}]{\includegraphics[width=\subfigurewidth]{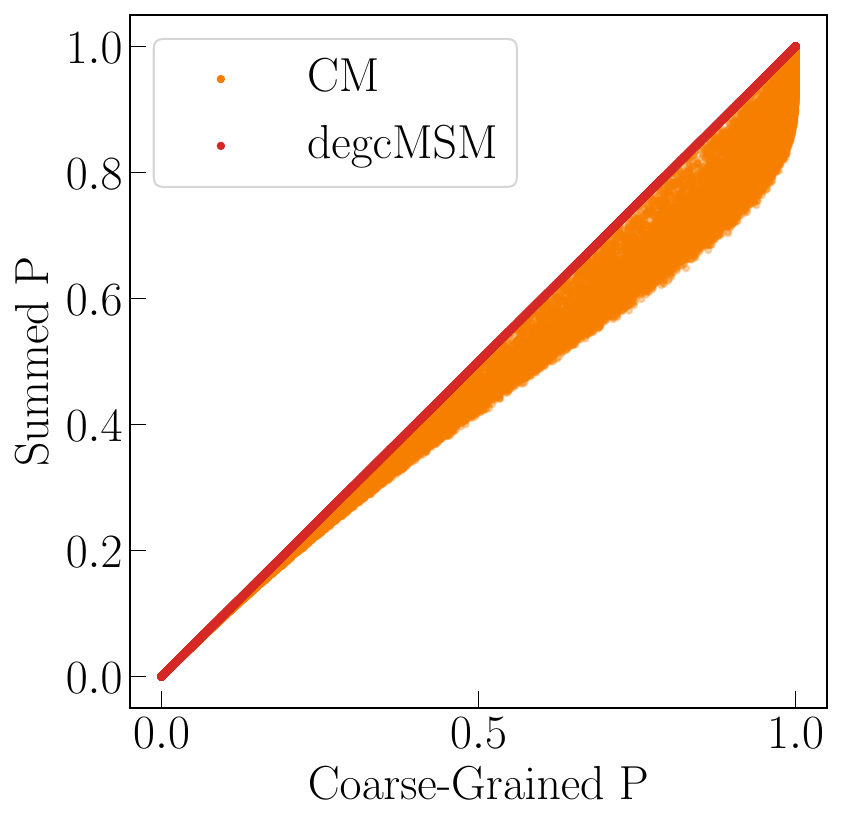}}
    \caption{Testing the scale-invariance principle on the single-scale models (fitnCM, CM) and the multi-scale ones (fitnMSM, degcMSM) at the coarser level 2. On the x-axis (y-axis), the RHS (LHS) of the \autoref{SI:eq:MSM_micro_psumVSpcg}.}
    \label{fig:ING_sum_vs_cg_pmatrix}
\end{figure}

\autoref{fig:ING_sum_vs_cg_pmatrix} illustrates the property of scale-invariance by plotting the left-hand side (LHS) of the equation in \autoref{SI:eq:MSM_micro_psumVSpcg} against its right-hand side (RHS). By construction, both the fitnMSM and degcMSM models perfectly recover the identity line, while the maxlMSM is not depicted since it uses the same probability as the degcMSM. In contrast, the summed versions of the fitnCM and CM models consistently underestimate their coarse-grained counterparts, forming a characteristic knife-like distribution that is typical of their functional form. This finding differs from that of \cite{2024_MSNE_milocco}, which reported a different relationship for the LogisticPCA model \cite{2021_symLPCA_Chanpuriya}. A simple example demonstrating why the fitnCM and CM models lack scale-invariance is provided in \autoref{SI:sec:CM}.

\subsection{Summed against Fitted parameters}
\setlength{\subfigurewidth}{.65\linewidth} 
\begin{figure*}[t]
    \subfloat[CM \label{ci1}]{\includegraphics[width=\subfigurewidth]{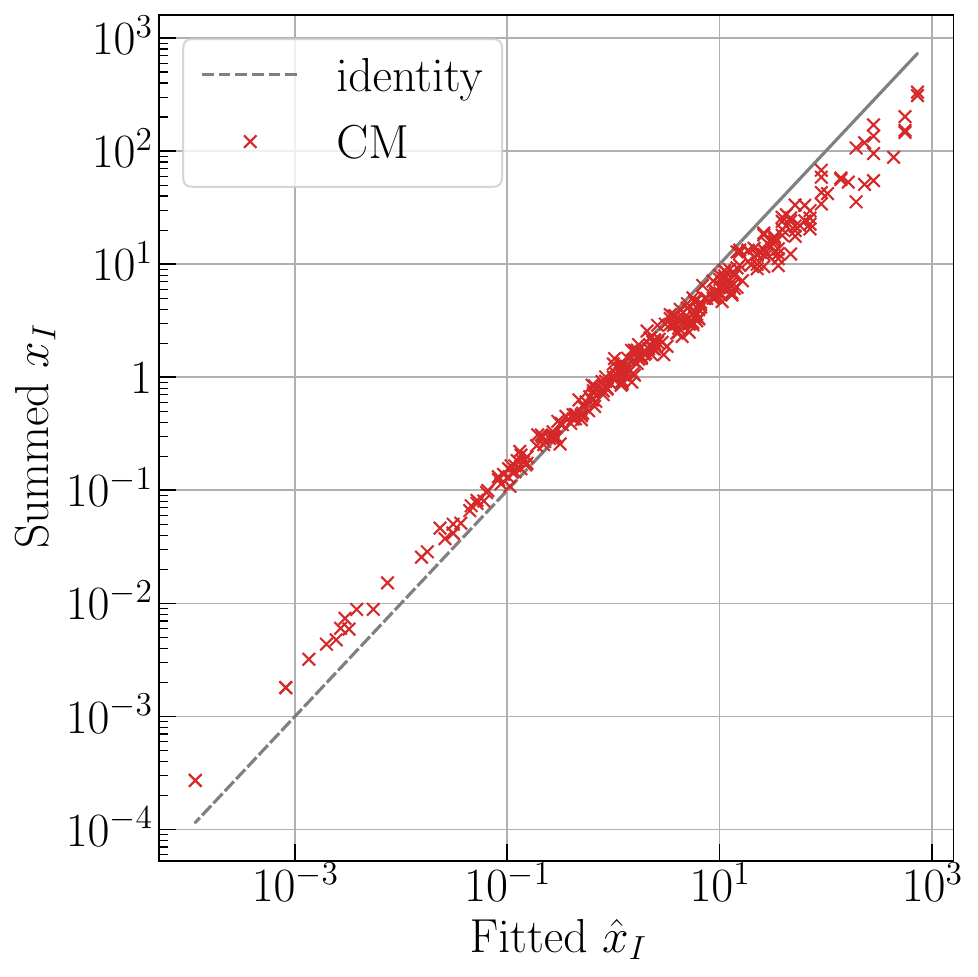}}
    \hfill
    \subfloat[degcMSM \label{ci2}]{\includegraphics[width=\subfigurewidth]{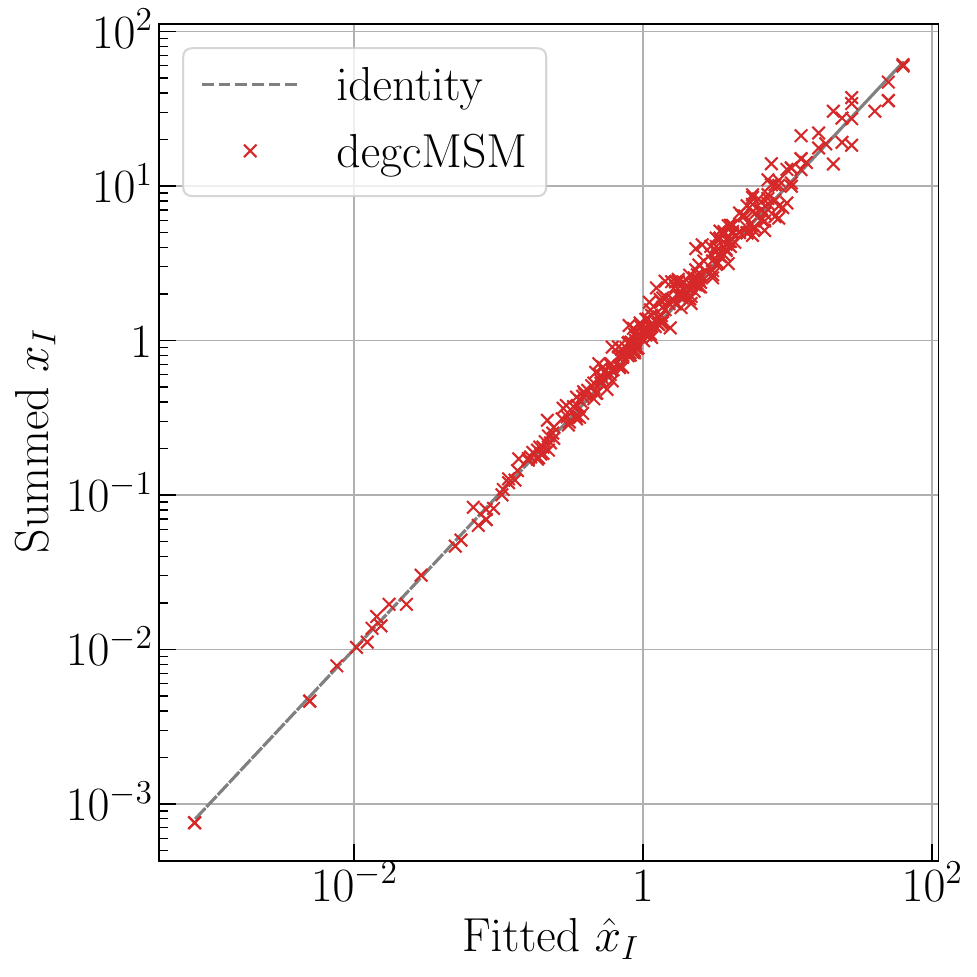}}
    \hfill
    \subfloat[maxlMSM \label{ci3}]{\includegraphics[width=\subfigurewidth]{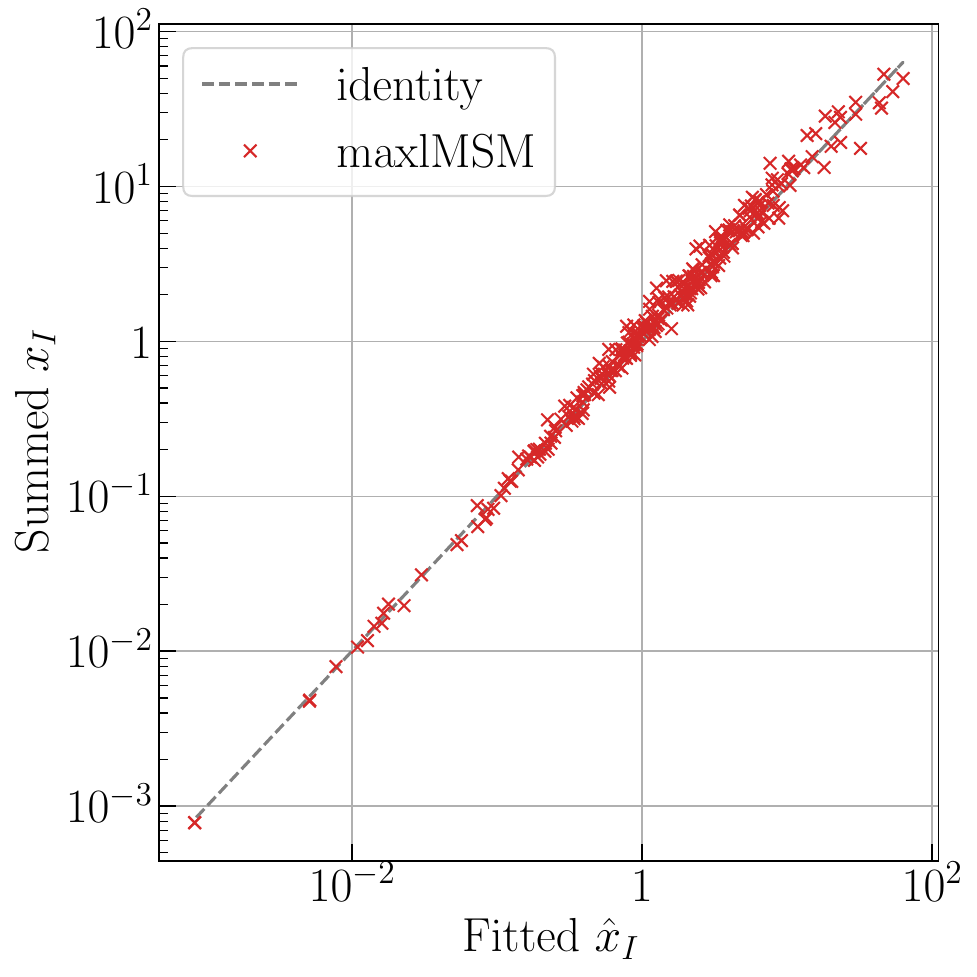}}
    \caption{
    Summed (y-axis) VS Fitted (x-axis) Parameters at level 2. From the left, we show the CM, degcMSM and maxlMSM. The identity line illustrates the perfect matching among the two.
    }
    \label{fig:ING_sumx_vs_fitx}
\end{figure*}

In the remainder of this paper, we evaluate how well the observed network measurements $Y_{i_\ell}$ align with the corresponding model predictions at different aggregation levels $ \ell $. Specifically, we compare them to the expected values obtained using the summed probabilities $\langle Y_{i_\ell} \rangle_{\text{sum}} $ and the re-fitted probabilities $ \langle Y_{i_\ell} \rangle_{\text{fit}} $.

In \autoref{fig:ING_sumx_vs_fitx}, we illustrate the relationship between the summed parameters $ x_{I} := \sum_{i_0 \in I} x_{i_0}, I := i_2 $, and the fitted parameters $ \hat{x}_{I} $ at aggregation level 2 for three models: the CM, degcMSM, and maxlMSM. The dashed identity line represents perfect agreement between the two.

For the CM (leftmost panel), the plot shows the characteristic ``cosine-like'' curve, indicating that low values of $x_I$ (from summation) tend to overestimate the fitted $\hat{x}_I$, while high values underestimate it. This bias stems from imposing a summation-based renormalization, despite the fact that the CM model is not scale-invariant and lacks a natural rule for renormalization (see \autoref{SI:sec:CM}). The panels on the right correspond to the MSMs (degcMSM and maxlMSM), where the summed and fitted parameters shows a better agreement at level 2 (see also \autoref{ci2} and \autoref{ci3}), indicating better consistency under the renormalization scheme.

This discrepancy can be attributed to the differences in functional forms between the models, as shown in \autoref{fig:function_comparison_CM_MSM}. In particular, the MSM connection probability satisfies $p^{\text{MSM}}(x) > p^{\text{CM}}(x) ; \forall x$. As a consequence, to reproduce the same low (or high) observed degrees, the CM typically requires smaller (or larger) fitted values $\hat{x}$ compared to the MSM. Additionally, since the CM function increases more gradually, i.e. it saturates more slowly, the optimizer is more sensitive to parameter variation, which allows it to reach lower values. As a result, the CM parameter range spans nearly two orders of magnitude more than the MSM's.

Interestingly, the close alignment between the summed and fitted parameters in the MSMs was not guaranteed a priori, although it was expected due to the multi-scale design. For a more in-depth discussion on this topic, refer to the following subsection \autoref{sec:ScaleInv_isnot_Fitting}.

\subsubsection{Interlude: Scale-Invariance Implications}
\label{sec:ScaleInv_isnot_Fitting}
The principle of scale-invariance, as defined in \autoref{SI:eq:MSM_micro_psumVSpcg_implicit}, requires that the network ensembles generated from summed and coarse-grained probabilities are statistically equivalent. This does not guarantee that the predictions derived from a summed model will match those from a re-fitted one. Therefore, true deviation from scale-invariance should be only assessed using the \autoref{SI:eq:MSM_micro_psumVSpcg} shown in \autoref{fig:ING_sum_vs_cg_pmatrix}, since other metrics can lead to misleading results that depend on the specific scale or quantity.

Nevertheless, we perform this broader comparisons to investigate the practical consequences of the scale-invariance principle, which is a crucial for real-world applications. For deeper insights on how the partitions affect this devation, refer to the companion paper \cite{2024_MSNE_milocco}.

\setlength{\temp}{\linewidth}
\begin{figure*}[t]
    \centering
    \subfloat[Level 0 \label{fig:level0_strengths}]{\includegraphics[width=\temp]{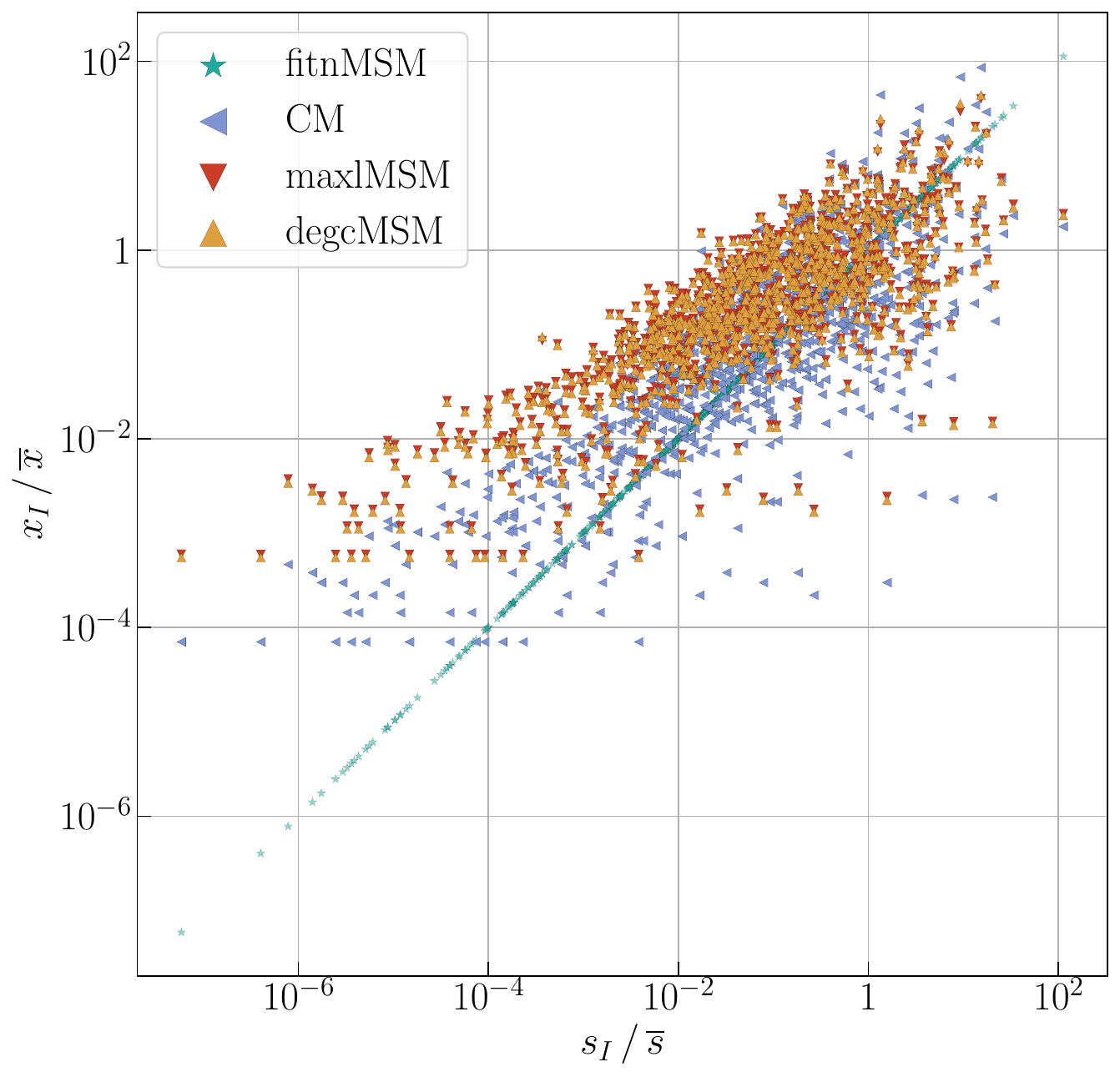}}
    \hfill
    \centering
    \subfloat[Level 2 \label{fig:level2_strengths}]{\includegraphics[width=\temp]{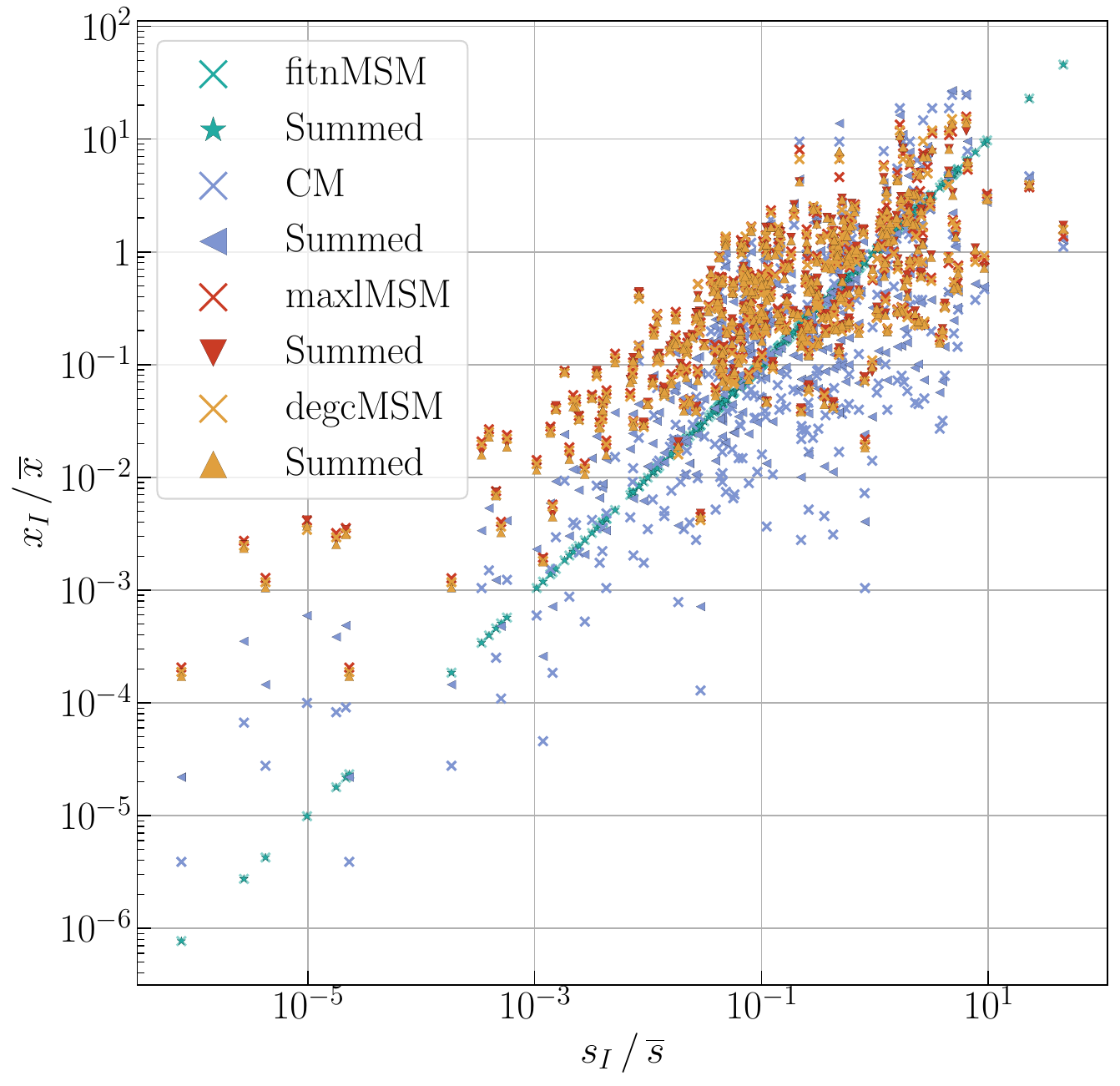}}
    \caption{Rescaled scalar parameters against the rescaled strengths at level $ 0$ and $ 2$. For level $ 0$, the weights are directly learned from the topology, whereas at level $ 2$ they are calculated by summing the microscopic parameters.}
    \label{fig:ING_sumXvsStrengths}
\end{figure*}

\subsection{Fitted VS Exogenous Variables}

The CM, degcMSM, and maxlMSM determine their parameters based on the network topology. For example, the CM constrains the degree sequence on average. Inspired by previous studies \cite{2004_Fitness_WTW_Garlaschelli, 2015_CimiModel_Cimini}, here we examine the relationship between the topologically-inferred parameters and the exogenous variables $f$, referred to as fitnesses, that characterize the underlying system\footnote{In the context of the WTW, GDP has previously been shown to play a dominant role \cite{2004_Fitness_WTW_Garlaschelli}. A similar fitness-based approach was later applied to the inter-firm network (IFN) in \cite{2015_CimiModel_Cimini}.}.

In \autoref{fig:ING_sumXvsStrengths}, we illustrated this relationship by plotting the node-specific parameters $x_i$ from the fitted and summed models on the y-axis, whereas the node strengths on the x-axis, measured at levels $\ell = 0$ and $\ell = 2$. To ensure additivity, at any resolution level, the strength of node $i$ is defined as: $s_i := \sum_j w_{ij}$ which includes self-loops (i.e., monetary flows from a node to itself). Remember that we excluded the self-loops at the firm level, prior to  coarse-graining, to prevent that self-payments (potentially arising from internal  accounting transactions) could distort the modeling of the production network.
To standardize the measurements, both parameters and fitnesses are normalized by their respective averages. Hence, the identity line in the figure represents a perfect agreement between fitness and model parameters up to a scaling factor. By construction, fitness models lie exactly on this line (with only the fitnMSM shown in the plot for clarity). Each model is represented using a distinct color, while crosses and circles indicate fitted and summed parameters, respectively, at the given resolution level.

\subsubsection{Level 0}

At level 0 (see \autoref{fig:level0_strengths}), node strengths closely approximate the parameters of the Configuration Model (CM), while they tend to underestimate those of the MSMs, especially for nodes with low values. As a result, the fitnCM is expected to yield better local-level predictions compared to the MSMs (refer to \autoref{fig:ING_rec_acc_by_level}).

As described in \autoref{sec:Methodology}, CM and degcMSM assign the same fitted parameter to all nodes with equal degree. This behavior becomes particularly noticeable for the lowest parameter values (related with the minimum degree), where the one parameter value corresponds to various strengths.

In the case of the maxlMSM, nodes with the same neighbors receive the same parameter. The plateau observed for low-regime weights, however, arises from very small variations in the fitted parameters (on the order of ( $\sim 10^{-3} $)), rather than from identical connectivity patterns, such as all nodes linking to the same hubs.

All models, except the maxlMSM\footnote{For the ION, the signed relative error of the predicted number of links is very low, approximately $ \frac{L_{\text{maxlMSM}} - L}{L} \approx 10^{-3}$. This holds true also in the context of the WTW.}, exactly match the observed total number of links at the fitting level. However, there is no straightforward relation between the node strengths and the expected network statistics, since strengths can either overestimate or underestimate the parameters.

For instance, the expected degree is a non-linear monotonic function $f$ of the parameter $x_i$, namely 
\begin{equation}
    \langle k_i \rangle := \sum_{j (\neq i)} p_{ij} \equiv f(x_{i})
\end{equation}
where $ f$ depends on all the parameters but $ x_{i}$. Thus, even if $s_i < x_i^{\text{degcMSM}}$, it does not necessarily imply that $\langle k_i \rangle{\text{fitnMSM}} < \langle k_i \rangle_{\text{degcMSM}}$.

For a more detailed analysis of node-level quantities, we refer the reader to the following sections.

\subsubsection{Level 2}

\autoref{fig:level2_strengths} illustrates the behavior of the summed variables $x_I$ with respect to the fitted values $\hat{x}_I$ for the CM, degcMSM, and maxlMSM. As expected, the discrepancy in the CM is larger than that observed in the two MSMs. This is because the CM is a single-scale model that does not support renormalization by summing the internal weights, or by any other procedure, as discussed in \autoref{SI:Inconsistency_CM}.

In contrast, the parameters of the MSMs vary within a much narrower range, roughly within one order of magnitude. Interestingly, however, these models are not explicitly designed to ensure that the summed values $x_I$ match the fitted values $\hat{x}_I$. Instead, they are constructed to remain \textit{self-consistent} as discussed in detail in \autoref{sec:ScaleInv_isnot_Fitting}. Additionally, when degree-corrected models are applied, for a single value of the fitted weight $\hat{x}_I$, there can be several distinct values of the corresponding summed parameter. This reflects the variability across internal strength distributions, of nodes with the same degree.

Another key observation is that most of the strength values tend to overestimate the MSM parameters, and this bias persists even when moving to coarser levels of aggregation. Surprisingly, the summed parameters in the CM, despite its lack of scale-invariance, are not drastically different from the fitted ones—highlighting an unexpected level of alignment.

\subsection{Expected Number of Links}

% \begin{table*}[t]
%     \centering
%     \begin{tabular}{c|c|c|c|c|c}
%         \toprule
%         Level $ \ell$ & fitnCM  & fitnMSM & CM & degcMSM & maxlMSM \\
%         \midrule
%         0 & 0 & 0 & 0 & 0 & 1e-6\\
%         1 & -4.3 & -3.8 & -1.8  & 1.8 &1.9\\
%         2 & -2.1 & 0.2 & -3.2  & 2.7 &2.8\\
%         3 & -3.8 & -2.1 & -3  & 1   &1.1\\
%         \bottomrule
%     \end{tabular}
% \end{table*}

\begin{figure}[htbp]
    \centering
    \includegraphics[width=\linewidth]{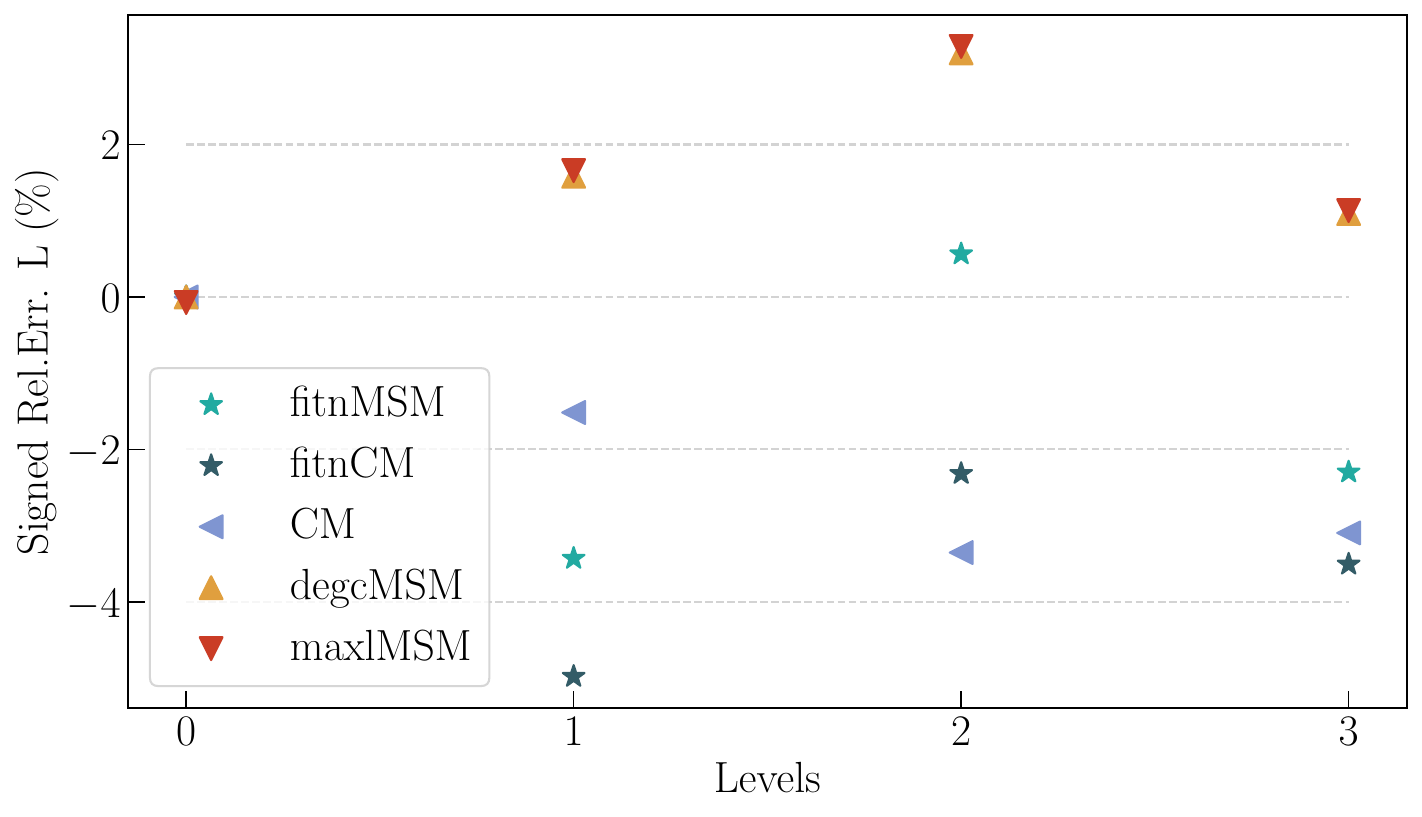}
    \caption{Evolution of the signed relative error accross scales. The error vanishes for level $ 0$, since the models have been fitted there. For the higher scales, we used the summed parameters as described in the main text.}
    \label{fig:ING_rel_err_n_edges_across_levels}
\end{figure}

In \autoref{fig:ING_rel_err_n_edges_across_levels}, we report the signed relative error $\sigma_m$ (on the y-axis) for each model $m$, across different resolution levels $\ell$ (on the x-axis). The signed relative error (SRE) is defined as
\begin{equation}
    \sigma^{(\ell)}_{m} := \frac{\langle L_{\ell} \rangle_m - L_{\ell}}{L_{\ell}}
\end{equation}
where $ \langle L_{\ell} \rangle_m$ is the expected number of links by the model $ m$ and $  L_{\ell}$ the corresponding empirical value at level $ \ell$. The reconstruction models (all but the maxlMSM) were fitted to retrieve at least the total number of links at the finest level. As a consequence, $\sigma_m^{(0)}$ vanishes for all models except the maxlMSM, which achieves a near-zero error $\sigma_{\text{maxlMSM}}^{(0)} \approx 10^{-6}$. For higher levels of aggregation, the expected number of links is determined using the models' summed probabilities, as defined in \autoref{eq:MSM_pIJ} and discussed in \autoref{SI:sec:sumCM}.

When averaging SRE across all levels, the fitnMSM results as the most accurate model overall, followed in order by degcMSM, maxlMSM, fitnCM, and finally the CM. However, if we focus on the absolute relative error, the CM and degcMSM yield similar performance up to the final level.

More broadly, while some metrics may suggest that the CM exhibits scale-invariance, this behavior only holds up to a certain degree of resolution, beyond which its single-scale nature becomes evident. Consequently, testing for scale-invariance through a specific metric should be done with care, as discussed in \autoref{sec:ScaleInv_isnot_Fitting}.

\subsection{Network Measurements}

In \autoref{fig:ING_net_meas_by_level_CM_degcMSM}, we show how three node metrics—Degree (DEG), Average Nearest Neighbor Degree (ANND), and Clustering Coefficient (CC)—evolve across different levels of resolution. Each panel illustrates the predictions from the empirical network $\mathbf{A}^{(\ell)}$, the summed model $\mathbf{P}{\text{sum}}^{(\ell)}$, and the refitted model $\mathbf{P}{\text{fit}}^{(\ell)}$. Small inset plots display the agreement between the summed and refitted probabilities, with point-colors representing the log-density $\rho$: from blue (low density, $\rho = 0$) to red (high density, $\rho = 1$).

The first set of plots (\autoref{fig:ING_net_meas_by_level_CM}) illustrates the behavior of the Configuration Model (CM), while the second (\autoref{fig:ING_net_meas_by_level_degcCM}) for the degree-constrained MSM (degcMSM). Similar plots were produced for fitnCM, fitnMSM, and maxlMSM, but they did not provide depper insights than the ones appreciable from these two representative cases.

Focusing on the ANND and CC, the CM consistenty underestimates the observed metrics as the aggregation levels increase. In contrast, the degcMSM predictions remain better aligned with the empirical measurements at all scales. Interestingly, the degcMSM seems to better approximate the observed metrics also at the fitting level $ 0$ (see also \autoref{fig:ING_rec_acc_by_level}) as the CM tends to overestimate ANND and CC for high-degree nodes.

These results are also reflected in the inset plots. At level 2, for example, the CM's summed probabilities $ p^{\text{CM}}{\text{sum}} $ tend to cluster near 0 and 1, underestimating the more widely distributed refitted values $ p^{\text{CM}}{\text{fit}} $. On the other hand, MSM summed probabilities $ p^{\text{MSM}}_{\text{sum}} $ are closely aligned with their refitted counterparts, scattered around the identity line. This means that at coarser levels, CM predictions tend to underestimate network statistics even more than MSM models do.

\setlength{\temp}{1.7\linewidth} % Adjust the 0.45 to whatever fraction of the line width you prefer
\begin{figure*}[p]
    \centering
    \subfloat[Multi-Level Configuration Model \label{fig:ING_net_meas_by_level_CM}]{\includegraphics[width=\temp]{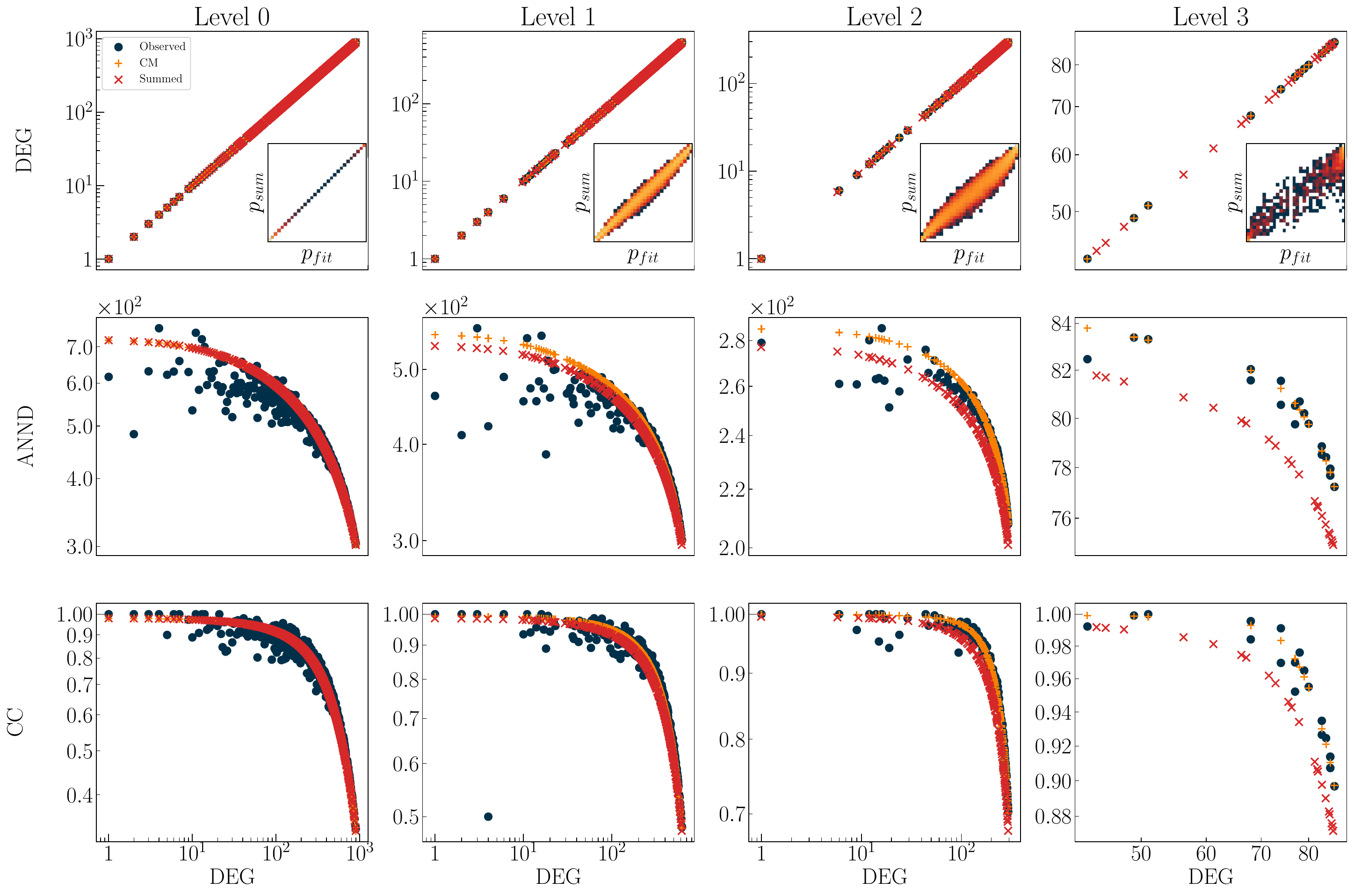}}
    \vfill
    % \vfill %\floatsep% normal separation between figures
    \subfloat[Multi-Level Degree-Corrected MSM \label{fig:ING_net_meas_by_level_degcCM}]{\includegraphics[width=\temp]{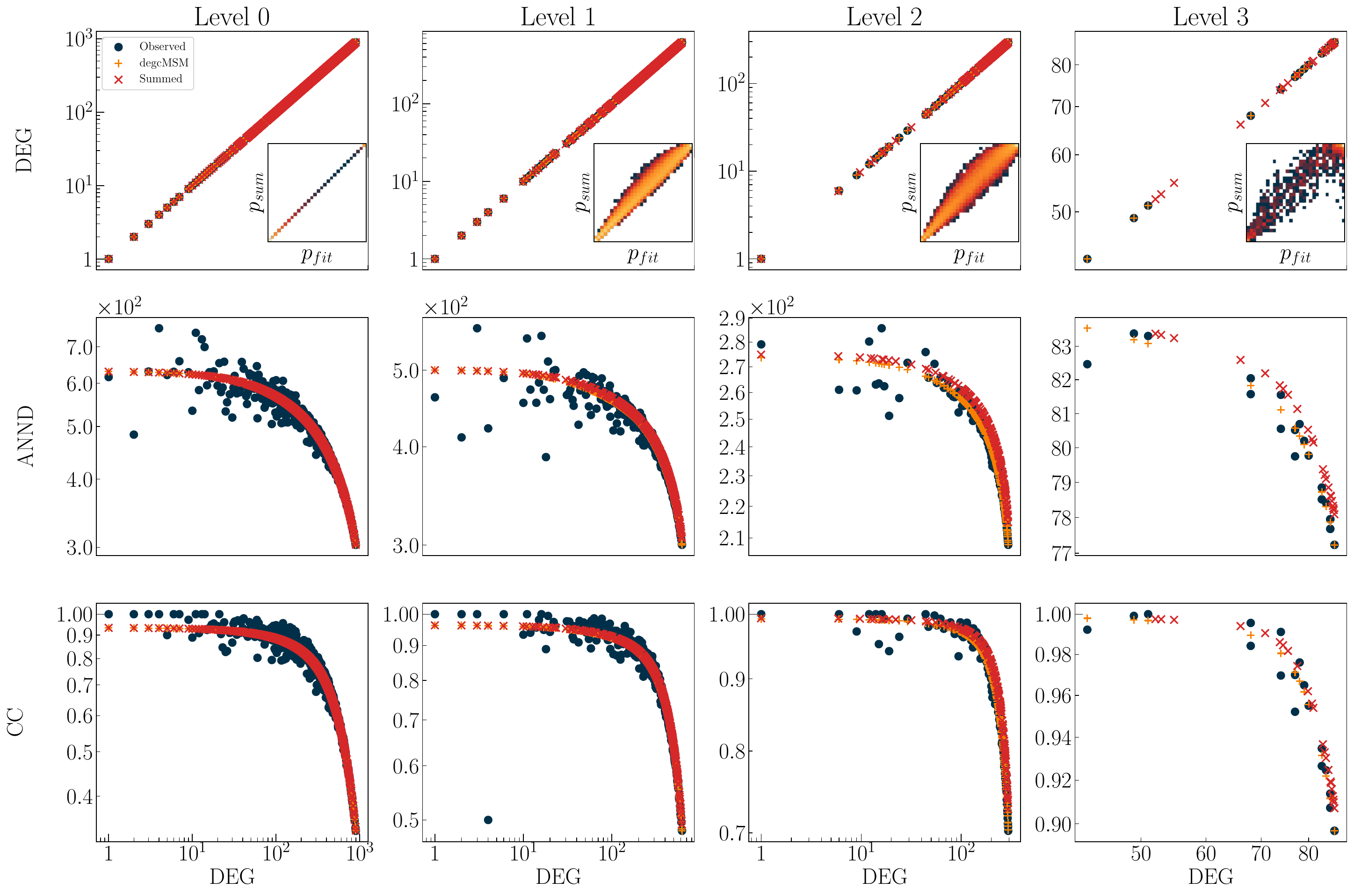}}
    \caption{Multi-Scale comparison of the Configuration Model and degree-corrected Multi Scale Model. Both have been fitted at level $ 0$, whereas the upper-scale parameters have been obtained undersummation. In the inset of DEG, we reported the values of the summed probability against the fitted ones. The DEG, ANND and CC metrics have been obtained by precisely using those probabilities.} 
    \label{fig:ING_net_meas_by_level_CM_degcMSM}
\end{figure*}

\subsection{Reconstruction Accuracy}

\begin{figure*}[t]
    \centering
    \subfloat[Reconstruction Accuracies of DEG, ANND and CC for all the models and levels. The models are fitted at level $ 0$, whereas at higher level we applied their summed version. \label{fig:ING_rec_acc_by_level}]{
        \includegraphics[height=.32\linewidth]{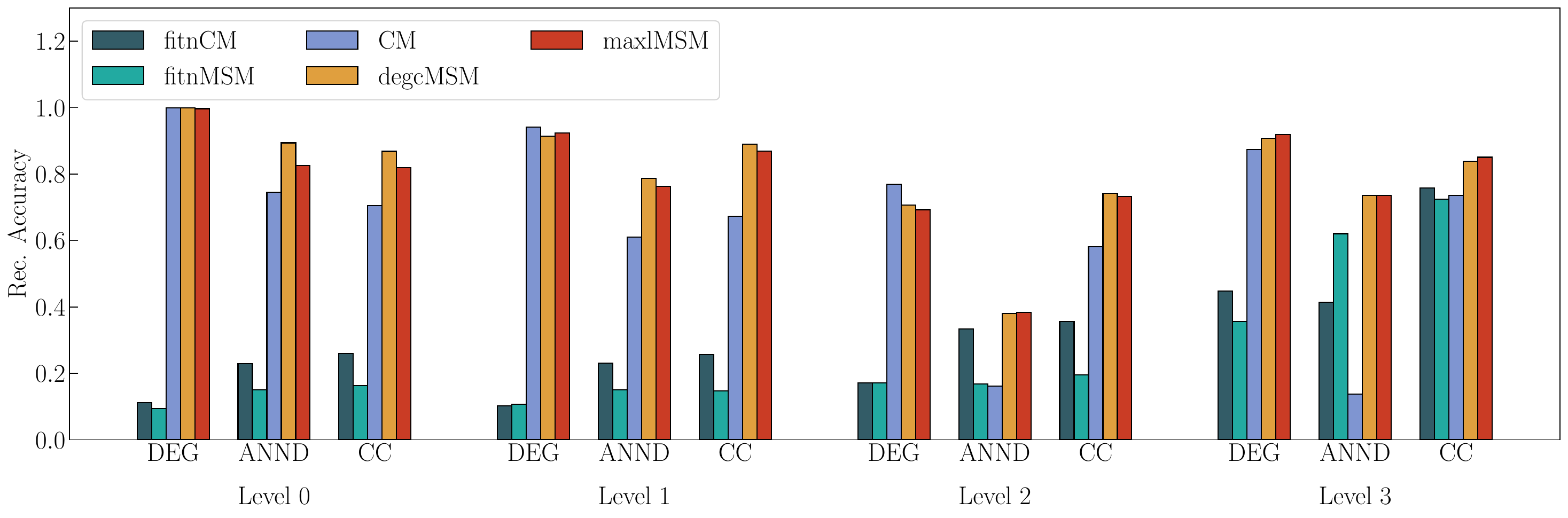}
    }
    \label{fig:ING_auc_rec_acc}
\end{figure*}

\autoref{fig:ING_rec_acc_by_level} shows the Reconstruction Accuracies (RA) for the ION dataset \cite{2023_RecEcon_DiVece}. In simple terms, RA represents the probability that an observed node-level metric $ Y_{i_{\ell}}$ falls within the uncertainty range predicted by the model. Put differently, it measures how well a model captures a given metric $ Y$ at resolution level $ \ell$. We denote this value as $RA^{(\ell)}_m\left(Y\right)$ and formally define it in \autoref{eq:RecAcc}.

Since all models are fitted at the microscopic level $\ell = 0$, most of the highest RA values are also found at this scale. However, some exceptions occur, where the best performance appears at higher aggregation levels. A notable example is the fitnCM model, which achieves its highest RA for the clustering coefficient (CC) at level 3. This level is especially peculiar, as it is the final one before the network becomes fully connected. Consequently, the structure is much more heterogeneous than in previous levels. Additionally, the number of nodes at this stage is the smallest among all layers, meaning that the accurate prediction of even a single node contributes more significantly to the overall score.

Although in \autoref{fig:ING_net_meas_by_level_CM} we showed that the Configuration Model (CM) consistently underestimates metrics like DEG, ANND, and CC, the corresponding RAs do not drop to zero. This is because the model's predicted uncertainty range is wide enough to include the observed values. In contrast, the degcMSM and maxlMSM—which are scale-consistent by design—achieve high RA values at all levels. The only significant exception occurs for the average nearest neighbor degree (ANND) at level 2. Here, the empirical ANND is not well captured by the MSMs. However, this appears to be a second-order effect, as we still observe strong agreement between the summed and fitted parameters (see \autoref{fig:ING_sumx_vs_fitx}) and high RA values for the degrees at the same level. As discussed in \autoref{sec:ScaleInv_isnot_Fitting}, this is not problematic: the principle of scale-invariance in MSMs enforces self-consistency across levels, rather than exact matching of observed values at each individual resolution (see also \autoref{SI:eq:MSM_micro_psumVSpcg}).

Among all models, the degree-constrained MSM (degcMSM) delivers the best performance at the fitted level $\ell = 0$, outperforming the CM as well. This outcome is somewhat surprising, as previous studies showed that logistic PCA—a Shannon Entropy-based, single-scale model—gave better results than MSMs at this resolution \cite{2021_symLPCA_Chanpuriya}. Although the objectives of that work (network modeling) differ from ours (network reconstruction), CM was still expected to be the top performer at the microscopic scale. Yet, our results do not confirm that assumption.

Another interesting result is the performance of the maximum likelihood MSM (maxlMSM). Its reconstruction accuracies are nearly the same as those of degcMSM, suggesting that constraining degree sequences and maximizing likelihood can yield comparable results—a pattern also observed in \autoref{fig:ING_sumx_vs_fitx}.

\autoref{fig:ING_rec_acc_by_level} also includes, for completeness, the reconstruction performance of the fitness-based models. Although neither fitnCM nor fitnMSM performs well in capturing local metrics, fitnCM outperforms fitnMSM. This result aligns with \autoref{fig:ING_sumXvsStrengths}, which shows a better agreement between node strengths and topological parameters in the fitnCM case.

That said, when focusing on global metrics—such as the total number of links $L$—the fitnMSM consistently captures $ L$ across scales (see \autoref{fig:ING_rel_err_n_edges_across_levels}), even though we fitted the unique parameter at level $0$. This behavior aligns with their objective, i.e. capturing global metrics, where local fluctuations are average out, rather than node-level details.

\subsection{ROC-PR Curves}

\begin{figure*}[htbp]
    \centering
    \subfloat[Evolution of the area Under the ROC and PR curves (y-axis) for the fitted models (level $ 0$) and summed ones (level $ 2$) as the scales increase (x-axis) or, equivalently, the numbers of nodes diminish. \label{fig:ING_auc_roc_prc}]
        {\includegraphics[height=.43\linewidth]{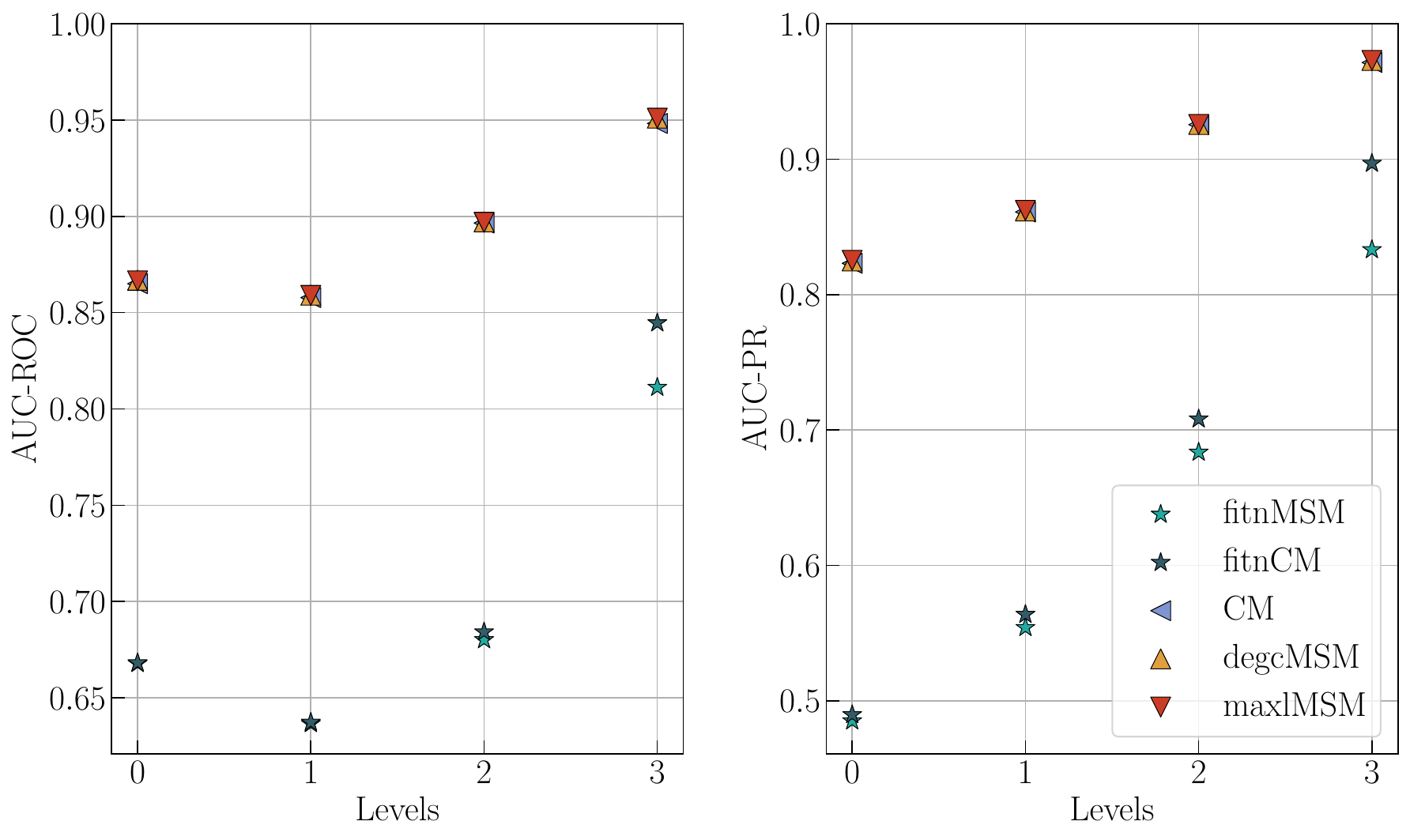}
    }
\end{figure*}

In \autoref{fig:ING_auc_roc_prc}, we reported the Area Under the Curve of the Receiver Operating Characteristic (ROC) and the Precision-Recall (PR) curves, defined in \autoref{sec:Scores}. The CM exhibits values that are often indistinguishable from its local multi-scale variants, making it hard to find the single-scale or scale-invariant one. Moreover, the fitnCM consistently outperforms the fitnMSM, as it benefits from a more effective fitness ansatz (see \autoref{fig:ING_sumXvsStrengths} and \autoref{fig:ING_rec_acc_by_level}). From these metrics, it becomes even more difficult to identify which models are scale-invariant.

Interestingly, all AUC-PR scores tend to (approximately) linearly increase with the resolution scale (see \autoref{fig:ING_level02_aucs}). This occurs even though both true positives and predicted positives decrease with scale; their ratio, however, increases. This trend was somewhat expected: as the network becomes denser at coarser scales, the probability of correctly predicting a link among the expected ones increases—leading to a higher precision score, i.e., a larger value of $P PV$.

A similar trend appears also in AUC-ROC scores. However, at intermediate level $1$, 
we observe a noticeable drop in AUC-ROC scores, as the models exhibit decreased sensitivity to true positives in absolute terms. In other words, while the models improve in precision—correctly identifying positives among those they predict—they become less effective when performance is measured against the set of all true positives (recall).

The key takeaway is that evaluation highlights the relative strength of local models compared to global ones. Novertheless, it remains difficult to differentiate single-scale from multi-scale models, as their AUC performances are often coinciding. Most importantly, across both evaluation metrics and all resolution levels, fitnCM consistently outperforms fitnMSM that we know it shouldn't be the case based on the results from \autoref{fig:ING_rel_err_n_edges_across_levels}.

\setlength{\temp}{\linewidth}
\begin{figure*}[htbp]
    \centering
    \subfloat[Level 0 \label{fig:ING_level0_aucs}]{
        \includegraphics[width=\temp, trim = {34cm 0 0 0}, clip]{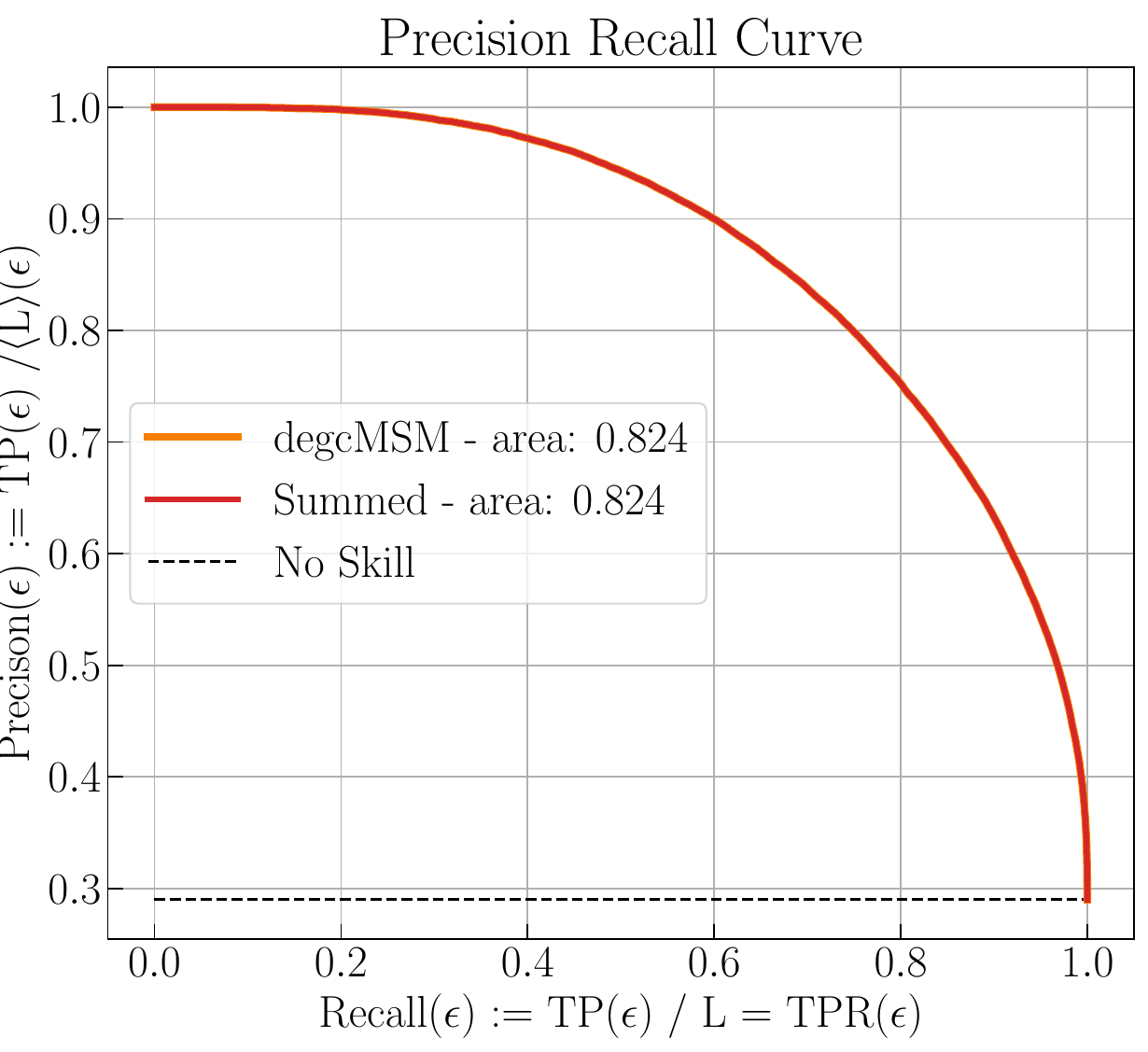}
        }
    \hfill
    \subfloat[Level 2 \label{fig:ING_level2_aucs}]{
        \includegraphics[width=\temp, trim = {34cm 0 0 0}, clip]{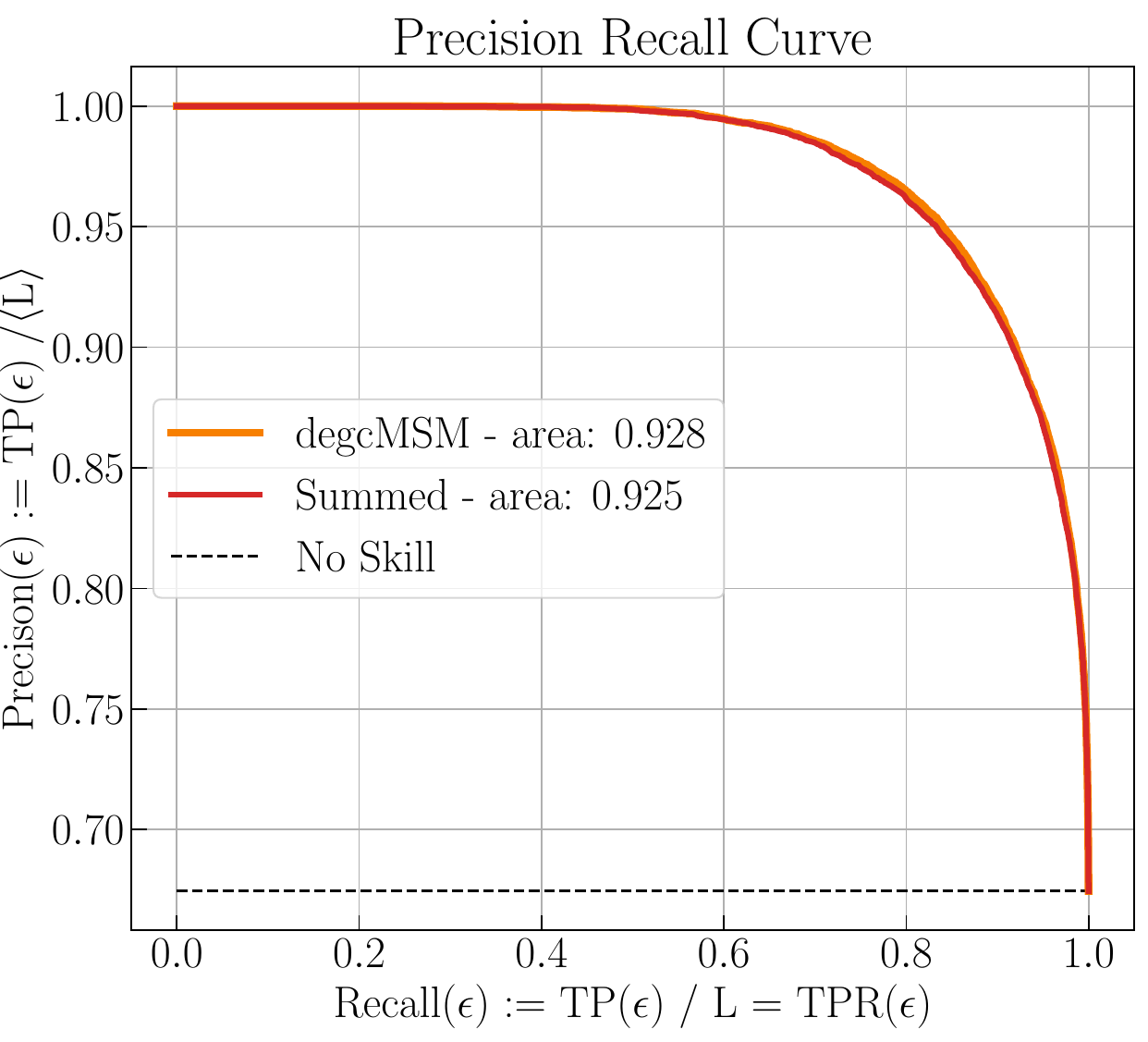}
        }
    \label{fig:ING_level02_aucs}
    \caption{Comparison of the PR curves for the degcMSM at scale $ 0$, where the parameters are fitted, and level $ 2$, with summed parameters.} 
\end{figure*}

\subsection{Expected Number of Triangles}

\setlength{\temp}{\linewidth}
\begin{figure*}[htbp]
    \centering
    \subfloat[Level 0 \label{fig:ING_level0_triangles}]{
        \includegraphics[width=\temp]{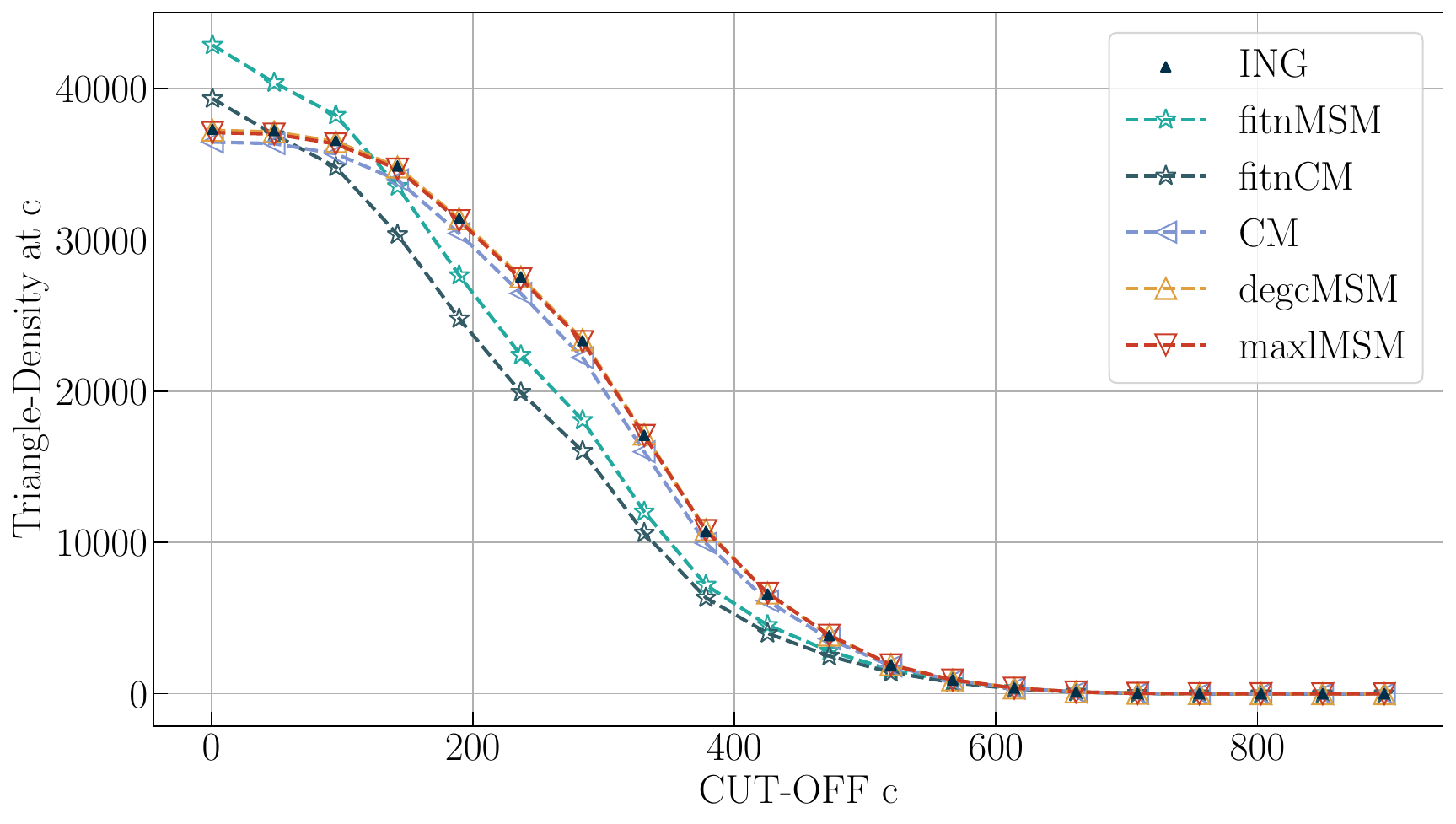}
        }
    \hfill
    \subfloat[Level 2 \label{fig:ING_level2_triangles}]{
        \includegraphics[width=\temp]{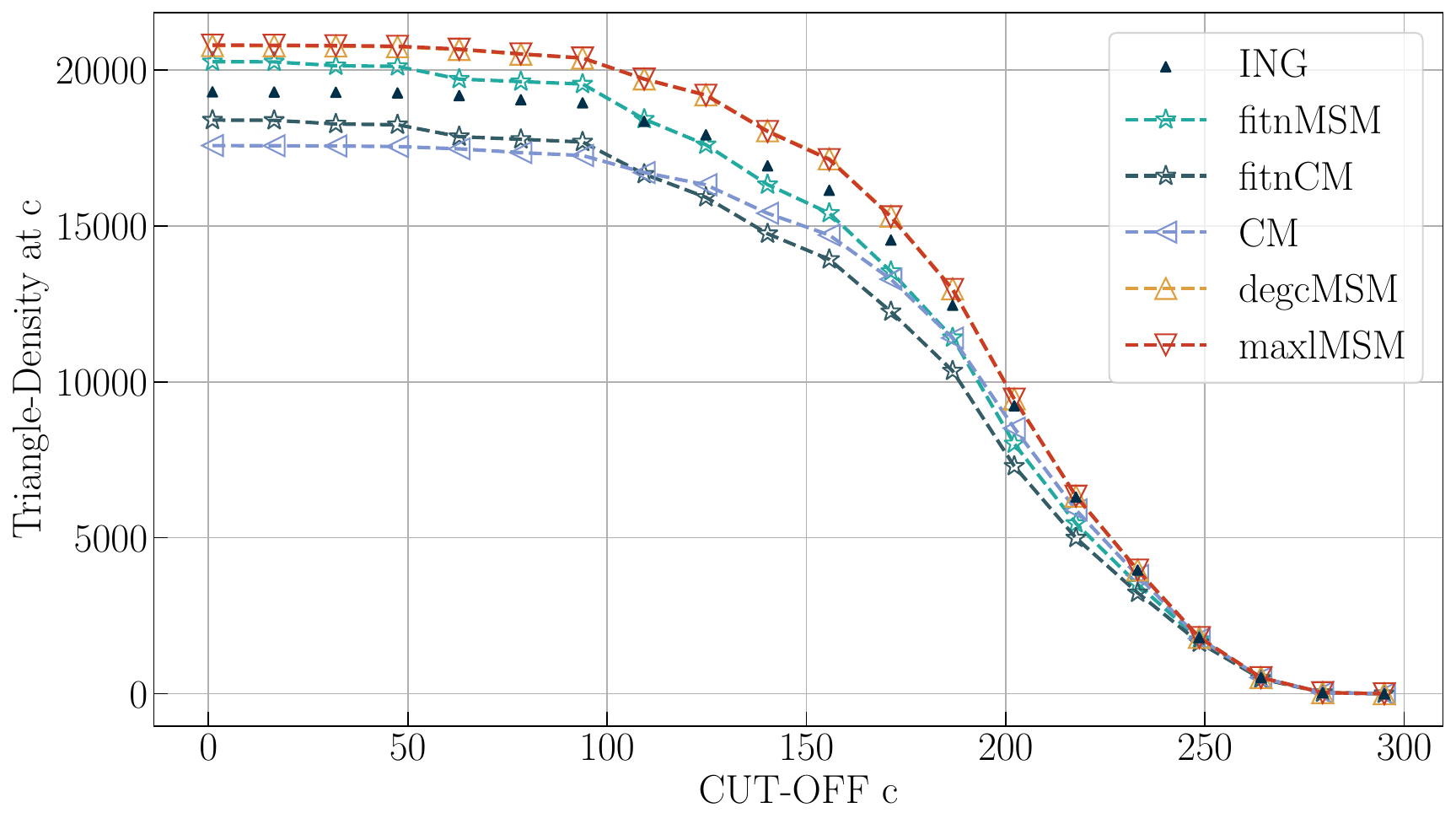}
        }
    \caption{Comparison of the triangle density on the subgraph obtained by filtering the nodes with a lower degree tha $ c$. The curves at scale $ 0$ involves models fitted at that level, whereas at resolution $ 2$ we used the summed model.}
    \label{fig:ING_levels02_triangles}
\end{figure*}

In \autoref{fig:ING_levels02_triangles}, we report the observed and expected triangle densities (TriDens), defined as $ \triangle(k_i \geq c) / N$ where the subgraph $\mathbf{G}{{i : k_i \geq c}}$ is obtained by extracting the nodes whose observed degree $ {k_i}{i \in [1, N]}$ is greater than or equal to a threshold ( c ). Here, ( N ) refers to the number of nodes at the given resolution level \cite{2021_symLPCA_Chanpuriya}.  

The panel \autoref{fig:ING_level0_triangles} shows results at level $\ell = 0$, where all models were originally fitted, whereas \autoref{fig:ING_level2_triangles} presents results at level $\ell = 2$ for the summed models. 

By construction, as $ c$ increases, the difference among the observed and expected densities vanishes until it coincides as $ c = N - 1$, i.e. for the whole graph. At very high thresholds $c \lessapprox N - 1$, only the highest-degree nodes (hubs) remain. However, these nodes tend to be weakly connected to each other, as the average nearest neighbor degree (ANND) decreases with increasing degree (see \autoref{fig:ING_net_meas_by_level_CM_degcMSM}). As a result, no triangle is closed among the ``richest'' nodes.

This outcome is not in contradiction with prior findings such as \cite{2020_impossibility_of_low_rank_red_Seshandhri, 2020_LPCA_Chanpuriya}, since we are analyzing the expected triangle densities rather than computing their exact values.

Our results show that, even with a scalar parameter (lowest-rank embedding), local models can faithfully reproduce triangle densities. In contrast, global fitness-based models have trouble replicating triangle patterns, as they are designed to capture aggregate properties of the network—such as $\sum_i \triangle(k_i \geq c) / N$—rather than node-level structural details. For this reason, results from the fitness models are included only for the sake of completeness.

\section{Conclusions}
\label{Conclusions}
The strength of graph-based models lies in their flexibility to represent different types of interactions by appropriately defining nodes and edges. By aggregating low-level (microscopic) nodes into larger block-nodes, a coarse-grained network is obtained—a simplified version of the original graph. Repeating this procedure across multiple levels produces a hierarchical, multi-scale unfolding of the observed system. We applied this approach to both the ION and WTW datasets (see \autoref{sec:DatasetDescription}), showing how a single underlying generative process can be represented consistently across different resolutions.

Most existing network reconstruction techniques \cite{2024_Rec_Supply_Chain_Mungo} are designed to operate at a single level of resolution, estimating link probabilities based solely on the data at that scale. As a result, they often overlook the importance of model consistency across scales. 

To highlight this limitation, we compared a widely used single-scale model—the Configuration Model (CM)—with our proposed multi-scale model (see \autoref{sec:Methodology}). For completeness, we also evaluated the performance of the fitness-based Configuration Model (fitnCM), the fitness-based MSM (fitnMSM), and the maximum-likelihood estimation of MSM parameters (maxlMSM). The key difference between these models lie in scale-invariance: MSMs are inherently renormalizable, aligning with the (hidden) generative process, whereas the CM are not. Specifically, the block-node parameters are defined as the sum of the parameters of their constituent nodes (see \autoref{eq:MSM_sum_graining_rule}), allowing for a principled interpretation of weights aggregation across scales. In contrast, CM-based models lack this scale consistency of its parameters and must be refitted independently at each level. More in general, from the single-scale model perspecive, each aggregated level is treated as an entirely new and unrelated structure—as if generated by a different underlying process—even though, in our case, all levels derive from the same multi-resolution aggregation of the original network.

As detailed in \autoref{sec:Results_and_Discussions}, the CM tends to perform well only at the resolution at which it is fitted, but struggles to generalize when applied to coarser levels. In contrast, our multi-scale model provides a consistent structural representation across resolutions, requiring only knowledge of how microscopic nodes are grouped into larger blocks. This advantage is most evident at higher aggregation levels, such as the final layer of the ION.

Together, these findings emphasize the importance of adopting a multi-scale perspective. This approach provides a robust and flexible framework for understanding complex systems by allowing analysis at various resolutions, depending on the granularity of the phenomenon under investigation.

\section{DATA AND CODE AVAILABILITY}
The dataset on sector transactions (ION) used in the paper is highly confidential and cannot be disclosed. However, the world trade is a long-standing public dataset, one may find at \href{http://ksgleditsch.com/exptradegdp.html}{Gleditsch}. Lastly, the code is freely available as the \href{https://github.com/RMilock/multi-scale-node-embeddings/tree/main}{multi-scale-node-embeddings} python package.

\section{ACKNOWLEDGMENTS}
We thank ING Bank N.V. for their support and active collaboration. A special thanks to the whole DataScience team at ING Bank for their advice that helped shape this research. This work is supported by the projects ``Reconstruction, Resilience and Recovery of Socio-Economic Networks'' (RECON-NET EP FAIR 005 - PE0000013 ``FAIR'' - PNRR M4C2 Investment 1.3) and ``SoBigData.it - Strengthening the Italian RI for Social Mining and Big Data Analytics'' (Grant IR0000013 n. 3264, 28/12/2021 -\url{https://pnrr.sobigdata.it/}), financed by the European Union - NextGenerationEU. This publication is part of the project ``Network renormalization: from theoretical physics to the resilience of societies’’ with file number NWA.1418.24.029 of the research programme NWA L3 - Innovative projects within routes 2024, which is (partly) financed by the Dutch Research Council (NWO) under the grant https://doi.org/10.61686/AOIJP05368.

\bibliography{bibliography}

\clearpage
\appendix
\setcounter{page}{1}
\onecolumngrid

{\center
\textbf{Supplementary Material}\\
$\quad$\\
accompanying the paper\\
\emph{``Multi-Scale Network Reconstruction''}\\
%by R. Milocco, F. Jansen and D. Garlaschelli\\
$\quad$\\
$\quad$\\
}

\section{Configuration Model}
\label{SI:sec:CM}
The configuration model is a generative random graph model which is ``maximally random'' out of the node-degrees which are enforced on average. Its functional form is obtained by finding the stationarity point of the Lagrangian
\begin{equation}
    \label{eq:Lagrangian_CM}
    -\sum_{\mathbf{G} \in \mathcal{G}} P(\mathbf{G}) \ln(P(\mathbf{G})) + \sum_{i \in [1, N]} \theta_i \left( k_i - \sum_{\mathbf{G} \in \mathcal{G}} P(\mathbf{G}) k_i(\mathbf{G}) \right)
\end{equation}
with respect to the probability distribution $ P$. The notation used above defines $ \mathcal{G}$ as the ensemble graphs $ \textbf{G}$ sampled from the distribution $ P(\textbf{G})$ and $ \theta_{i}$ is the i-th Lagrange parameter connected to the observed degree $ k_{i}$. Moreover, the first term of the summation provides the ``maximal uncertainty'', namely the maximum of Shannon Entropy, whereas the second imposes the expected degrees (on average) to be equal to the observed ones.

The stationarity (functional) point for \autoref{eq:Lagrangian_CM} reads 
\begin{equation}
    \label{eq:CM_probability_theta}
    p_{ij} := \frac{e^{-\left(\theta_{i} + \theta_{j}\right)}}{1 + e^{- \left(\theta_{i} + \theta_{j}\right) }} 
\end{equation}
where the Lagrangian parameters $ \theta \in (- \infty, \infty)$ are fixed via \textit{maximum likelihood} - the least unbiased functional for these kinds of models \cite{2008_maxlike_unbiased_Garlaschelli}. Due to exponential form, this class of models is generally addressed in Physics as Exponential Random Graph models (ERG). In our context, the \autoref{eq:CM_probability} may describe either the connection probability of a link or the likelihood that a link-fermion particle has energy $ x_{i}x_{j}$ \cite{2004_StatMecNet_Park}. Indeed, since the links are either present or not (0 or 1 in the adjacency matrix), they must be described by the same statistics of a fermion particle.

Referring to \autoref{eq:CM_likelihood}, the hessian of the likelihood is positive defined \cite{2011_AnalMax_Squartini} and, therefore, the stationarity point (\autoref{eq:CM_maximum_point}) is the unique maximum of the likelihood. Moreover, it requires only the knowledge of the observed degrees -the observed quantity that one assumed to know from the beginning. Hence, this method is the ``most unbiased'' one \cite{2008_maxlike_unbiased_Garlaschelli} -no more information needed- where the degrees stand as the \textit{sufficient statistics} $ \left\{k_{i}\right\}_{i\in[1,N]}$ to enforce the $ \vec{x}$. In general, fixing the first moment of the probability distribution -method of moments- requires less information with respect to the knowledge of the full adjacency matrix -the input of the maximum likelihood principle. However, this is enough for the CM.

\subsection{Inconsistency of the CM: a trivial example}
\label{SI:Inconsistency_CM}
\begin{figure}[t]
    \begin{tikzpicture}[baseline=(base), node distance={2cm}, thick, main/.style = {fill=blue!20, draw, circle}] 
        \centering
        \node (base) at (0,-.5ex) {};
        \node[main] (0) {$0$}; 
        \node (3) [left of = 0] {\textbf{level 0} };
        \node[main] (1) [below of=0] {$1$};
        \node[main, fill=red!20] (2) [right of=0] {$2$}; 
        \draw[-] (1) to (2);          
    \end{tikzpicture}
    $ \qquad \stackrel{ \textnormal{Coarse-Graining} }{\longrightarrow} \qquad$ 
    \begin{tikzpicture}[baseline=(base), node distance={2cm}, thick, main/.style = {fill=blue!20, draw, circle}] 
        \centering
        \node (base) at (0,-.5ex) {};
        \node[main] (0) {A}; 
        \node[main, fill=red!20] (1) [right of=0] {B}; 
        \node (2) [right of = 1] {\textbf{level 1} };
        \draw[-] (0) to (1);
    \end{tikzpicture}
    \caption{Simple multi-scale network: the blue nodes belong to block $ A$ whereas the red node to the block $ B$.}
    \label{fig:trivial_cg_network}
\end{figure}
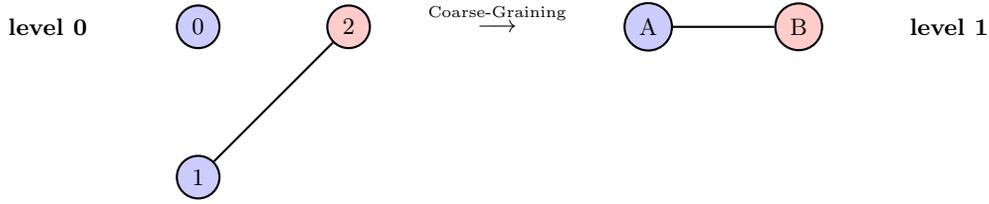

By referring to \autoref{fig:trivial_cg_network}, the microscopic network of 3 nodes $0,1,2$ merges into the coarse-grained network with blocks $ A,B$, containing respectively the nodes $ 0,1$ and the node $ 2$, namely $ A = \Omega(0) = \Omega(1), B = \Omega(2)$.

Assuming having fixed the CM parameters at the bottom level, the upper-level model is obtained by means of the RHS of the \autoref{SI:eq:MSM_micro_psumVSpcg}. For example, the probability among $ A, B$ block-nodes reads
\begin{align*}
    \label{eq:P_cg_AB}
    P_{cog}(\left\{x_{i}\right\}_{\left\{\left\{0,1\right\} \in A\right\}}, x_{2})
    &=  1 - \frac{1}{1+ x_{0}x_{2}} \frac{1}{1 + x_{1}x_{2}} \\
    &= \frac{x_{2} (x_{0} + x_{1}) 
    + x^{2}_{2} x_{0} x_{1}}{1 + x_{2} (x_{0} + x_{1})
    + x^{2}_{2} x_{0} x_{1}}.
\end{align*}
Hence, the functional dependence of $ P_{cog}$ is different from the actual CM one \autoref{eq:CM_probability} since there is no renormalization rule that can aggregate the parameters in such a way to recover the \autoref{eq:CM_probability}. 
In turns, this implies that the CM is not \textit{scale-invariant}. Indeed, the LHS of the \autoref{SI:eq:MSM_micro_psumVSpcg} is
\begin{equation*}
    \label{eq:P_AB}
    p(x_{A}, x_{B}) = \frac{x_{A}x_{B}}{1 + x_{A} x_{B}},
\end{equation*}
which by no means equal to $ P_{cog}$. Therefore, the \textit{scale-invariant} principle has to be enforced as a grounding premises, changing the model functional form accordingly.
Therefore, the model is not renormalizable.

\subsection{Summed (renormalized) CM}
\label{SI:sec:sumCM}
To enforce the \textit{self-consistency} (scale-invariance), i.e. the activation function to be the same at every scale, we impose the block-parameters are the sum of their members parameters $ x_{A} := (x_{0} + x_{1}) \textnormal{ and } x_{B} := x_{2}$ as for the MSM. With this choice, the \textit{summed} probability is
\begin{equation}
    p_{IJ} := \frac{x_{I} x_{J}}{1 + x_{I} x_{J}}
\end{equation}
where $ x_{I} := \sum_{i \in I} x_{i}, x_{J} := \sum_{j \in J} x_{j}$. Therefore, the comparison between the CM and the MSM would be about how well their expected measurements can predict the observed ones.

\section{Multi-Scale Model}

\subsection{Model Derivation}
\label{SI:sec:derivation_MSM_probability}
By referring to the full derivation of the multi-scale model (MSM) \cite{2023_MSNR_Garuccio}, we report here only the crucial passages for our treatment. The observed network is assumed to lay at the lowest level $ 0$, whereas the higher scales $ \ell \geq 1$ identify the coarser representations obtained by aggregating the level $ 0$ graph. However, the treatment would hold for every pair $m (\geq 0), \ell \geq m$ as pointed out in \cite{2023_MSNR_Garuccio}, to which we refer for a more rigorous derivation.

Let us consider the \textit{binary undirected} adjacency matrix $\mathbf{A}^{(\ell = 0)}$ describing the microscopic interactions among the 0-nodes. To obtain the coarser graphs, one needs a \textit{hierarchical and non-overlapping} partition of the microscopic nodes $ \left\{\mathbf{\Omega}_{\ell}\right\}_{\ell \geq 0}$ into block nodes. More concretely, the block-nodes $I := i_{\ell}$, at level $ \ell+1$, is defined as $I := \mathbf{\Omega}_{0 \to \ell - 1}(i_0)$
reached after either $ \ell$ subsequent partitions or, equivalently, after one bigger aggregation $\Omega_{0 \to \ell-1} := \Omega_{\ell-1} \circ \dots \circ \Omega_{0}$. Once the groups $ I$ have been identified, one assigns a link among the blocks via
\begin{equation}
    \label{SI:eq:cg_rule_nextl}
    a_{IJ} = 1 - \prod_{i_0 \in I, j_0 \in J} (1-a_{i_0j_0})
\end{equation}
where 
% all the capital indexes refer to the scale $ \ell$ the smaller ones refer to level $0$, and 
$a_{IJ} := a^{(\ell)}_{i_{\ell+1}j_{\ell+1}}, a_{i_{0}j_{0}} := a^{(0)}_{i_{0}j_{0}}$ and $J := j_{\ell} = \mathbf{\Omega}_{0 \to \ell-1}(j_{0})$. In other words, a link among the block-nodes exists if there is at least one micro-link among their members. In turns, by iterating the procedure $ \ell$ times, it is possible to create $ \mathbf{A}^{(\ell)}$, the low-resolution graph describing the \textit{original} phenomenon the $ \ell$th scale.

In order to model this architecture, one needs several assumptions. The \textit{first one} requires that the MSM must describe the microscopic matrix $\textbf{A}^{(0)}$, namely 
$\textbf{A}^{(0)} \sim P^{(0)}\big(\mathbf{A}^{(0)},\mathcal{X}^{(0)}\big)$ 
subject to 
$P^{(0)}\big(\cdot,\mathcal{X}^{(0)}\big) \stackrel{!}{=} (P^{(0)})^{T}\big(\cdot,\mathcal{X}^{(0)}\big)$ 
and $\sum_{\textbf{S} \in  \mathscr{A}^{(0)} } P^{(0)} \big(\textbf{S},\mathcal{X}^{(0)}\big) \stackrel{!}{=} 1$ where $\mathscr{A}^{(0)}$ is the ensemble of all the binary symmetric graphs with $ N_{0}$ nodes.

In line with \cite{2023_MSNR_Garuccio}, we further assume that the parameters
\begin{equation}
    \mathcal{X}^{(0)}_{i_0j_0} := 
    \begin{cases}
        x_{i_0} x_{j_0} & \textnormal{if } i_0 \neq j_0 \\[2ex]
        \frac{1}{2} x_{i_0}^{2} + w_{i_0} & \textnormal{if } i_0 = j_0
    \end{cases}
\end{equation}
are given by the product between $ \left\{x_{i_0}\right\}_{i_0 \in [1, N_0]}$ with additional node-wise parameters $ \left\{w_{i_0}\right\}_{i_0 \in [1, N_0]}$ only active in the self-loop part ($ i_0 = j_0$).
In particular, $ x_{i_0}$ encodes the capability of node $ i_0$ to connect to other nodes; whereas $w_{i_0}$ its propensity for a \textit{self}-interaction. Since the principles leading to an edge are different to the self-loop ones, we explicitly introduced the $ w$ parameters. The $ \frac{1}{2}$ in the self-loop part is introduced to avoid the double-counting of the self-interaction. We discarded a possible dependence of $\mathcal{X}^{(0)}$ on a \textit{dyadic} $ d_{i_0j_0}$ factor and the higher order terms. For further details we refer to \cite{2023_MSNR_Garuccio}.

In the following, we will use the notation $ P^{(0)}\big(\mathbf{A}^{(0)},\mathcal{X}^{(0)}\big) := P^{(0)}\big(\mathbf{A}^{(0)}, \mathbf{X}_{w}^{(0)} \big)$ where 
$
\mathbf{X}^{(0)}_w := [x^{(0)}, \vec{w}^{(0)}], 
x^{(0)} := \left[x_{1_0}, \dots, x_{N_0}\right]^{T} \in N_0 \times \mathbb{R}_{\geq 0}, 
\vec{w}^{(0)} := \left\{w_{i_0}\right\}_{i_0 \in [1, N_0]} \in N_0 \times [-\frac{1}{2} \left(\vec{x}_{i_0}\right) ^{2}, \infty)
$ 
are the fitted parameters at level $ 0$.

At the lower resolution $ 0$, the generated ensemble $ \mathscr{A}^{(0)}$ contains multiple configurations that, after coarse-graining, lead to the observed macroscopic $ \mathbf{A}^{(\ell)}$, i.e. $ \{\mathbf{A}^{(0)}\}\xrightarrow{\mathbf{\Omega_{0 \to \ell-1}}}\mathbf{A}^{(\ell)}$ (for a nice visualization of this technical point see \cite{2023_MSNR_Garuccio}). In turns, this induces the probability of observing $\mathbf{A}^{(\ell)}$ as
\begin{equation}
    P_{\ell}\big(\mathbf{A}^{(\ell)},\mathbf{X}^{(0)}_w\big) :=
    \!\!\!\!\!\!\!\!\!\!\!\!\!\sum_{\{\mathbf{A}^{(0)}\}
    \xrightarrow{\Omega_{0 \to \ell-1}}\mathbf{A}^{(\ell)}}
    \!\!\!\!\!\!\!\!\!\!\!\!\!\!P_{0}\big(\mathbf{A}^{(0)},\mathbf{X}^{(0)}_w\big).
    \label{eq:induced_prob}
\end{equation}
The \textit{scale-invariance} principle requires that the \textit{functional form} of the MSM has to be independent of the chosen scale, i.e. $ P_{\ell}\big(\cdot,\cdot) \stackrel{!}{=} P_{0}\big(\cdot,\cdot) \quad \forall \ell \geq 0,$ and that the model can generate the $\ell$-graphs in two equivalent ways, either \emph{hierarchically} or \textit{directly}. The \textit{hierarchical} procedure (see \autoref{eq:induced_prob}) prescribes to generate the $0$-graph ensemble with probability $P\big(\mathbf{A}^{(0)},\mathbf{X}^{(0)}_w\big)$ and, then, coarse-graining them $\ell$ times via the partitions $\{\mathbf{\Omega}_k\}^{\ell-1}_{k= 0}$. The \textit{direct} method, instead, requires to calculate the \textit{renormalize} parameters $ \widetilde{\mathbf{X}}_{w}^{(\ell)}$ and, then, directly model $\mathbf{A}^{(\ell)}$ via $P\big(\mathbf{A}^{(\ell)},\widetilde{\mathbf{X}}_{w}^{(\ell)})$.
Imposing both requirements
\begin{align}
    P\big(\mathbf{A}^{(\ell)}, \widetilde{\mathbf{X}}_{w}^{(\ell)} \big) 
    &\stackrel{!}{=}
    \!\!\!\!\!\!\!\!\!\!\!\!\!
    \sum_{\{\mathbf{A}^{(0)}\}
    \xrightarrow{\mathbf{\Omega}_{0 \to \ell-1}}\mathbf{A}^{(\ell)}}
    \!\!\!\!\!\!\!\!\!\!\!\!\!\!
    P\big(\mathbf{A}^{(0)}, \mathbf{X}_{w}^{(0)} \big)
    \quad \Leftrightarrow \quad
    \mathbf{P}_{sum}^{(\ell)} \stackrel{!}{=} \mathbf{P}_{cog}^{(\ell)} 
\end{align} 
where we have defined the LHS and RHS of the first equation as $\mathbf{P}_{sum}^{(\ell)} \textnormal{ and } \mathbf{P}_{cog}^{(\ell)} $ respectively.
In other words, the form of the $ P(\cdot, \cdot)$ will depend on the scale $ \ell$ only through the renormalized parameters $ \widetilde{\mathbf{X}}_{w}^{(\ell)} $. By assuming that the links are statistically independent, the previous equation yields 
\begin{align}
    \label{SI:eq:MSM_micro_psumVSpcg_implicit}
    p_{IJ} &\stackrel{!}{=} p^{cog}_{IJ}
\end{align}
where $p_{IJ} := p_{IJ}(\widetilde{\mathbf{X}}_{w}^{(\ell)})$
and
\begin{equation}
    \label{eq:MSM_pIJ_cg}
    p^{cog}_{IJ} := 1 - \prod_{i_0 \in I, j_0 \in J} (1-p\left(x_{i_0}, x_{j_0}, w_{i_0}\right))
\end{equation}
depend, respectively, on the renormalized and the fitted parameters at level $\ell = 0$. 
The interpretation of \autoref{SI:eq:MSM_micro_psumVSpcg_implicit}: the probability $ p_{IJ}$ that two blocks $ I, J$ are connected is given by the probability that there is \textit{at least one link} among their microscopic members $ i_0 \in I, j_0 \in J$ (see \autoref{SI:eq:cg_rule_nextl}). 

The missing ``sum'' superscript over $ p_{IJ}$ in \autoref{SI:eq:MSM_micro_psumVSpcg_implicit} reflects the \textit{scale-invariant} nature of functional form, whereas the subscript $ IJ$ refers to the $ x_{I}, x_{J}$ parameters that are, indeed, \textit{varying} at coarser resolutions under \textit{renormalization}.

By taking the logarithm of both sides of the \autoref{SI:eq:MSM_micro_psumVSpcg_implicit}, the only functional form compatible with that constraint 
\begin{equation*}
    \ln(1-p_{IJ}) = -  g(x_I), g( x_J) 
\end{equation*}
where $ g(x)$ is a positive, monotonic function that has to satisfy $ g( x_I) := \sum_{i_0 \in I} g( x_{i_0})$. For simplicity, we take $ g$ as the identity, namely $ g( y) := y$, which yields to . 
\begin{equation}
    \label{SI:eq:MSM_sumX}
    x_{I} \stackrel{!}{=} \sum_{i_0 \in I} x_{i_0}
\end{equation}
By applying the exponential to the above equation, it yields
\begin{equation*}
    \ln(1-p_{IJ}) = -  x_I x_J
\end{equation*}
that yields the (off-diagonal) \textit{scale-invariant} probability in \autoref{eq:MSM_pij}. 

For the self-loops part, the steps are similar to the previous ones, except for the factor of $ \frac{1}{2}$ that is introduced to avoid double counting. Indeed, referring to the network represented in \autoref{fig:trivial_cg_network}, \autoref{SI:eq:MSM_micro_psumVSpcg} yields
\begin{align*}
    1 - p_{AA} & = 1 - (1 - p_{00})(1 - p_{11})(1 - p_{01}) 
    \\ &\textnormal{ iff }
    \\
    \frac{1}{2}  x_{A}^{2} 
    %= \frac{1}{2}  x_{0} + x_{1} ^{2} 
    = \frac{1}{2}  x_{0} ^{2} + \frac{1}{2}  x_{1} ^{2} + \frac{2}{2}  x_{0}, x_{1}  
    &=
    \frac{1}{2}  x_{0}^{2} +  x_{1} ^{2} +  x_{0}, x_{1} 
\end{align*}
which is solved only because the authors of \cite{2023_MSNR_Garuccio} imposed the $ \frac{1}{2}$ in the self-loop part.

Given the results we have just obtain, the defining relationship \autoref{SI:eq:MSM_micro_psumVSpcg_implicit} reads
\begin{equation}
    \label{SI:eq:MSM_micro_psumVSpcg}
    p( x_{I} x_{J}, w_{I} ) := p( \sum_{i_0 \in I} x_{i_0} \sum_{j_0 \in J} x_{j_0}, \sum_{i_0 \in I} w_{i_0} ) \stackrel{!}{=} 1 - \prod_{i_0 \in I, j_0 \in J} (1-p\left(x_{i_0}, x_{j_0}, w_{i_0}\right))
\end{equation}
where the $ p$ functional form is reported in \autoref{eq:MSM_pij}.

\subsection{Parameter Estimation}
In order to tackle the \textit{reconstruction} problem -inferring the link probability starting from aggregated inputs-, the MSM parameters can be estimated by enforcing the degree sequence on average as done for the CM in \autoref{SI:sec:CM}. For this reason, the output model is called the \textit{degree-corrected} MSM (degcMSM). However, as the MSM is not an ERG model, the MoM and the maximum-likelihood estimation of the parameters may differ. For this reason, we will also consider the \textit{maximum-likelihood} MSM (maxlMSM) to evaluate if, knowing the full structure of network, the latter procedure leads to a better modelling of the underlying structure. To stress the point, the comparison among the degcMSM and maxlMSM is a theoretical exercise the former assumes the availability of the degree sequence at best, whereas the maxlMSM needs the whole adjacency matrix, which implies the knowledge also of the degree sequence, but it is often unavailable due to policy restrictions \cite{2022_RecNet_ION_Ialongo}.

\subsubsection{Maximum Likelihood MSM}
Assuming to know the adjacency matrix, the likelihood for the MSM reads
\begin{equation}
    \label{eq:MSM_full_likelihood}
    \mathcal{L}(\vec{x},\vec{w}|\textbf{A}) =
    \sum_{i \leq j} a_{ij} \ln(p_{ij}) + (1-a_{ij}) \ln(1-p_{ij})
\end{equation}
Its gradient can be calculated by deriving the \autoref{eq:MSM_full_likelihood} with respect to $ x_i$ and $ w_{i}$, namely
\newlength{\itsp}
\setlength{\itsp}{3ex}
\begin{align}
    \begin{cases}
        \partial_{x_i} \mathcal{L}
        &= \sum_{j} \left( \frac{a_{ij}}{p_{ij}} - 1 \right) \frac{ \partial_i p_{ij} }{q_{ij}} \\[\itsp]
        &= \sum_{j} \left(\frac{a_{ij}}{p_{ij}} - 1\right) x_j \\[\itsp]
        \partial_{w_i} \mathcal{L}
        &= \frac{a_{i i}}{p_{i i}} - 1
    \end{cases}
\end{align}
which vanishes when the parameters are at the optimal point $ \vec{x}^{*}, \vec{w}^{*}$, i.e.
\begin{align}
    \label{eq:MSM_full_grad0}
    \begin{cases}
        \partial_{x_i} \mathcal{L} \vert_{\vec{x} = \vec{x}^{*}} = \sum_{j} \left(\frac{a_{ij}}{p^{*}_{ij}} - 1\right)  x^{*}_j &= 0 \\[\itsp]
        \partial_{w_i} \mathcal{L} \vert_{\vec{w} = \vec{w}^{*}} = \frac{a_{ii}}{p^{*}_{ii}} - 1 &= 0
    \end{cases}
\end{align}
where $ p^{*}_{ij} := p(\vec{x}^{*}_{i}\vec{x}^{*}_{j}) \textnormal{ and } p^{*}_{ij} := p(\vec{x}^{*}_{i}, \vec{w}^{*}_{i})$. 
In particular, the second equation suggests that $ \left\{w_{i}^{*}\right\}_{i \in [1, N]}$ are a function of $ \vec{x}$ at the stationarity point, namely
\settowidth{\temp}{$\to \infty \qquad$}
\begin{equation}
    w_{i} =
    \begin{cases}
        \FixedSize{\temp}{\to \infty} \textnormal{ if } a_{i i} = 1 \\[1ex]
        \FixedSize{\temp}{-\frac{1}{2} x_{i}^{2}} \textnormal{ if } a_{i i} = 0.
    \end{cases}
\end{equation}
In addition, one can reduce the number of parameters by inserting the second equation into the first one, which leads to
\begin{equation}
    \partial_{x_i} \mathcal{L} \vert_{\vec{x} = \vec{x}^{*}} = \sum_{j (\neq i)} \left(\frac{a_{ij}}{p^{*}_{ij}} - 1\right)  x^{*}_j = 0
\end{equation}
and, by integrating over $ x_{i}$, to \autoref{eq:MSM_likelihood}. 

In conclusion, it is enough to find the maximum of the \textit{off-diagonal} likelihood \autoref{eq:MSM_likelihood}, and, then, fix the $ \left\{w_{i}\right\}_{i \in [1, N]}$. This implies that the \textit{diagonal} part of \autoref{eq:MSM_full_likelihood} ($ i=j$) vanishes, whereas the off-diagonal part is already maximized by means of the optimization procedure. Note that for a fair comparison among the CM (no self-loops modelling) and the MSM, the evaluation metrics (DEG, ANND, CC, $\dots$) don't take into account the self-interactions.

The hessian was not calculated since it was used neither for a theoretical analysis nor in the numerical optimization -the SciPy methods can approximate it in a sophisticated way.

\subsubsection{No natural statistical equivalence from structural equivalence}
\label{SI:sec:StructE_not_StatE}
\begin{figure}[t]
    \begin{tikzpicture}[baseline=(base), node distance={2cm}, thick, main/.style = {fill=blue!20, draw, circle}] 
        \centering
        \node (base) at (0,-.5ex) {};
        \node[main] (1) {$1$}; 
        \node[main, fill=red!20] (2) [below of=1] {$2$};
        \node[main] (3) [right of=2] {$3$};
        \node[main, fill=red!20] (4) [right of=1] {$4$};
        % \node[main, ] (2) [right of=0] {$2$}; 
        % Edges
        \path (2) edge[violet, loop below] (2); % Self-loop for node 1
        \path (1) edge (2); % Connection between node 1 and node 2
        \path (1) edge (4); % Connection between node 1 and node 4
        \path (2) edge (3); % Connection between node 2 and node 3
        \path (4) edge[dashed, violet, loop above] (4); % Self-loop for node 3
        \path (3) edge (4);
    \end{tikzpicture}
    $ \qquad \stackrel{ \textnormal{Adjacency Matrix} }{\longrightarrow} \qquad$
    \raisebox{-.8\height}{ % This will vertically center the matrix
        \scalebox{1.5}{
        $
        \begin{bmatrix}
            1 & 0 & 1 & 0 \\
            0 & 1 & 0 & 1 \\
            1 & 0 & 1 & 0 \\
            0 & 1 & 0 & 0
        \end{bmatrix}
        $
        }
    }
    \caption{External Structural Equivalence example: node 1 and 2 are, respectively, equivalent to node 3 and 4, even thought node 2 has a self-loop which is not present in node 4. This is reflected in the adjacency matrix, reported on the right, since the 1st and 3rd (or 2nd and 4th) rows have the same \textit{external} connections, but not the \textit{internal} ones.}
    \label{SI:fig:StructE_StatE}
\end{figure}
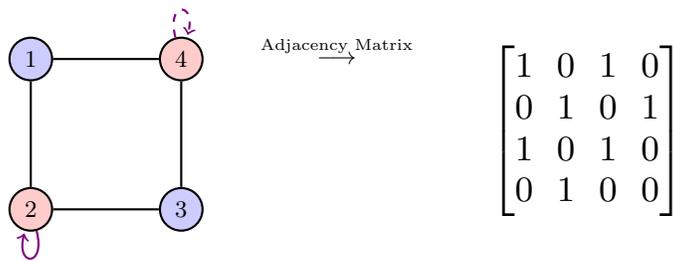

In the \autoref{sec:degcMSM}, we have seen that the degcMSM assigns the same $ x$'s to nodes with the same degree. Here, since the likelihood sums over the links, one would expect that nodes with the same neighbors (structural equivalence - StructE) should have the same parameters (statistical Equivalence - StatE) at the optimal point \autoref{SI:eq:MSM_grad0_StructE}. 
However, this would not be the case (see next section), and one would need to enforce the equality among the parameters $ x_{i}$ ultimately leading to the statistical equivalence of the StructE nodes. Roughly, by imposing the StatE, one discards a possible source of redundancy and lets the model focus on the backbone of the observed network, i.e. the inequivalent classes.

To show that the stationarity condition is not sufficient to guarantee the StatE, let us assume that the node $ i,s$ shares the same neighbors, namely $ N(i) \equiv  N(s)$ where $ N(i) := \left\{j : j \in [1, N], a_{ij} = 1, j \neq i\right\}$ (self-loop included). This implies that their adjacency entries are the same, i.e. $ a_{ij} \equiv a_{sj} \, \forall j$ as nodes $ 1,3$ in the \autoref{SI:fig:StructE_StatE}.
Furthermore, from the stationarity condition, one can obtain the following relationships among $ x^{*}_{i}, x^{*}_{s}$, formally
\begin{align}
    % \sum_{j} \left(\frac{a_{ij}}{p_{ij}} - 1\right)  x_j 
    % &= 
    % \sum_{j} \left(\frac{a_{sj}}{p_{sj}} - 1\right)  x_j\\[1em]
    \label{SI:eq:MSM_grad0_StructE}
        \sum_{j \in N(i)} 
        \frac{x^{*}_{j} }{p(x^{*}_{i}x^{*}_{j})}
        + x^{*}_{i}
        &=
        \sum_{j \in N(s)} 
        \frac{x^{*}_{j}}{p(x^{*}_{s}x^{*}_{j})}
        + x^{*}_{s}
\end{align}
where we highlighted the parameters' dependence of the model. The previous equations show that, even thought the nodes are \textit{structurally} equivalent, their \textit{statistical} equivalence is not guaranteed from the previous equations. Therefore, the StatE has to be imposed by hand for all the StructE nodes, namely 
\begin{equation*}
    x_{i} \equiv x_{e} \quad \forall i \in S(e)  
\end{equation*}
where $ S(e) := \left\{j : N(j) \equiv N(e)\right\}$.

As final remarks, note that the degcMSM assign the same parameter $ x_{i}$ (no self-loop parameter is involved) for nodes with the same degree. This equivalence class is larger than the StructE used so far, and so the degcMSM will have a smaller set of available parameters than the maximum-likelihood ones. However, this is not the case for the CM since the degree sequence is its \textit{sufficient statistics}, i.e. the maximum of the likelihood is recovered by imposing the degree sequence on average. Hence, the cardinality of the set of parameters is the same of fitting the CM via maximum likelihood or fitting the node degrees.

\subsection{Removal of deterministic nodes}
Network reconstruction assumes that the observed $ \mathbf{A}$ is a realization of a Bernoullian random process. However, it may happen that some nodes are \textit{deterministic}, i.e. fully-connected (FC) or disconnected (D) to the other nodes. Therefore, their behavior is \textit{trivially} recovered by setting the hidden variable of the FC nodes to infinity or the D nodes to zeros (see \autoref{eq:MSM_pij}). Their optimized parameters would be $ x_{FC} \to \infty$ and $ x_{D} = 0$, respectively implying that $\left\{p_{FC,j} \equiv 1\right\}_{ j \in [1, N] }$ and $\left\{p_{FC,j} \equiv 0\right\}_{ j \in [1, N] }$. In turn, $ \textnormal{ Var}(a_{FC,j}) = \textnormal{ Var}(a_{D,j}) = 0 \quad \, \forall j$ implying that their sampled edges are fixed, and not contributing to the ensemble fluctuation. Concretely, we have hard coded their variables, as seen before, to account for their roles in the graph. 

A similar reasoning applies to the nodes connected only to FC, the so called \textit{only2fc} nodes ($ \nu$). In particular, in the degree-corrected case (both Configuration Model and degree-corrected MSM), by assuming there is only one FC node and focusing on the FC/only2fc equations,
\begin{equation}
    \begin{cases}
        \langle k_{FC} \rangle := \sum_{j (\neq FC)} p_{FC,j} \stackrel{!}{=} N \\[1ex]
        \langle k_{\nu} \rangle := p_{\nu,FC} + \sum_{j (\neq \nu, FC)} p_{\nu j} \stackrel{!}{=} 1
    \end{cases}
\end{equation}
which solution reads $ p_{FC,j} \equiv 1 \, \forall j (\neq FC), p_{\nu j} \equiv 0 \, \forall j (\neq \nu, FC)$. 
Similarly, in the maximum likelihood setting, the likelihood of only2fc reads
\begin{align}
    \begin{cases}
        \mathcal{L}_{FC} = \sum_{j (\geq FC)} \ln\left(p_{FC,j}\right)
        \mathcal{L}_{\nu} = \ln\left(p_{\nu,FC}\right) - x_{\nu}  \sum_{j (\neq FC,\nu)} x_{j} = - x_{\nu} \sum_{j (\neq FC,\nu)} x_{j}
    \end{cases}
\end{align}
where $ \ln\left(p_{\nu, FC}\right) = 0$ for the unconstrained optimization of $ \mathcal{L}_{FC}$. 

In both cases, the only2fc parameters are undetermined. Setting for convenience $ x_{FC} = \tau^{2} $ with $ \tau \to \infty $, they require that
\begin{equation*}
    \begin{cases}
        x_{o} x_{FC}  \sim \tau  \\ 
        x_{o} x_{j}  \sim 1/\tau \qquad \forall j \neq FC
    \end{cases}
\end{equation*}
which are trivially solved by setting $ x_{o} \stackrel{!}{=} 1/\tau$ as the $ x_{j} \approx \textnormal{ const}$ with respect to $ x_{o}$. As the FC-parameters are \textit{degenerate}, we decided to fix them after the optimization. In addition, this results in a more stable maximization of the likelihood, since one is manually setting the values for the trivial (disconnected entities) or degenerate nodes (FC and only2fc).

Lastly, this procedure doesn't spoil the \textit{ renormalization rule}. Including an FC node to a block-node produces an FC block-node; whereas a vanishing parameter (disconnected or only2fc) gives the freedom to the other terms in the summation (see \autoref{eq:MSM_sum_graining_rule}) to determine the group parameter.

\subsection{Algorithmic Complexity}
\label{SI:sec:Algorithmic_Complexity}
To describe the higher levels, one needs a lower computational complexity by means of a \textit{scale-invariant} model than a single-scale (SSM) one. Specifically, the computational complexity of the RHS of \autoref{SI:eq:MSM_micro_psumVSpcg} (the only one possible to coarse-grain an SSM)
reads
\begin{align}
    \mathcal{C}_{SSM} = c \sum_{I > J} N_{I}N_{J} = \frac{c}{2} \sum_{I,J} N_{I}N_{J} = c \frac{N^{2}_{0}}{2}
\end{align}
where $ c$ is the complexity for the calculation of the connection probability $ p$ for the referred model, e.g. CM, and $ N_{0}$ are the number of structural inequivalent nodes\footnote{Due to this, $ N_{0}$ will be model dependent.} at the microscopic level $ \ell = 0$.
In contrast, assuming that the ``sum'' operation is $ \mathcal{O}(1)$, the complexity of the LHS reads
\begin{equation}
    \mathcal{C}_{MSM} \approx N_\ell + c \frac{N_{\ell}(N_{\ell}-1)}{2}
\end{equation}        
since there are $ N_\ell$ summation to get  $ x_{I}$\footnote{To account for the summation over the $ w_{I}$, one should add a factor of $ N_{\ell}$} and all the $ \binom{N_{\ell}}{2}$ pairs at the level $ \ell$.     
Hence, the ratio reads
\begin{equation}
    \frac{\mathcal{C}_{MSM}}{\mathcal{C}_{SSM}} \approx 
    \left(\frac{N_{\ell}}{N_{0}}\right)^{2}
\end{equation}        
For example at level $\ell = 3$ for the ING network and maxlMSM, $ N_{3} = 67, N_0 = 972$,
\begin{equation}
    \frac{\mathcal{C}_{MSM}}{\mathcal{C}_{SSM}} \approx 0.0048
\end{equation}        
giving an improvement of 3 order of magnitude using the summed MSM rather that the full evaluation of the coarse-graining probability.

\subsection{Numerical Comparison between the CM and MSM functional form}

\begin{figure}[t]
    \centering
    \begin{tikzpicture}
        \begin{axis}[
            legend pos=south east, % Changed from south east
            xlabel = $x$,
            ylabel = {$p(x)$},
            xmin = 0,
            xmax = 10,
            ymin = 0,
            ymax = 1.1,
            xtick pos=left,
            ytick pos=left,
            legend style={
                /tikz/every even column/.append style={column sep=0.5cm},
                legend plot pos=left,
                cells={anchor=west},
                nodes={align=left, inner xsep=2pt},
                yshift = 0.2cm,
                row sep = 0.2cm
            }
        ]
        
        % Define the first function (p^{MSM}(x))
        \addplot[domain=0:10, blue, samples = 100, thick] {1 - exp(-x)};
        \addlegendentry{\eqmakebox[fnlabel][l]{$p_{MSM}(x)$} $= 1 - e^{-x}$}
        
        \addplot[domain=0:10, dashed, forget plot] {1};

        % Define the second function (p^{CM}(x))
        \addplot[domain=0:10, red, samples=100, thick] {x / (1+x)};
        \addlegendentry{\eqmakebox[fnlabel][l]{$p_{CM}(x)$} $= \frac{x}{1+x}$}
        
        % added the x approximation
        \addplot[domain=0:.8, orange, style = very thick, dashed] {x}; 
        \addlegendentry{$x$}
        
        \end{axis}
    \end{tikzpicture}
    \caption{Plot of $p^{MSM}(x)$ and $p^{CM}(x)$}
    \label{fig:function_comparison_CM_MSM}
\end{figure}
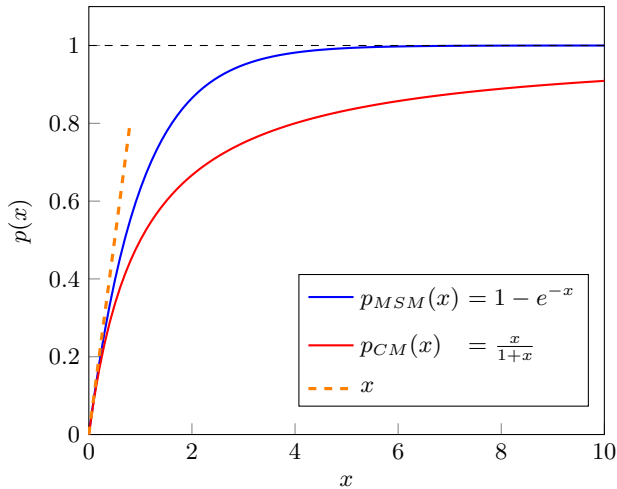

Here, we evaluated the CM and MSM functional forms for $ x \in [0, 10]$. The crucial point is that, even if the CM is not \textit{theoretically} scale-invariant (see \autoref{SI:Inconsistency_CM}), it may be \textit{numerically} similar to the MSM. Actually, for small $ x$, $ p_{CM} \approx p_{MSM} \approx x$. Hence, the CM expected quantities may be similar to the MSM ones, spoiling the idea the CM is a single-scale model, which should be generalizable.

\section{Scores}
\label{sec:Scores}

\subsection{Comparison Between Probabilities}
\label{sec:ComparisonAmongProbabilities}
For every level $ \ell$, both CM and MSM give rise to 3 probability functions: the \textit{fitted} $ \mathbf{P}^{(\ell)}_{fit} $, \textit{summed} $ \mathbf{P}_{sum}^{(\ell)} $ and the \textit{coarse-grained} $ \mathbf{P}_{cog}^{(\ell)} $. Technically, the first twos have the same functional form, while depending on the fitted and the summed parameters, respectively. Yet, $ \mathbf{P}_{cog}^{(\ell)} $ has a different connection probability for the CM-like models (\autoref{SI:Inconsistency_CM}), whereas it is equal to the summed probability for the MSMs (\autoref{SI:sec:derivation_MSM_probability}). To display these differences, we cross compared them to understand their hallmarks.

Firstly, in \autoref{fig:ING_sum_vs_cg_pmatrix}, one can find how much $ \mathbf{P}_{sum}^{(\ell)} $ differs from $ \mathbf{P}_{cog}^{(\ell)} $ both for the CM-like and MSM models. Secondly, the comparison among the summed $ \mathbf{P}_{sum}^{(\ell)} $ against fitted $\mathbf{P}_{fit}^{(\ell)}$ probabilities is shown in the insets of \autoref{fig:ING_rec_acc_by_level}. In particular, we plotted the $ 2D$ histogram of the density $ \rho_{bin}$  of points inside each bin (the total number of bins is $30^2$), i.e.
\begin{equation}
    \rho^{(\ell)}_{bin} = 2\frac{n_{\textnormal{bin}}}{N_\ell (N_\ell - 1)} \in [0,1]
\end{equation}
where $ n_{\textnormal{bin}}$ is the total number of points inside that bin and $N_\ell (N_\ell - 1)$ the number of pairs.
Note that we colored the bins according to $ \rho_{bin}$ (creating a heatmap) - the bigger the value, the brighter the color.
Lastly, we didn't evaluate $\mathbf{P}_{fit}^{(\ell)} $ against $ \mathbf{P}_{cog}^{(\ell)} $ since the \textit{coarse-grained} probability is functionally different from the others for the CM-like models (see \autoref{sec:sum_CM}) whereas $ \mathbf{P}_{cog}^{(\ell)} = \mathbf{P}_{sum}^{(\ell)}$ for the MSM. Therefore, we used $ \mathbf{P}_{cog}^{(\ell)}$ only in the first comparison.

\subsection{Degree, Average-Nearest Neighbor Degree and Clustering Coefficient}
\label{sec:NetworkMeasurements}
The fundamental topological properties of a network are the degree, the average nearest neighbor degree (ANND) and the clustering coefficient (CC) \cite{2004_StatMecNet_Park}. 
Formally, each of such measurements is a function $ Y(\mathbf{A})$ of an $ N \times N$ adjacency matrix representing a graph $ \mathbf{G}$.

Here, we compute them both in the observed network $ \mathbf{A}$ and as expected by the model $ \mathbf{P}(\mathbf{A}|\textbf{X})$. More precisely, the \textit{degree} counts the number of edges that are incident to a node $ i$, i.e.
\begin{equation}
    \label{eq:degree_observed}
    k_{i}(\mathbf{A}) := \sum_{j (\neq i)} a_{ij}
\end{equation}
and its expected value is given by
\begin{equation}
    \langle k_{i} \rangle = \sum_{j (\neq i)} p_{ij}
\end{equation}
where $ \langle \cdot \rangle$ denotes the expected value over the ensemble of graphs sampled from $ \mathbf{P}(\mathbf{A}|\textbf{X})$.
Moving \textit{two-hops} away from $ i$, the \textit{ANND} reports the average degree of the neighbors of the node $ i$, i.e.
\begin{align}
    \label{eq:ANND_observed}
    k^{n n}_{i}(\mathbf{A}) :&= \sum_{j (\neq i)} \frac{a_{ij}k_{j}}{k_{i}} 
    \\ &= \frac{\sum_{j (\neq i), k (\neq j)} a_{ij}a_{jk}}{ \sum_{j (\neq i)} a_{ij}}
\end{align}
whereas its expected value reads
\begin{align}
    \langle k^{n n}_{i} \rangle &:= \langle \frac{\sum_{j (\neq i), k (\neq j)} a_{ij}a_{jk}}{ \sum_{j (\neq i)} a_{ij}} \rangle
    \\ &\approx 1 + \frac{\sum_{j (\neq i), k (\neq j, i)} p_{ij}p_{jk}}{ \sum_{j (\neq i)} p_{ij}}
\end{align}
where in the second passage we took advantage on the first order approximation $ \mathbb{E}[ \frac{X}{Y} ] \approx \frac{\mathbb{E}[X]}{\mathbb{E}[Y]}$  (\textit{delta approximation}) \cite{2011_AnalMax_Squartini}.
Lastly, the \textit{CC} is defined as the ratio among the number of triangles of node $ i$ and its number of wedges, namely 
\begin{align}
    \label{eq:CC_observed}
    c_{i}(\mathbf{A}) 
    &:= \frac{\triangle_i}{\wedge_i} \\
    &= \frac{\sum_{i \neq j \neq k} a_{ij} a_{jk} a_{ki} }{\sum_{j \neq k} a_{ij} a_{ik}}
\end{align}
whereas the expected one is
\begin{align}
    \label{eq:CC_expected}
    \langle c_{i} \rangle &:= \langle \frac{\triangle_i}{\wedge_i} \rangle \\
    \label{eq:CC_expected_approx}
    &\approx \frac{\langle \triangle_i \rangle}{\langle \wedge_i \rangle} \\
    &\approx \frac{\sum_{i \neq j \neq k} p_{ij} p_{jk} p_{ki} }{\sum_{j \neq k} p_{ij} p_{ik}}
\end{align}

\subsection{Triangle Density}
Inspired by \cite{2020_LPCA_Chanpuriya}, we computed the \textit{expected} number of triangles for every model at disposal\footnote{In this essay, our objective was to model \textit{probabilistically} the observed network rather than describing it \textit{exactly}, namely the limit where $ p_{ij} \equiv a_{ij} \, \forall i > j$.}.
Specifically, the expected density of triangles at a certain level $ \ell$ is defined as
\begin{equation}
    \label{eq:observed_TriDens}
    \rho^{(\ell)}(c) := \frac{ \triangle(\mathbf{G}^{(\ell)}_{k_{i_{\ell}} \geq c})}{2N_{\ell}} = \frac{ \sum_{i \neq j \neq k} g_{ij} g_{jk} g_{ki} }{2N_{\ell}}
\end{equation}
where $\triangle( \mathbf{G}^{(\ell)}_{k_{i_{\ell}} \geq c})$ is the number of observed triangles (see \autoref{eq:CC_observed}) calculated on the subgraph $ \mathbf{G}^{(\ell)}_{k_{i_{\ell}} \geq c}$ composed by the nodes $I_c := {\left\{i_\ell : k_{i_\ell} \geq c\right\}}$ with degree lower (or equal) than a threshold $ c$ \cite{2020_impossibility_of_low_rank_red_Seshandhri, 2020_LPCA_Chanpuriya}.
Its expected value reads
\begin{equation}
    \label{eq:expected_TriDens}
    \langle \rho^{(\ell)}(c) \rangle = \frac{ \langle \triangle(\mathbf{G}^{(\ell)}_{k_{i_{\ell}} \geq c}) \rangle}{2N_{\ell}} \approx \frac{ \sum_{i < j < k} \tilde{p}_{ij} \tilde{p}_{jk} \tilde{p}_{ki} }{N_{\ell}}
\end{equation}
where $ i \in I_c, j \in I_c, k \in I_c$ and the probabilities $ \tilde{p}_{ij}$ refers to the summed model $ \mathbf{P}_{sum}^{(\ell)} $.

\subsection{On the ensemble estimation}
For more complicated measurements, e.g. the variance of the \textit{ANND}, the \textit{delta approximation} won't be valid and one has to estimate them as the average over a \textit{sufficiently large} ensemble $\mathcal{A} := \left\{\mathbf{A}_{s}\right\}_{s \in [1, \mathcal{S} ]}$ where $ \mathcal{S}$ is the number of graphs. 
In particular, having optimized the parameters of the model, we can generate unbiased realizations $ \mathcal{A}$ by sampling each $ a_{ij}$ independently with probability $ p_{ij}$ \cite{2011_AnalMax_Squartini,2023_MSNR_Garuccio}.

In the limit of $ \mathcal{S} \to \infty$, the \textit{sampled} average of any measure $ Y_i$ meets its \textit{analytical} estimations $ \langle Y_i \rangle$ \cite{2023_MSNR_Garuccio}, i.e.
\begin{align}
    \label{eq:ensemble_average_analytical}
    \bar{Y_i} 
    :&= \frac{1}{\left\vert \mathcal{A}_{N}\right\vert}
    \sum_{\mathbf{\hat{A}} \in \mathcal{A}}  Y_i(\mathbf{\hat{A}})
    \\[1ex]
    \to 
    \langle Y_i \rangle &= \sum_{\mathbf{B} \in \mathcal{A}_{N}} \mathbf{P}(\mathbf{B}|\textbf{X}) Y_i(\mathbf{B})
\end{align}
where $\mathbf{B} \in \mathcal{A}_{N}$ is a matrix drawn from the set of the undirected binary graphs $ \mathcal{A}_{N}$ of $ N$ nodes. 

Lastly, to estimate the uncertainty of the model over the sampled realizations, we calculated the \textit{dispersion intervals} $ \Delta_c( \langle Y_i \rangle ) := \langle Y_i \rangle \pm 2\sigma_{\langle Y_i \rangle}$ \cite{2023_RecEcon_DiVece}. In other words, an interval centered in $ \langle Y_i \rangle$, but $ 4\sigma_{\langle Y_i \rangle}$ wide.

\subsection{Reconstruction Accuracy}
In order to have a cross-comparison among all the levels and models, we exploited the \textit{reconstruction accuracy} \cite{2023_RecEcon_DiVece}. This measure is defined as the fraction of times an observed statistics $ Y_{i}$ falls within the \textit{dispersion interval} $ \Delta_c( \langle Y_i \rangle )$ (see \autoref{sec:NetworkMeasurements}). More formally, the reconstruction accuracy at level $ \ell$ for the statistics $ Y$ is defined as
\begin{equation}
    \label{eq:RecAcc}
    RA^{(\ell)}_s := \frac{1}{N_{\ell}} \sum_{i = 0}^{N_{\ell} - 1} \mathbb{I}\left\{Y_i \in \Delta( \langle Y_i \rangle )\right\}
\end{equation}
where $ \mathbb{I}$ is the indicator function\footnote{Further refinements are possible, but we stick with this definition for the sake of simplicity.}. 
Roughly, it counts the frequency at which the sampled ensemble includes the observed statistics. If all the observed statistics, e.g. degrees, were included in the interval, the accuracy would be $1$, whereas the accuracy would be $0$ if none of them were included.

\subsection{Rescaled ROC and PR Curves}
By considering the introduced models as \textit{binary classifiers} it seems natural to inspect their \textit{expected} confusion matrix, Receiver Operating Characteristic (ROC) and the Precision-Recall (PR) curves \cite{2020_ROC_PR_creation_Cook,2006_ROC_PR_skewed_data_Davis}. In particular, the \textit{expected} confusion matrix is a $ 2 \times 2$ matrix that reports the expected value of True Positives (TP), i.e. $ \langle TP \rangle := \sum_{i < j} a_{ij} p_{ij}$, False Positives (FP), i.e. $ \langle FP \rangle := \sum_{i < j} (1-a_{ij}) p_{ij}$, True Negatives (TN), i.e. $ \langle TN \rangle := \sum_{i < j} \left(1-a_{ij}\right) \left((1-p_{ij})\right)$, and False Negatives (FN), i.e. $ \langle FN \rangle := \sum_{i < j} a_{ij} \left(1 - p_{ij}\right)$ \cite{2017_ECAPM_Squartini}. By combining these scores, one recovers the True Positive Rate ($TPR$), the False Positive Rate ($FPR$) and the Positive Predictive Value ($PPV$) \cite{2006_ROC_PR_skewed_data_Davis}, namely
\begin{align}
    TPR :&= \frac{ TP }{Pos}
    = \frac{\sum_{i < j} a_{ij} p_{ij}}{ L }
    \\[1ex] FPR  :&= \frac{ FP }{Neg} =
    \\ &= \frac{\sum_{i < j} (1 - a_{ij}) p_{ij}}{ \binom{N}{2} - L }
    \\[1ex] PPV  :&= \frac{TP}{P P} = \frac{\sum_{i < j} a_{ij} p_{ij}}{ \sum_{i < j} p_{ij}}
    \\ &= \frac{\sum_{i < j} a_{ij} p_{ij}}{ \langle L \rangle}
\end{align}
{
\raggedright
where $ Pos := \sum_{i < j} a_{ij} = L$, $Neg := \sum_{i < j} (1 - a_{ij}) = \binom{N}{2} - L$.
} 
Instead of taking these expected values, another common practice is to set the entries s.t. $ p_{ij} \leq \epsilon$ for a fixed threshold $ \epsilon$, and otherwise to $ 0$. For convenience, we will call them $  TPR (\epsilon),  FPR (\epsilon),  P PV (\epsilon)$. Furthermore, the ROC and PR curves are obtained by spanning $ \epsilon \in [0,1]$ \cite{2020_ROC_PR_creation_Cook} and plotting the TPR-VS-FPR, and the TPR-VS-PPV, respectively (see \autoref{fig:ING_auc_roc_prc}). Note that if $ \epsilon \to 1$, then $TPR(\epsilon \to \infty) = FPR(\epsilon \to \infty) = 0, P PV(\epsilon \to \infty) := 1$, whereas if $ \epsilon = 0$, $TPR(\epsilon = 0) = FPR(\epsilon = 0) = 1, P PV(\epsilon = 0) = p := \frac{Pos}{Pos + Neg} = \frac{L}{\binom{N}{2}}$.

Since the last point of the $P PV$ is the density, the ``naive classifier'' (NC) - always predicting the majority class - can have a high AUC-PR (On the ROC curve the NC always lies on the diagonal). For this reason, one can rescale the Area Under the Curve (AUC) of each model with respect to the AUC predicted by the NC. Specifically, one defines
\begin{align}
    AUC-ROC_{norm} &= \frac{AUC_{ROC} - 0.5}{0.5}
    \\ AUC-PR_{norm} &= \frac{AUC_{PR} - p}{1 - p}.
\end{align}
Therefore, the perfect classifier still have $ \textnormal{ AUC-ROC } = \textnormal{ AUC-PR } = 1$ but the random one $ \textnormal{ AUC-ROC } = \textnormal{ AUC-PR } = 0$. The \textit{new} AUCs can be negative, as a signal of a worse performance than the random classifier. Furthermore, this novelty does not spoil the ranking of the \textit{summed} models which was the ultimate objective of the AUC scores.

\section{World Trade Web results}

In this section, we report the same results presented in the main text, but for the World Trade Web. The figure analysis are reported in the caption of each figure.

\setlength{\subfigurewidth}{.4\linewidth}
\begin{figure*}[t]
    \centering
    \subfloat[CM]{\includegraphics[width=\subfigurewidth,]{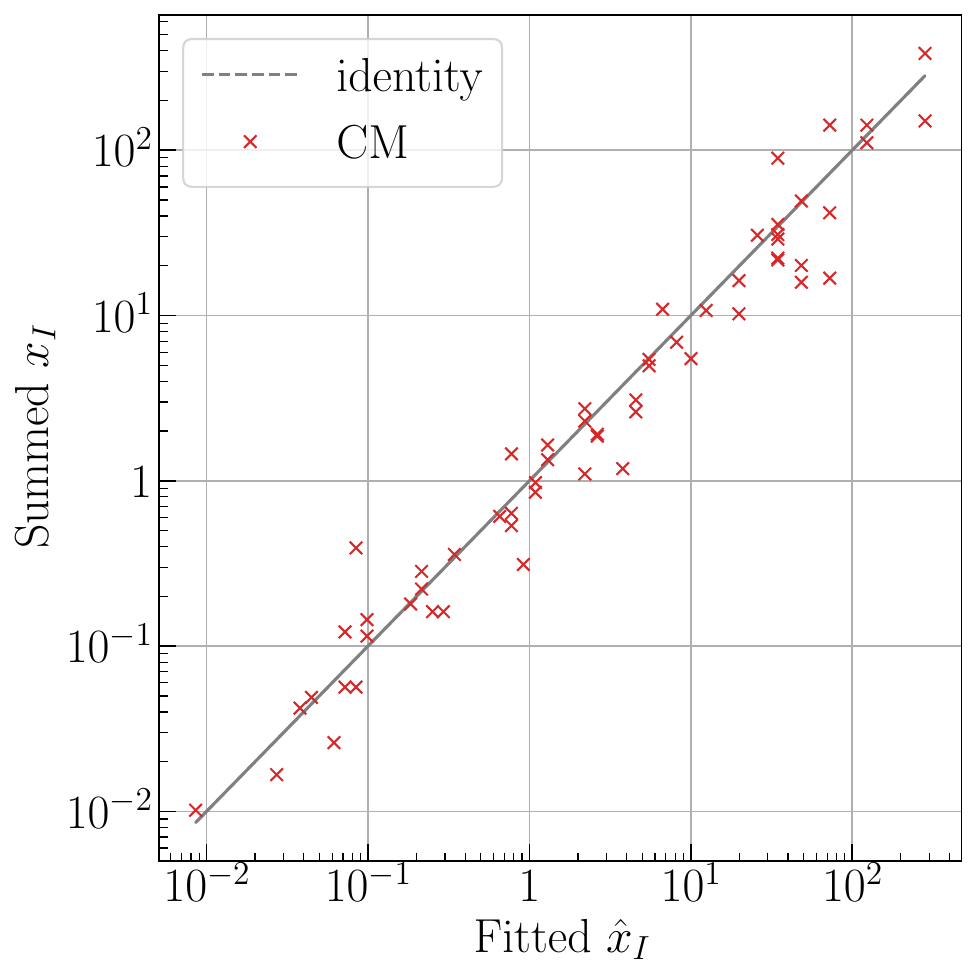}}
    \hfill
    \subfloat[degcMSM]{\includegraphics[width=\subfigurewidth,]{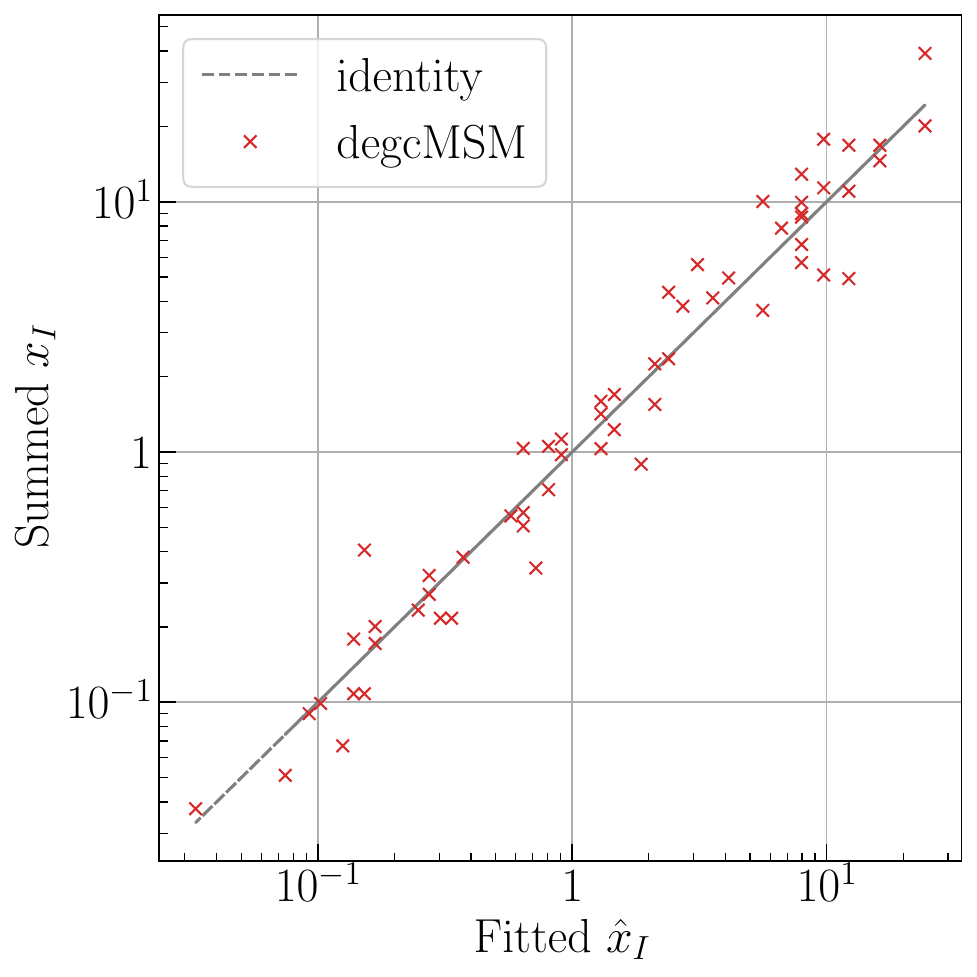}
        }
    \hfill
    \subfloat[maxlMSM]{\includegraphics[width=\subfigurewidth,]{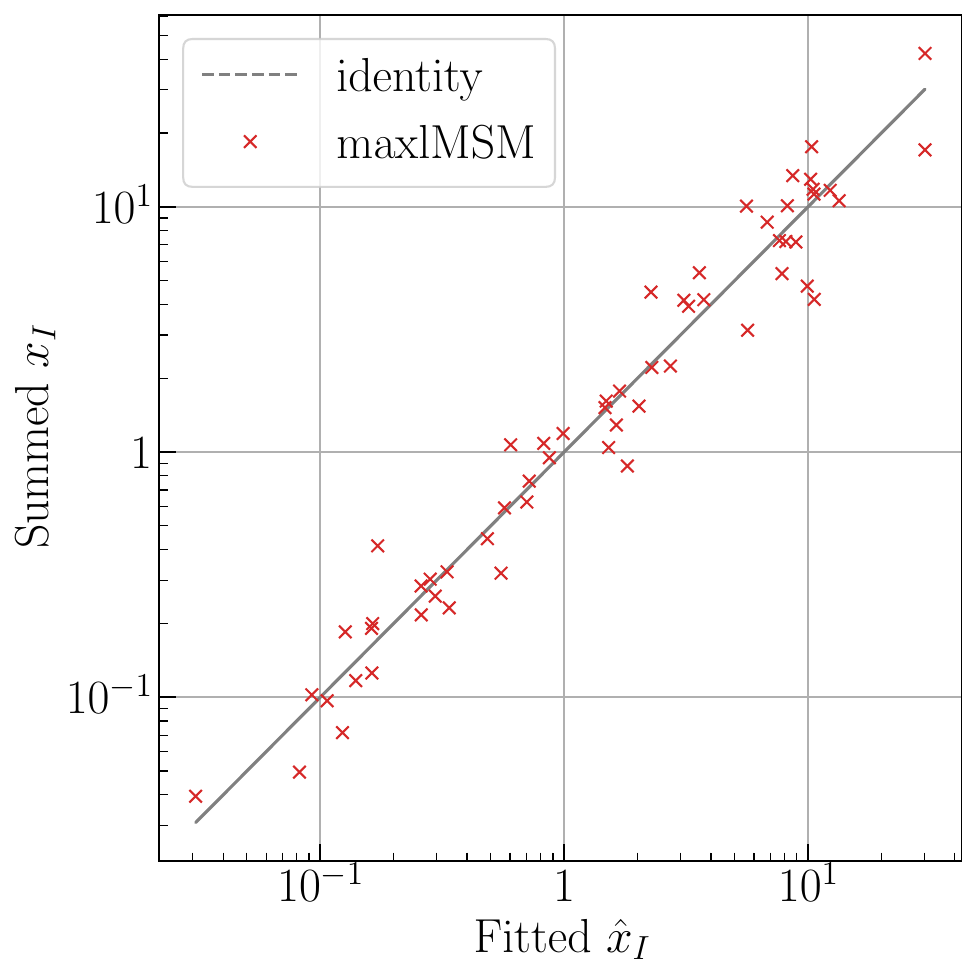}
        }
    \caption{
    Summed (y-axis) VS Fitted (x-axis) Parameters at level 4. From the left, we show the CM, degcMSM and maxlMSM. The identity line illustrates the perfect matching among the two. In summary, the CM resummed parameters at level $ 4$ are underestimating the fitted counterpart, whereas the MSM-based models are showing a better agreement among summed and fitted weights.
    }
    \label{fig:WTW_sumx_vs_fitx}
\end{figure*}

\setlength{\temp}{.5\linewidth}
\begin{figure}[t]
    \subfloat[Level 0]{\includegraphics[width=\temp]{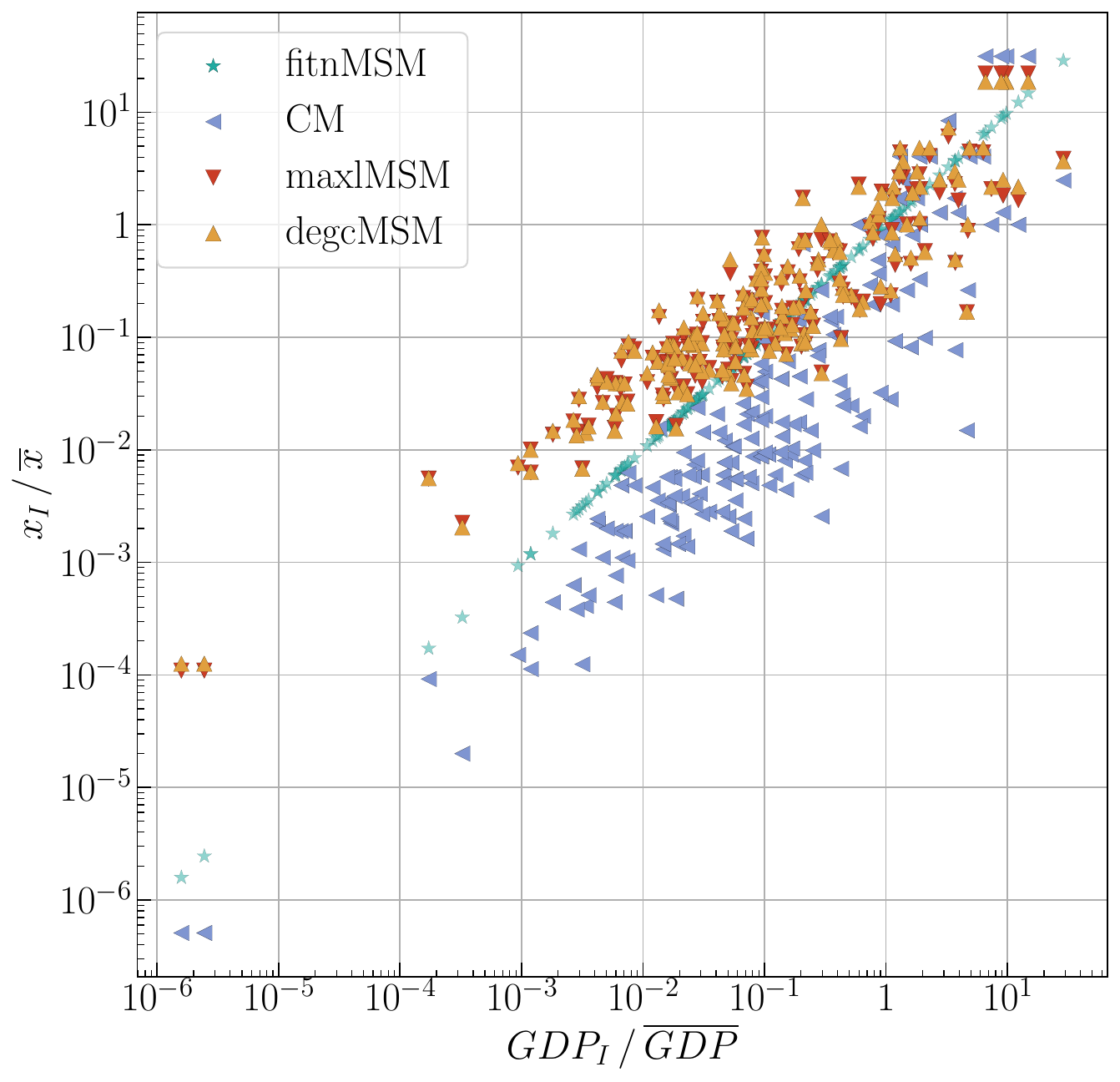}
        }
    \hfill
    \subfloat[Level 2]{\includegraphics[width=\temp]{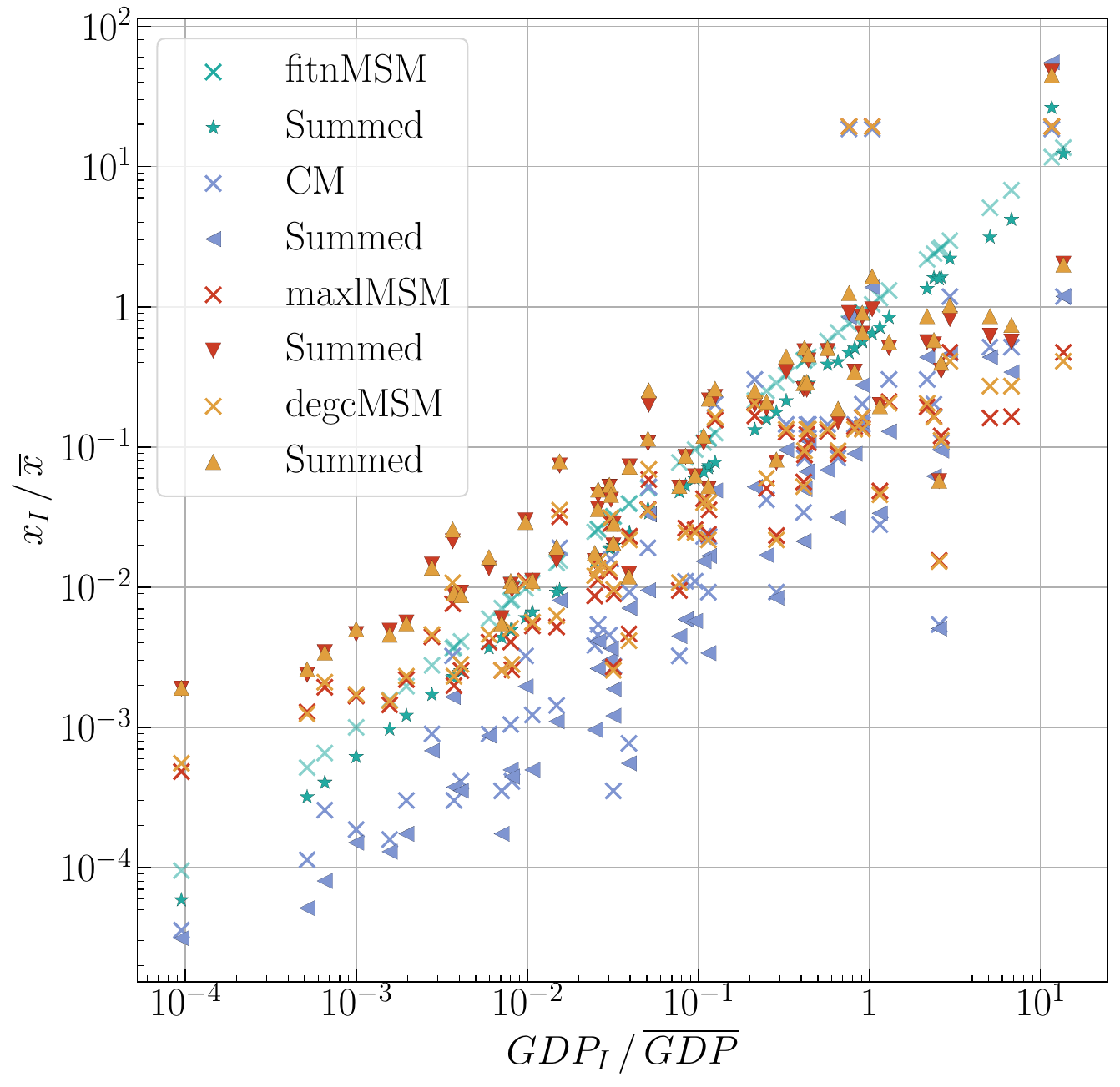}
        }
\caption{
    Rescaled scalar parameters against the rescaled GDPs at level $ 0$ and $ 2$. For level $ 0$, the weights are directly learned from the topology, whereas at level $ 2$ they are calculated by summing the microscopic parameters. Differently from the ION, here the GDPs are not well approximating the topological weights for the CM or MSM. So, the fitness ansatz is a weak approximation for both models.
    }
    \label{fig:WTW_sumXvsStrengths}
\end{figure}

\begin{figure}[t]
    \centering
    \includegraphics[width=\temp]{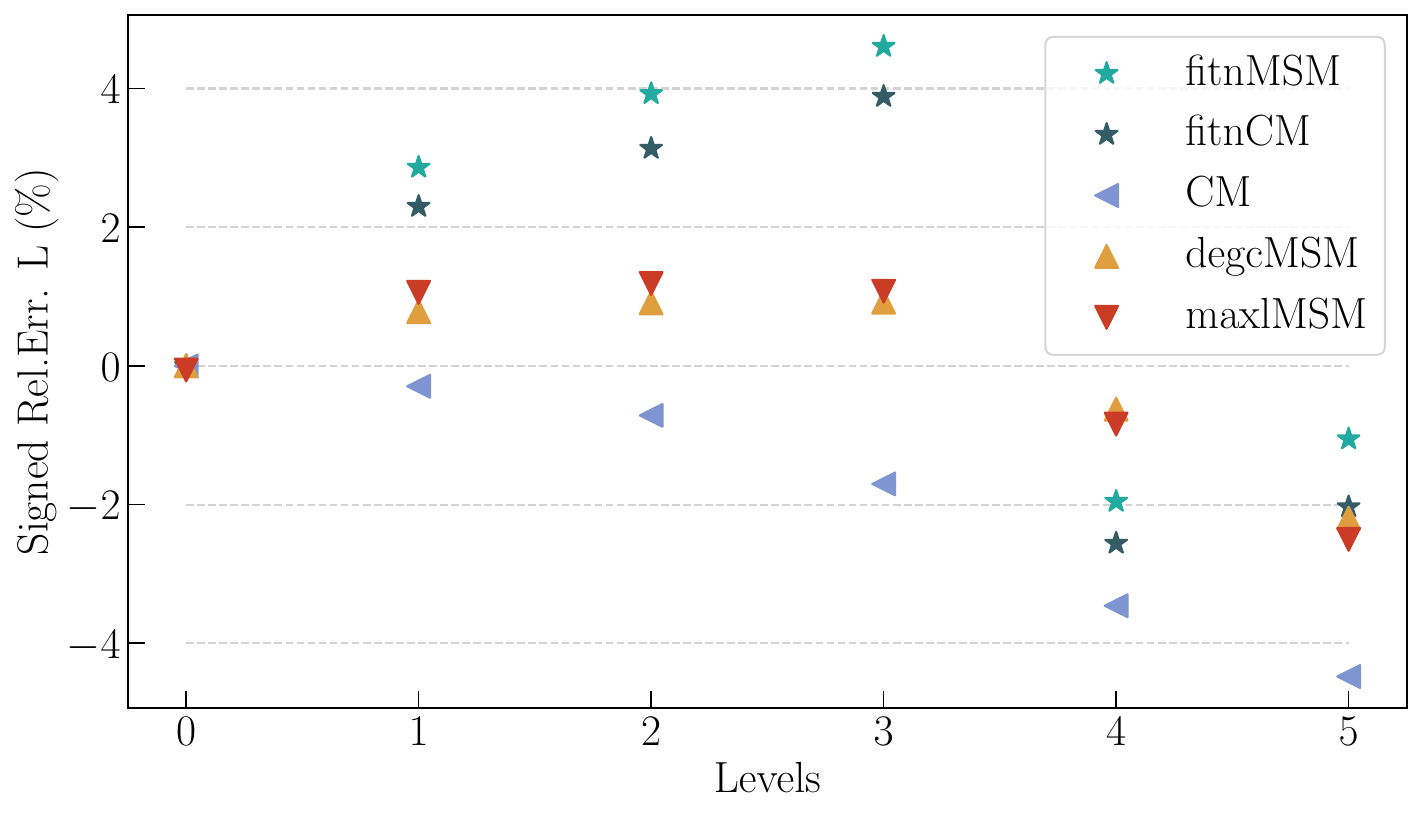}
    \caption{Evolution of the signed relative error accross scales. The error vanishes for level $ 0$, where the models have been fitted. For the higher scales, we used the summed parameters as described in the main text. The CM consistently underestimates the number of links at higher levels, whereas the degcMSM reaches better estimations in the long run. This suggests that a single-scale model may lead to better results for subsequent levels, while the improvement of MSM emerges eventually. Both the fitness-based model, overestimates the number of links of at least the $ 2\%$ in absolute values. This behavior was expected as all the summed fitness were lower than the fitted counterpart (see \autoref{fig:WTW_sumx_vs_fitx}). Overall, the degcMSM has the lowest value of the signed relative error.}
    \label{fig:WTW_rel_err_n_edgess_across_levels}
\end{figure}

\setlength{\temp}{.8\linewidth}
\begin{figure*}[p]
    \centering
    \subfloat[Multi-Level Configuration Model \label{fig:WTW_net_meas_by_level_CM}]{\includegraphics[width=\temp]{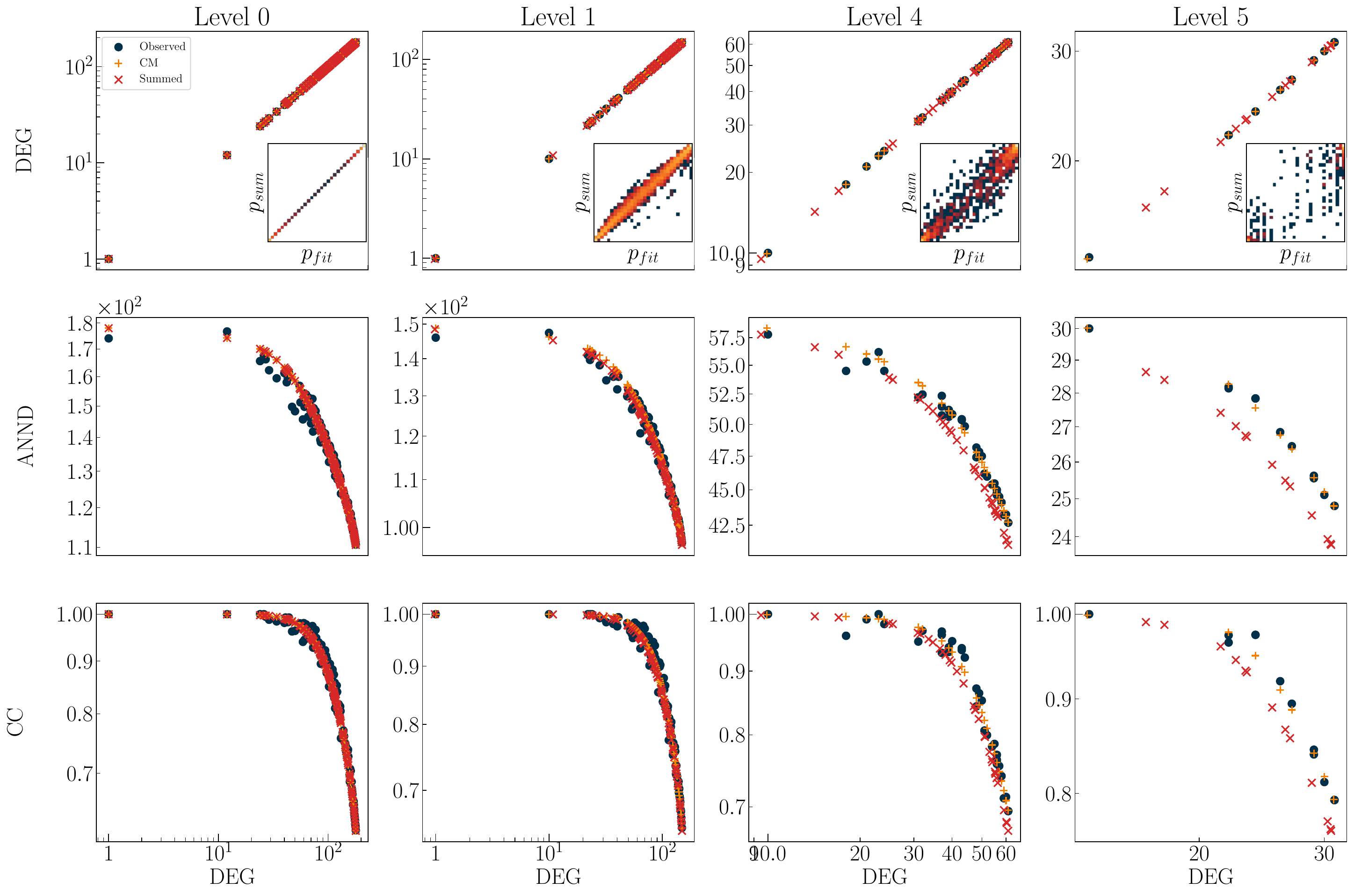}}
    \\[4ex]
    % \vfill %\floatsep% normal separation between figures
   \subfloat[Multi-Level Degree-Corrected MSM \label{fig:WTW_net_meas_by_level_degcCM}]{\includegraphics[width=\temp]{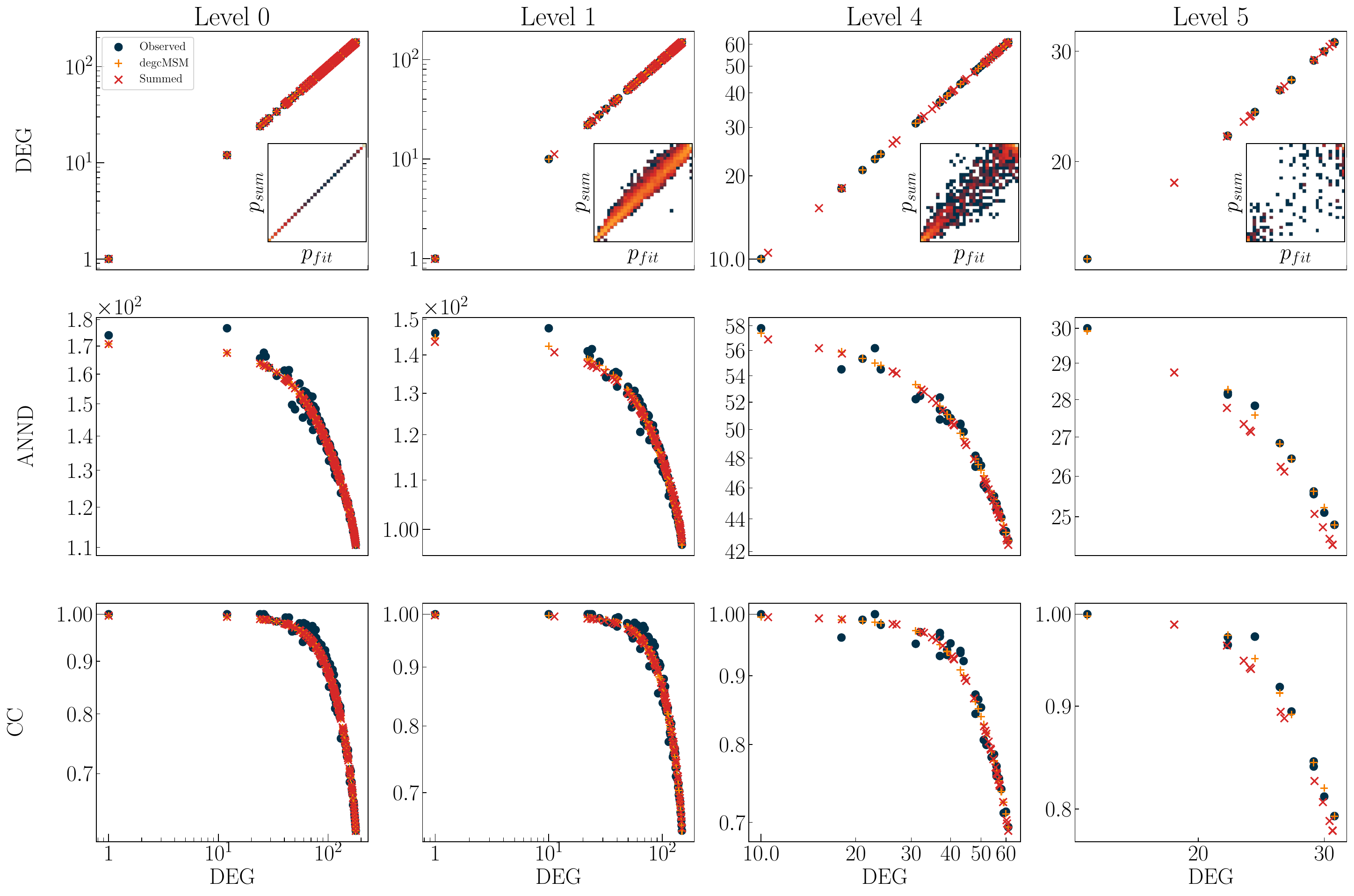}}
    \caption{Multi-Scale comparison of the Configuration Model and degree-corrected Multi Scale Model. Both have been fitted at level $ 0$, whereas the upper-scale parameters have been obtained undersummation. In the inset of DEG, we reported the values of the summed probability against the fitted ones. The DEG, ANND and CC metrics have been obtained by precisely using those probabilities. The degcMSM is a more consistent model also at higher levels as it deviates less from the observed metrics (see ANND at level 4 or 5). However, the summed CM still produces good estimates of the observed metrics, although it was not inherently renormalizable. This ``apparent'' scale-invariance is a side effect of similar probability functions as described in the main text.
    } 
    \label{fig:WTW_net_meas_by_level_CM_degcMSM}
\end{figure*}

\begin{figure*}[htbp]
    \centering
    \includegraphics[width=\linewidth]{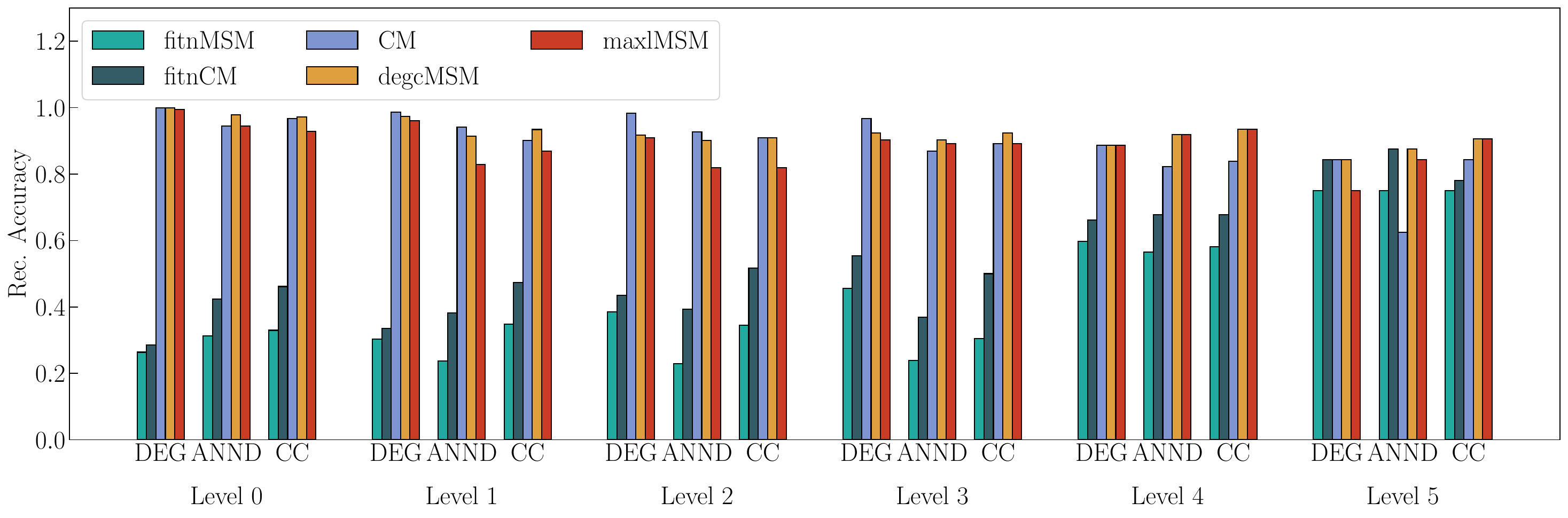}
    \caption{Reconstruction Accuracies of DEG, ANND and CC for all the models and levels. The models are fitted at level $ 0$, whereas at higher level we applied their summed version. Similarly to the ION case, the dispersion intervals are centered in the ensemble average and wide $ 2 \sigma$ from positive and negative side. The fitness-based models are not capable of describing the observed measures in a satisfying way. Moreover, the increasing of reconstruction accuracy with the scale is due to two effects: (1) one matched observation counts more and (2) there is less variability to be grasped, in other words, the system is less complex. Overall, both the CM and the degcMSM are good candidate for the Gleditsch WTW. On the other hand, based on the previous results, the degcMSM is the best performer between the twos.}
    \label{fig:WTW_rec_acc_by_level}
\end{figure*}

\begin{figure*}[htbp]
    \centering
    \subfloat[Evolution of the area Under the ROC and PR curves (y-axis) for the fitted models (level $ 0$) and summed ones (level $ 2$) as the scales increase (x-axis) or, equivalently, the numbers of nodes diminish. Differently from the ION case, the AUCs of the CM are distinguishable from the MSM-based models and the AUCs are not always increasing. This suggests that aggregated levels are less predictable than the lower resolution ones, although the number of observable (links) diminish.
    \label{fig:WTW_auc_roc_prc}]
    {\includegraphics[width=.9\linewidth]{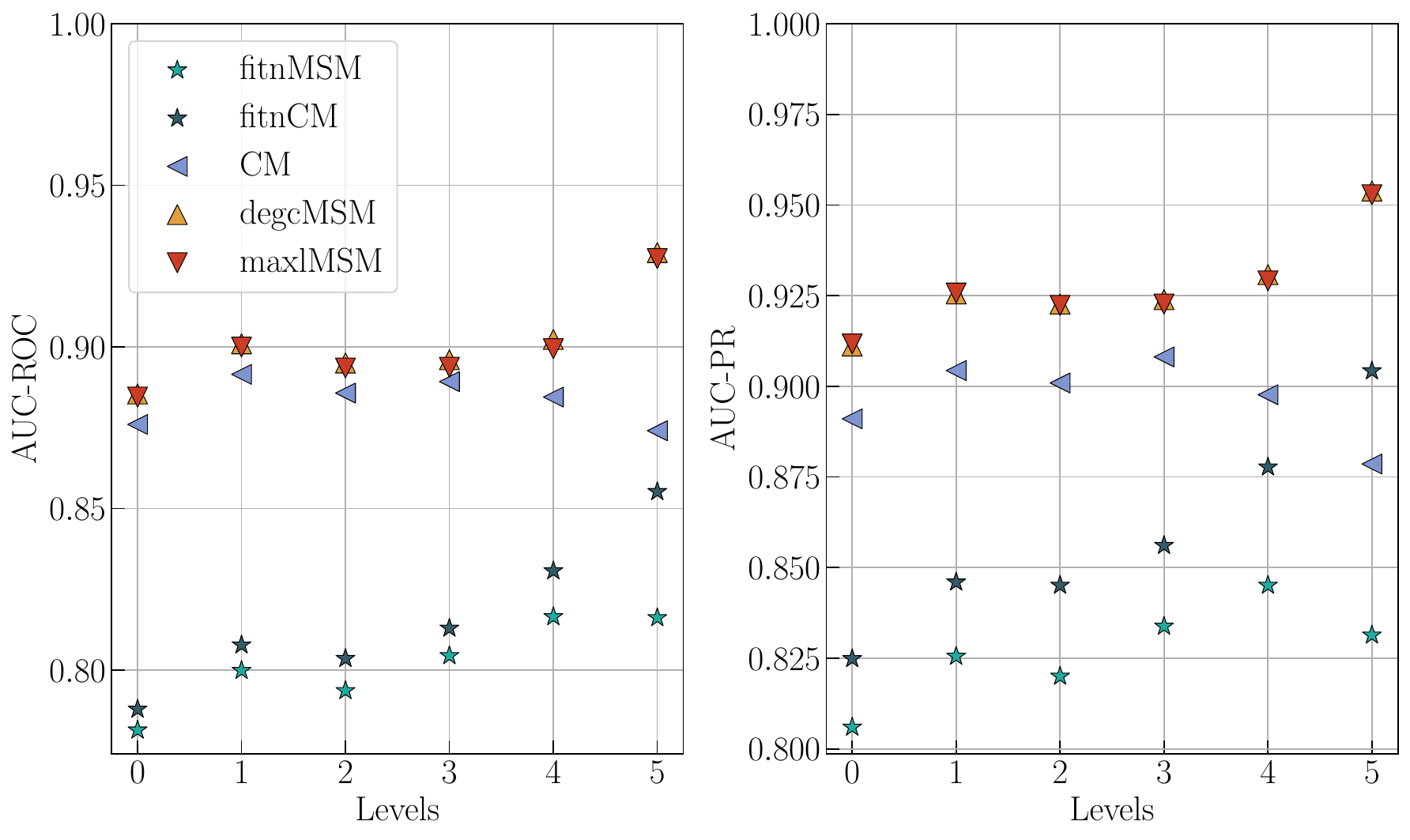}
    }
\end{figure*}

\setlength{\temp}{.8\linewidth}
\begin{figure*}[htbp]
    \centering
    \subfloat[Level 0 \label{fig:WTW_level0_triangles}]{\includegraphics[width=\temp]{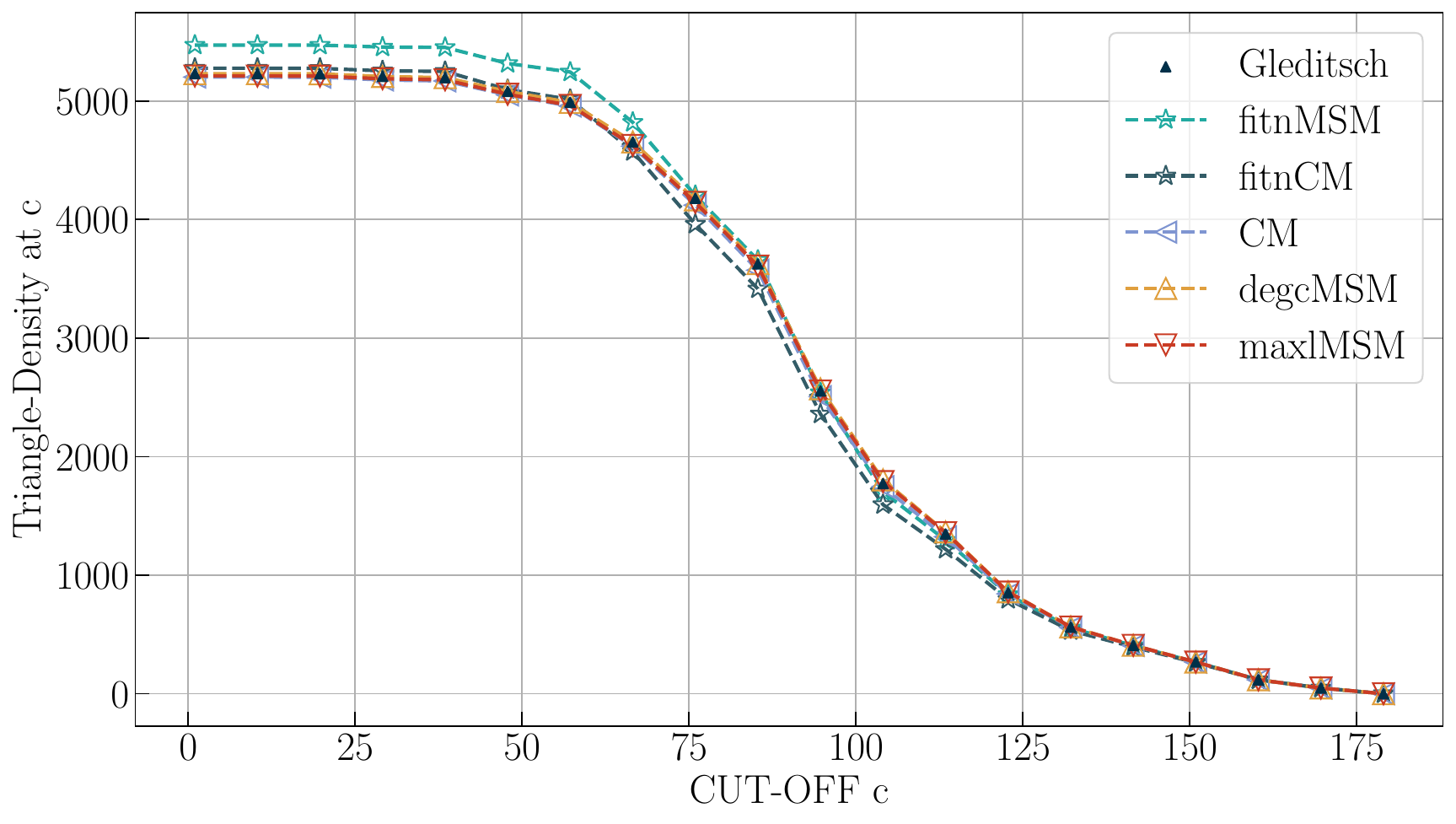}}
    \hfill
    \subfloat[Level 4 \label{fig:WTW_level4_triangles}]{\includegraphics[width=\temp]{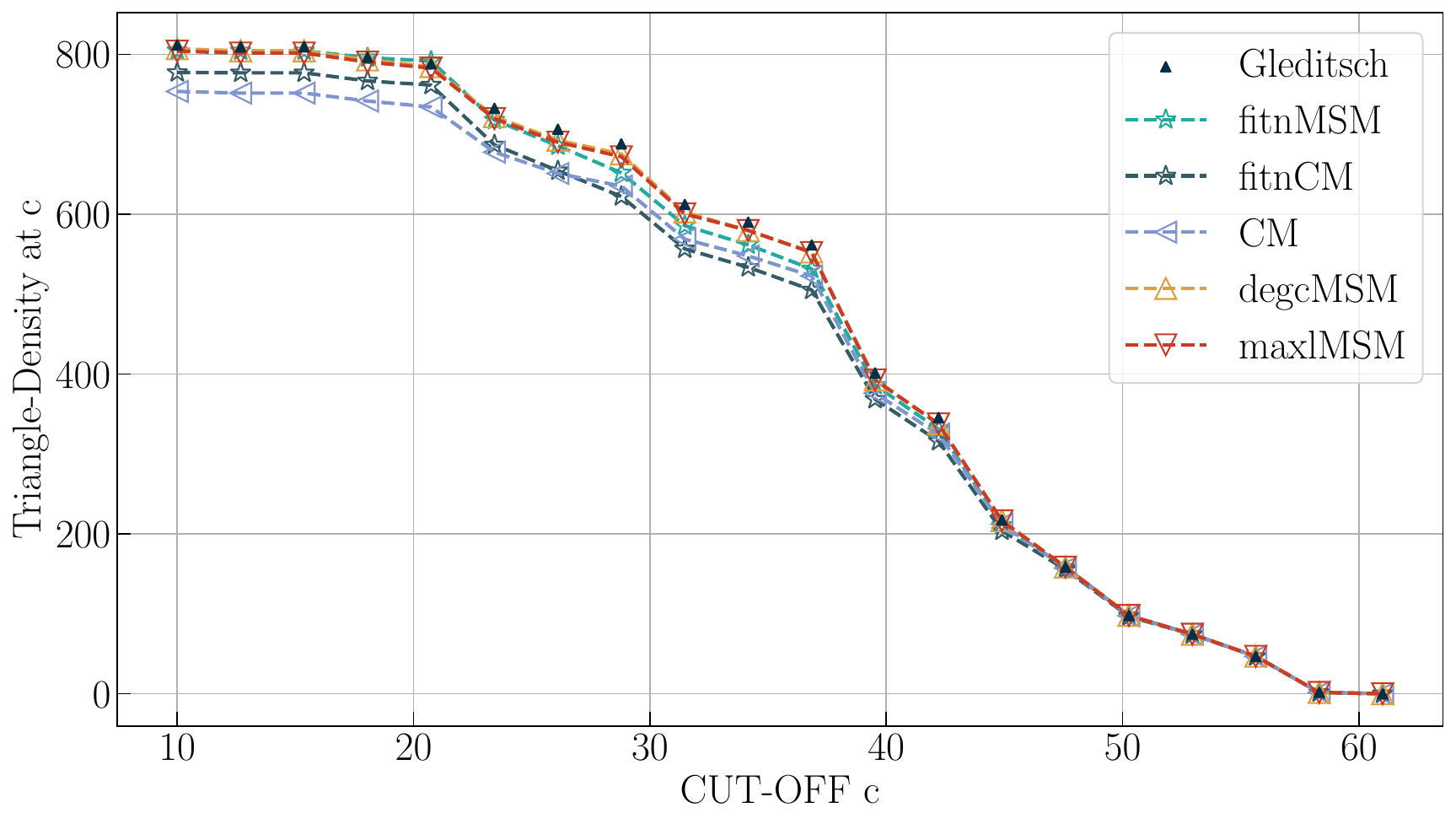}}
    \caption{
        Comparison of the triangle density on the subgraph obtained by filtering the nodes with a lower degree tha $ c$. The curves at scale $ 0$ involves models fitted at that level, whereas at resolution $ 2$ we used the summed model.
        }
    \label{fig:WTW_levels04_triangles}
\end{figure*}

\end{document}